\documentclass{pasa}
\usepackage{rotating}
\usepackage{epsfig}
\usepackage{graphicx}
\title[$R_{\rm e}$. I. Galaxy sizes]
{$R_{\rm e}$. I. Understanding galaxy sizes, associated luminosity
  densities, and the artificial division of the early-type galaxy population}

\author[A.\ W.\ Graham]{Alister W.\ Graham\thanks{AGraham@swin.edu.au}, 
\affil{Centre for Astrophysics and Supercomputing, Swinburne 
University of Technology, Hawthorn, Victoria 3122, Australia.}
}

\jid{PASA}
\doi{10.1017/pas.\the\year.xxx}
\jyear{\the\year}

\begin{document}
\begin{abstract}


For decades, the deceptive simplicity of the radius $R_{\rm e}$, enclosing an
arbitrary 50 percent of a galaxy's light, has hamstrung the understanding of
early-type galaxies (ETGs).  Half a century ago, using these ``effective
half-light'' radii from de Vaucouleurs' $R^{1/4}$ model, S\'ersic reported
that bright ETGs follow the relation $\mathfrak{M}_B\propto2.5\log R_{\rm e}$;
and consequently one has that $\langle\mu\rangle_{\rm e}\propto2.5\log R_{\rm
  e}$ and $\mu_{\rm e}\propto2.5\log R_{\rm e}$, where $\mu_{\rm e}$ and
$\langle\mu\rangle_{\rm e}$ are the effective surface brightness at $R_{\rm
  e}$ and the mean effective surface brightness within $R_{\rm e}$,
respectively.  S\'ersic additionally observed an apparent transition which led
him to advocate for a division between what he called dwarf and giant ETGs; a
belief frequently restated to occur at $\mathfrak{M}_B \approx -18$~mag or
$n\approx 2.5$.  Here, the location of this false dichotomy in diagrams using
``effective'' parameters is shown to change by more than 3~mag simply
depending on the arbitrary percentage of light used to quantify a galaxy's
size.  A range of alternative radii are explored, including where the
projected intensity has dropped by a fixed percentage, plus a battery of
internal radii, further revealing that the transition at $\mathfrak{M}_B
\approx -18$~mag is artificial and does not demark a boundary between
different physical processes operating on the ETG population.

The above understanding surrounding 
half-light radii (either model-independent or 
effective radii $R_{\rm e}$ from models such as the $R^{1/n}$ model) 
is of further importance because quantities such as dynamical mass $\sigma^2R/G$,
gravitational binding energy $GM^2/R$, acceleration $GM/R^2$, and the
``Fundamental Plane'' depend systematically on the arbitrary percentage of light used to
define $R$, with implications for dark matter estimates, galaxy formation
theories, compact massive galaxies, studies of peculiar velocity flows, and
more.  Finally, some of the vast literature which has advocated for
segregating the ETG population at $\mathfrak{M}_B \approx -18$~mag
($M\approx1$--$2\times10^{10}\,M_{\odot}$) is addressed, and it is revealed how this
pervasive mindset has spilled-over to influence both the classical bulge
versus pseudobulge debate and recently also correlations involving
supermassive black hole masses.

\end{abstract}


\maketitle

\section{Introduction}\label{Sec_intro}

During the first half of the 20$^{\rm th}$ century, astronomers developed 
several empirical functions to describe the observed, i.e.\ projected on the
plane of the sky, radial distribution of light in external galaxies.  These
functions provided physical measurements which enabled astronomers to better 
answer simple questions such as, How big is it, and, How bright is it?  This
helped to place extragalactic astronomy on a more scientific footing, elevating
some sky surveys above the somewhat derogatory status of ``stamp collecting''.

For both early-type galaxies (ETGs), and late-type galaxies (LTGs), these
mathematical functions had two parameters: one stretched the model light
profile along the horizontal (radial) axis, and the other stretched it along
the vertical (intensity) axis.  One could arbitrarily set the scale radius to
be where the intensity had dropped by some fixed factor from the central
intensity, or it could be set as the radius effectively enclosing some fixed
fraction of the total light, such as 50 or 90 percent.  Due to the
homologous nature of these two-parameter models, defining the scale radius or
the scale intensity in a different way would shift all galaxies
equally in diagrams involving the logarithm of these model-determined
quantities.  As such, trends and patterns in such diagrams were not dependent
on how these scale parameters were set.  However, if galaxies are not
adequately described by these two-parameter functions, then the galaxies'
distribution in the scaling diagrams can become a function of the
arbitrarily defined scale-radius and scale-intensity.

The above fact, and the implications of the above fact, have not been
adequately realised in the literature, and countless papers have drawn
questionable scientific conclusions based upon the distribution of galaxies in
diagrams involving a galaxies' arbitrary 50 percent radius and the intensity
associated with this radius.  Given that this has gone on for decades, this
paper goes to some length to try and carefully explain the curved distribution of ETGs
in diagrams involving effective half light parameters. These curved
distributions have been used many times in the literature to argue for a
distinct divide among the ETG population into dwarf and giant\footnote{In the
  literature, ``giant'' ETGs are also referred to as ``ordinary'' or
  ``normal'' ETGs, or often simply as ``elliptical'' (E) galaxies, as distinct
  from the ``dwarf elliptical'' (dE) galaxies fainter than $\mathfrak{M}_B =
  -18$~mag (Sandage \& Binggeli 1984, using $H_0=50$ km s$^{-1}$ Mpc$^{-1}$),
  or two-thirds of a magnitude fainter using $H_0=67.8$ km s$^{-1}$ Mpc$^{-1}$
  (Planck et al.\ 2016).} ETGs separated at the bend midpoint: $\mathfrak{M}_B
\approx -18$~mag (e.g.\ Kormendy et al.\ 2009, hereafter K09; Kormendy \&
Bender 2012; Kormendy 2016; Tolstoy et al.\ (2009); Somerville \& Dav\'e
2015).  In order to help better appreciate this issue, and more fully
understand galaxy structure, the curved distributions of ETGs in diagrams
involving radii that enclose different percentages of the total light are
presented, and it is revealed how the absolute magnitude associated with the
midpoint of the bend changes considerably.

Advocates for an ETG dichotomy have alleged that the formation physics must be
dramatically different for ETGs fainter and brighter than $\mathfrak{M}_B
\approx -18$~mag, because the slope of certain scaling relations is different
at magnitudes fainter and brighter than this.  For example, Kormendy \&
Djorgovski (1989; their section~8) wrote, ``A fundamental application of
parameter correlations has been the demonstration that diffuse dwarf
spheroidal\footnote{Kormendy \& Djorgovski (1989) consider ``dwarf
  spheroidal'' galaxies to appear at $\mathfrak{M}_B \gtrsim -18$~mag, rather
  than the more commonly used values of $\approx -13\pm$~mag.  In
  order to better emphasize their view of a distinct population from the
  ``ordinary elliptical'' galaxies brighter than $\mathfrak{M}_B \approx
  -18$~mag, they use this term to describe what others call ``dwarf
  elliptical'' galaxies ($-13 \gtrsim \mathfrak{M}_B \gtrsim -18$~mag).}  
galaxies are a family of objects unrelated to
ellipticals''. This claim was, however, at odds with other research that did
not use effective half light parameters and which instead advocated for a
continuity among the ETG population at $\mathfrak{M}_B \approx -18$~mag
(e.g.\ Caldwell 1983a, his Figure 6; Binggeli et al.\ 1984; Sandage et
al.\ 1985; Binggeli 1985; Bothun et al.\ 1986, their Figure~7; Caldwell \&
Bothun 1987).

As noted by James (1994), the shape of ETG light profiles had also been
considered one of the principal differences separating dwarf and ordinary ETGs
--- with ``dwarf'' ETGs having exponential light profiles
(similar to the disks of LTGs), and ``ordinary'' ETGs having $R^{1/4}$ profiles
--- emboldening those interpreting transitions in certain scaling diagrams as
evidence of different formation physics at magnitudes fainter and brighter
than $\mathfrak{M}_B \approx -18$~mag.  However, as we shall see, the
systematically changing (with absolute magnitude) shape of the ETG light
profile, i.e.\ structural non-homology, is key to understanding the
unification of dwarf and ordinary ETGs.

To understand the mechanics of the structural parameter scaling diagrams,
Section~2 of this paper provides a context-setting background using de
Vaucouleurs' $R^{1/4}$ model and S\'ersic's $R^{1/n}$ model, and provides a
familiarity with the model parameters $R_{\rm e}$ and both the surface
brightness at $R_{\rm e}$, denoted by $\mu_{\rm e}$, and the average surface
brightness within $R_{\rm e}$, denoted by $\langle\mu\rangle_{\rm e}$.
Section~3 then presents two key empirical relations, providing the foundation
for the insight which follows.

Equipped with the above background knowledge, Section~4 presents an array of
scaling relations based on radii and surface brightnesses which effectively
enclose different fixed percentages of the galaxy light.  It soon becomes
apparent why the $\mu_{\rm e}$--$R_{\rm e}$ relation is itself quite tight for
bright ETGs, but not for faint ETGs.  Section~\ref{Sec_Alt} then goes on to
explore a range of alternative radii, and surface brightnesses.  In
particular, radii where the intensity has dropped by a fixed percentage are
introduced, and the use of isophotal radii is revisited in
Section~\ref{Sec_iso}.  Section 5 expands on the analysis using internal radii
that define spheres which effectively enclose a fixed percentage of the galaxy
light.  These internal radii include ``effective'' radii, plus the new radii where
the internal density has declined by a fixed amount, isodensity radii, virial
radii, and new Petrosian-like radii.  The changing location of the bend
midpoint in various scaling relations reveals that it has nothing to do with
changing physical processes but is instead merely a result of the arbitrary
definition used two quantify the sizes of ETGs.

Section~\ref{Sec_Case} presents ETG data from Ferrarese et al.\ (2006) and
K09, and resolves the different interpretations given in
those papers.  Finally, a discussion in Section~\ref{Sec_Dis} 
broaches some of the literature which has 
advocated for a dichotomy of the ETG population at
$\mathfrak{M}_B \approx -18$~mag.  Considerable 
historical context is included to aid
the reader in understanding how the topic evolved.  This is also partly necessary
because support for interpreting these curved relations, in terms of
different formation processes at magnitudes brighter and fainter than the bend
midpoint at $\mathfrak{M}_B \approx -18$~mag, attracted a range of bright ideas over
the years and many of these are sometimes heralded without adequate
qualification.  Some of the literature surrounding the
similar separation of bulges into ``classical'' or ``pseudobulge'' is also
discussed.  
Bulge scaling relations, as distinct from ETG scaling relations, are also
discussed in 
the context of high-$z$ compact massive systems, which by all accounts appear
to be the bulges of massive local galaxies.  In addition, 
subsection~\ref{Sec_planes} reveals why the ``Fundamental Plane'', 
involving the velocity 
dispersion $\sigma$ (Djorgovski \& Davis 1987; see also Fish 1963), is tighter
than the $\mu_{\rm e}$--$R_{\rm e}$ relation for ordinary
ETGs, and a warning about fitting and interpreting 2D planes to
curved distributions involving supermassive black hole mass and ``effective''
parameters is also issued.

\section{Mathematical background} 

\subsection{de Vaucouleurs' $R^{1/4}$ model}

First in French (de Vaucouleurs 1948), then in English, de Vaucouleurs (1953) 
presented an empirical function that was to 
become known as the $R^{1/4}$ model due to how the projected
(on the plane of the sky) 
intensity profile $I(R)$ depends on the projected radius $R$ raised to the 1/4
power.  This mathematical model can be expressed as, 
\begin{equation}
I(R)=I_0\exp\left[ -b\left(\frac{R}{R_{\rm s}}\right) ^{1/4}\right]
 = \frac{I_0}{\left( {\rm e}^{b}\right)^{\left(R/R_{\rm s} \right)^{1/4}}}, 
\label{Eq_deV} 
\end{equation}
where $R_{\rm s}$ is a scale radius, $I_0$ is a scale
intensity at $R=0$, and $b$ is a constant that shall be explained below. 
Given that galaxies do not have clear edges --- and in the middle of
the 20th century it was not
known how their radial profiles behaved at large radii --- the practice was to
extrapolate one's adopted model to infinity in order to determine a galaxy's
total luminosity. 

The projected luminosity (from three-dimensions to two-dimensions, 
onto the plane of the sky)
interior to a circle of radius $R$, is determined by integrating
the intensity over the enclosed area, such that
\begin{equation}
L(<R) = \int_0^R I(R^{\prime})2\pi R^{\prime} {\rm d}R^{\prime}. 
\end{equation} 
Using the substitution $x=b(R/R_{\rm s})^{1/4}$ in equation~\ref{Eq_deV}, the
above integral reduces to 
\begin{equation}
\label{Eq_L4}
L(<R)=\frac{I_0 R_{\rm s}^{2}8\pi}{b^8}\gamma (8,x),
\end{equation}
where $\gamma $(8,x) is the incomplete gamma function defined by
\begin{equation}
\gamma (8,x)=\int ^{x}_{0} {\rm e}^{-t}t^{8-1}{\rm d}t.
\label{gamFunc}
\end{equation}
As noted, the total luminosity is obtained by integrating to infinity, in which case
$\gamma (8,x)$ is replaced with the complete gamma function, $\Gamma (8)$, and
one has that 
\begin{equation}
\label{Eq_L4T}
L_{\rm tot} = \frac{I_0 R_{\rm s}^{2}8\pi}{b^8}\Gamma (8).
\end{equation}

Now, here is where things can, and did, become arbitrary. 
Gerard de Vaucouleurs elected to define the radius $R_{\rm s}$ such that it
enclosed 50 percent of the total light $L_{\rm tot}$. 
He did this by determining the value of $b$
required to balance the equation 
\begin{equation}
\gamma (8,b) = 0.5\,\Gamma (8).
\label{Eq_g8}
\end{equation}
With $b=-7.669$, the projected radius $R_{\rm s}$ effectively 
encloses half of the model's total light, and it was subsequently denoted $R_{\rm e}$ and
referred to as the ``effective half light radius''.  
The $R^{1/4}$ model's central surface brightness, $\mu_0$, is given by $-2.5\log I_0$, and 
the projected intensity at $R=R_s \equiv R_{\rm e}$ is given by 
\begin{equation}
I_{\rm e}=I_0{\rm e}^{-b} = I_0/2141.
\label{Eq_2141}
\end{equation}
The average intensity $\langle I \rangle_{\rm e}$ within $R_{\rm e}$ is such that 
\begin{equation}
0.5\,L_{\rm tot} = \pi R_{\rm e}^2 \langle I \rangle_{\rm e}, 
\end{equation}
and it can be shown that 
\begin{equation}
\langle I \rangle_{\rm e} = 3.61 I_{\rm e} = I_0/594
\end{equation}
(Graham \& Driver 2005, their equations~7 and 9). 

As alluded to above, 
de Vaucouleurs {\it could} have chosen a radius enclosing any fraction of the
light, and his two-parameter model would still have the same functional form
(equation~\ref{Eq_deV}).  
That is, one could use
a radius $R_X$ containing any percentage of the total light, and one
could use the intensity $I_Y$ taken from any (similar or different) fixed radius (in units of
$R_{\rm e}$). 
The homology of the $R^{1/4}$ model is such that $R_X = C_1 R_{\rm e}$, 
$I_Y = C_2 I_{\rm e}$, 
and $\langle I \rangle _Y  = C_3 \langle I \rangle_{\rm e}$, 
where $C_1$, $C_2$ and $C_3$ are constants. 
In trying to understand the behavior of, and connections between, galaxies,
astronomers could plot $\log R_X$ versus $-2.5\log I_Y$, and 
versus $-2.5\log \langle I \rangle _Y$, and the trends would be the same as
obtained when using $R_{\rm e}$, $I_{\rm e}$ and $\langle I \rangle_{\rm e}$, 
just shifted vertically or horizontally in one's diagram.  
As such, the arbitrary selection of 50 percent by de Vaucouleurs did not
appear to matter.  
To give a more concrete example, de Vaucouleurs could have set the scale radius $R_s
= R_{10}$, i.e.\ enclosing 10 percent of the total light (e.g.\ Farouki et
al.\ 1983). The mean intensity
$\langle I \rangle_{10}$ within this radius is related by the expression 
\begin{equation}
\label{Eq_L1}
0.1\,L_{\rm tot} = \pi R_{10}^2 \langle I \rangle_{10}, 
\end{equation}
and the associated value of $b$ is obtained by solving the equation
\begin{equation}
\gamma (8,b) = \Gamma (8)/10, 
\label{eq_gamma8}
\end{equation}
to give $b=4.656$ and $I_{10}=I_0{\rm e}^{-b} = I_0/105.2$   
(cf.\ equation~\ref{Eq_2141}).  In this example, 
de Vaucouleurs' model would then read
\begin{equation}
I(R)=I_0\exp\left[ \frac{-4.656}{R_{10}^{1/4}} R^{1/4}\right], 
\label{Eq_deV10}
\end{equation}
where $R_{10} = (4.656/7.669)^4\, R_{\rm e} = R_{\rm e}/7.361$, and 
$\langle I \rangle_{10}=I_0/54.77$.  

However, and this is the crux of the matter: ETGs, and also the
bulges of spiral galaxies, do
not follow the $R^{1/4}$ model; that is, there is not structural homology.
This has important consequences when
using radii enclosing a fixed percentage of the total light, and when using the
associated surface brightness terms.  

It is noted that the $R^{1/4}$ model 
had become so entrenched during the second half of the 20th century that it
was invariably referred to as the $R^{1/4}$ {\em law}. 
That is, this empirical model was 
effectively elevated to the status of a physical law because it was thought
that all ETGs did have $R^{1/4}$ light profiles.  
Indeed, it was not uncommon for astronomers to vary the sky-background in
order to make their light profiles more $R^{1/4}$-like (e.g.\ 
Tonry et al.\ 1997; see also the ``Seven Samurai'' team data from Burstein et al.\ 1987 as
presented in D'Onofrio et al.\ 1994, their Figure~4).  
This belief 
was in part because de Vaucouleurs (1959) had shown that the $R^{1/4}$ model fit better
than the popular Reynolds' (1913) model\footnote{Reynolds' (1913) model was 
later referred to as Hubble's (1930) model.}, and because of de Vaucouleurs 
\& Capaccioli's (1979) study of NGC~3379 which revealed that its light profile is
remarkably well fit by the $R^{1/4}$ model over an extensive range in surface
brightness (see also Fish 1964 in the case of M87 and M105). 
However, Caon et al.\ (1990, 1993, 1994) and
D'Onofrio et al.\ (1994), both co-authored by Capaccioli, subsequently revealed 
that other ETGs, with different absolute magnitudes, are equally well fit
down to $B$-band surface brightnesses of $\sim$28 mag arcsec$^{-2}$ when 
using exponents in the light profile model that are different to the value of
1/4.

\subsection{S\'ersic's $R^{1/n}$ model} 
 
Today, it is widely recognised\footnote{See the reviews by Graham 2013, 2016,
  and references therein.}  that ETGs --- and the bulges of spiral galaxies
--- display a range of light profile shapes that are better represented by a
generalised version of the $R^{1/4}$ model, referred to as S\'ersic's (1963)
$R^{1/n}$ model, in which the exponent $1/n$ can take on a range of values
other than just 1/4. This realisation applies to not just the ordinary ETGs
(e.g.\ Caon et al.\ 1993; D'Onofrio et al.\ 1994), but also the dwarf ETGs
(e.g.\ Davies et al.\ 1988; Cellone, Forte, \& Geisler 1994; James 1994; Vennik
\& Richter 1994; Young \& Currie 1994, 1995) which had previously been fit
with an exponential model (e.g.\ Faber \& Lin 1983; Binggeli et al.\ 1984).
Despite this, the early assumption of structural homology for dwarf ETGs
versus a different structural homology for giant ETGs had been sown into the
astronomical literature and psyche.  Moreover, the implications of a varying
exponent upon the use of the arbitrary 50 percent half light radius, and the
associated surface brightness terms, remained poorly recognised. 

Jos\'e S\'ersic's (1963, 1968a) $R^{1/n}$ model, which was introduced in
Spanish, is a generalisation of de Vaucouleurs' $R^{1/4}$ model 
such that 
\begin{eqnarray}
I(R) & = & 
I_0\exp\left[ -b_n\left(\frac{R}{R_{\rm e}}\right) ^{1/n}\right] 
 = \frac{I_0}{\left( {\rm e}^{b_n}\right)^{\left(R/R_{\rm e} \right)^{1/n}}}
\nonumber
\\
   & = & I_{\rm e}\exp\left\{ -b_n\left[\left( \frac{R}{R_{\rm e}}\right) ^{1/n}
  -1\right]\right\}.
\label{Eq_Ser}
\end{eqnarray}
The exponent $1/n$, or its inverse $n$, describes the curvature of the light profile.  Within
$\approx 1 R_{\rm e}$, a larger value of $n$ results in a more centrally
concentrated distribution of light, while beyond $\approx 1 R_{\rm e}$, a 
larger value of $n$ results in a less steeply declining light profile.  The
quantity $b_n$ was defined such that $I_{\rm e}$ is, again,  the intensity at the
``effective half light'' radius $R_{\rm e}$ that encloses half of the total 
light (Capaccioli 1989; Ciotti 1991; Caon et al.\ 1993).
The value of $b_n$ is solved via the equation 
\begin{equation}
\gamma(2n,b_n) = 0.5\,\Gamma(2n) 
\label{bsubn}
\end{equation}
(cf.\ equation~\ref{Eq_g8}), and the total luminosity, giving the
total magnitude, is given by
\begin{equation}
\label{Eq_LnT}
L_{\rm tot} = \frac{I_0 R_{\rm e}^{2}2n\pi}{(b_n)^{2n}}\Gamma (2n) 
\end{equation}
(cf.\ equation~\ref{Eq_L4T}). 
For $0.5 < n < 10$, $b_n \approx 1.9992n-0.3271$ (Capaccioli 1989).

However, what was initially
(for the $R^{1/4}$ model) an inconsequential selection of an arbitrary scale
radius enclosing 50 percent of the light now has considerable
consequences given that galaxies do not all have the same light profile shape,
i.e.\ the same value of $n$.  Crucially, 
the ratio between radii containing different fixed percentages of the
projected galaxy light is no longer a constant value --- as we just saw it was for the
$R^{1/4}$ model --- but rather changes with the
S\'ersic index $n$.  Given that ETGs and bulges possess a range of light
profile shapes that are described well by the $R^{1/n}$ model (e.g., Caon et
al.\ 1993; D'Onofrio et al.\ 1994), this remark
about the changing ratio of radii holds even if one does not fit an
$R^{1/n}$ model but instead measures the radii independently of any light
profile model.

\begin{figure*}[ht]
\begin{center}
\includegraphics[angle=270, trim=7.5cm 2.3cm 2.0cm 2.0cm, width=0.8\textwidth]{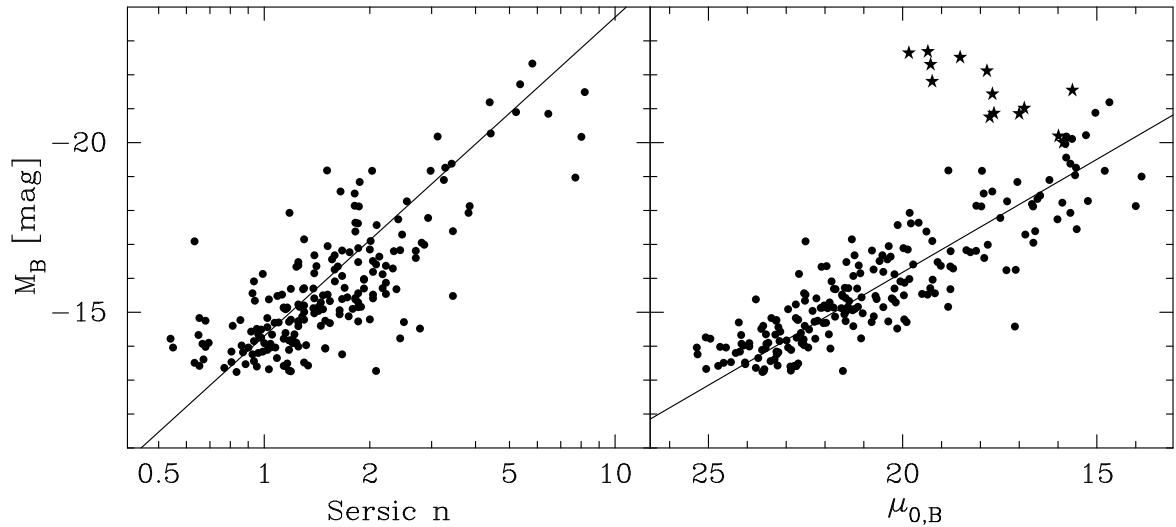}
\caption{Left panel: Absolute $B$-band magnitude (Vega) versus the logarithm
  of the $B$-band S\'ersic index $n$ for ETGs.  Right panel: Absolute
  magnitude versus the $B$-band central surface brightness $\mu_{0,B}$.
  Figure adapted from Graham (2013), with data from Binggeli \& Jerjen (1998),
  Stiavelli et al.\ (2001), Graham \& Guzm\'an (2003), Caon et al.\ (1993),
  D'Onofrio et al.\ (1994), and Faber et al.\ (1997, with stars representing
  their ``core-S\'ersic'' galaxies).  The core-S\'ersic galaxies have partially depleted
  cores with fainter central surface brightnesses than the relation shown (Equation~\ref{Eq_Mmu}).
  However, the inward extrapolation of these galaxies' outer S\'ersic
  profile yields $\mu_{0,B}$ values which follow the relation, as 
  noted by Jerjen \& Binggeli (1997).  }
\label{fig0}
\end{center}
\end{figure*}

What this means is that the distribution of points in scaling
diagrams involving the logarithm of scale radii and scale intensity will look different
depending on what scale radius is used.  That
is, the arbitrary choice of radius, which to date has been the 50 percent
radius,  produces a somewhat arbitrary pattern in diagrams
using $\log R_{\rm e}$, $\mu_{\rm e}$ and $\langle \mu \rangle_{\rm e}$.  
Also apparent, from equation~\ref{Eq_Ser}, is that the scale radius no longer
occurs where the intensity has declined by the same fixed amount, but rather
by different amounts depending on the value of ${\rm e}^{b_n}$ and thus on the
value of $n$.  To quantify this, 
Section~\ref{Sec_eff} will explore scaling diagrams using projected radii
containing fixed percentages of the total light, including 50 percent,
revealing how the bend in scaling relations using ``effective'' parameters changes.
Section~\ref{Sec_Alt} will explore the use of scale radii where the intensity
has dropped by the same amount, yielding monotonic 
size-luminosity relations without the strong bends seen in
Section~\ref{Sec_eff}.

\section{Two key empirical relations: 
$\mathfrak{M}$--$\log\, n$ and $\mathfrak{M}$--$\mu_0$}\label{Sec_TWO}

\begin{figure}[ht]
\begin{center}
\includegraphics[angle=270, trim=2.0cm 4.0cm 2.6cm 10.6cm, width=\columnwidth]{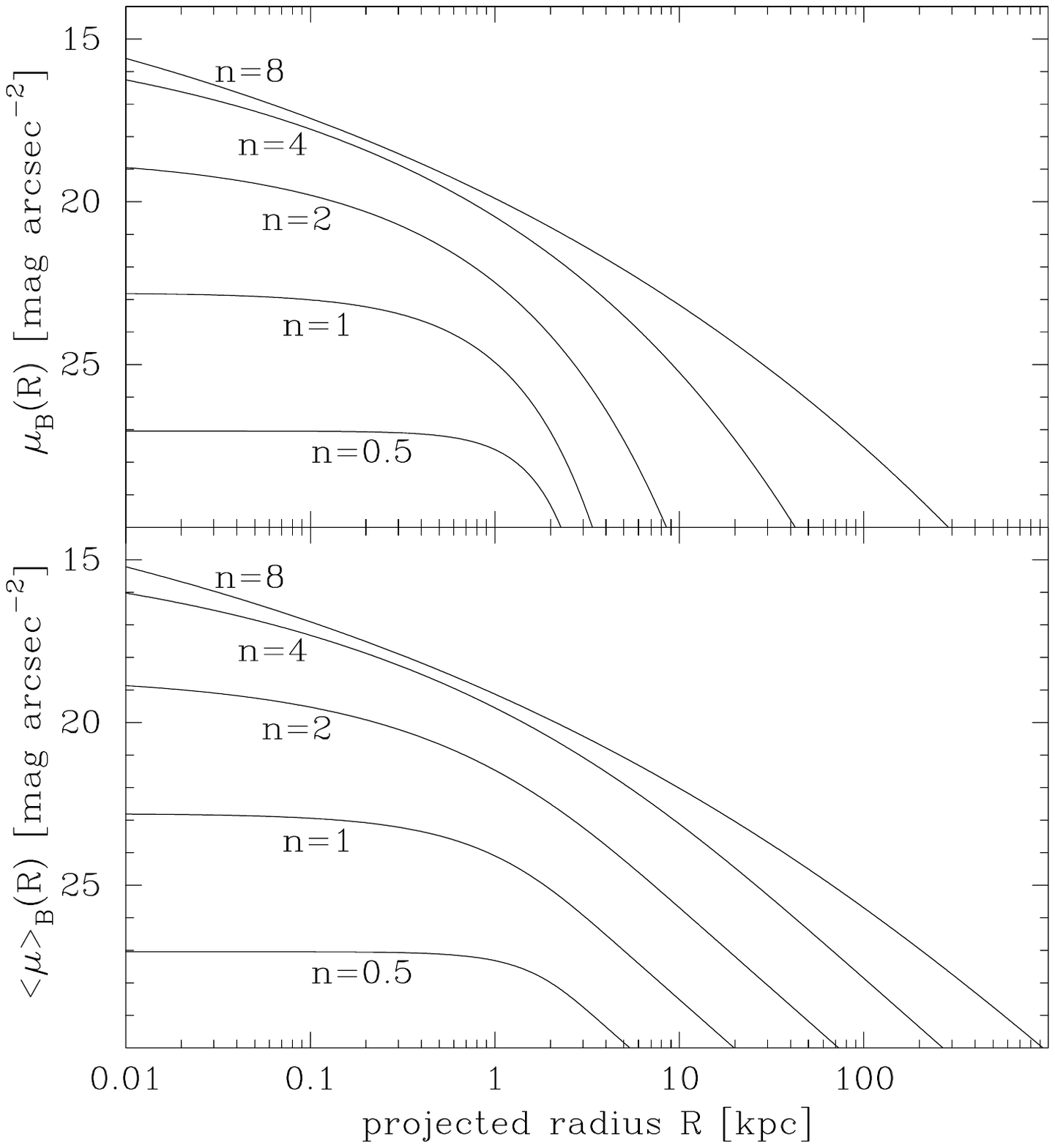}
\caption{Upper panel: S\'ersic light profiles ($B$-band,
  Vega mag), for a range of S\'ersic indices $n$, that are representative of the
  ETG population at large.  Lower panel: Associated set of representative mean
  surface brightness profiles.  These stem from equations~\ref{Eq_Mn} and
  \ref{Eq_Mmu} and the $R^{1/n}$ model.} 
\label{Fig1}
\end{center}
\end{figure}

Two key linear scaling relations describe the structural properties of ETGs.
These have been known for decades and were common in the 1960s, 70s and early
80s before somewhat falling from favour as the ``effective'' parameters from the
$R^{1/4}$ model started to dominate the landscape.

The first relation 
relates to the central concentration of the galaxy light.\footnote{Of note,
  Lundmark (1925, and references therein), was using concentration to classify
  galaxies 100 years ago.}   This was the
primary criteria of the concentration classes in the Yerkes system (e.g., Morgan
1958, 1959, 1962), although introduced to match the changing spectra 
along the Aitken-Jeans-Lundmark-Hubble\footnote{Aitken (1906), Jeans
  (1919, 1928), Lundmark(1925), Hubble (1926, 1936).} 
sequence (Graham 2019) as observed by Morgan \& Mayall (1957).  Fraser
(1972) subsequently quantified the concentration using $C_{21}$, the
ratio of radii containing 50 and 25 percent of the total light, and $C_{32}$,
the ratio of radii containing 75 and 50 percent of the total
light\footnote{This followed in the footsteps of the concentration ratio
  introduced by King (1962, his equation~17) for globular clusters.}.
Subsequently, de
Vaucouleurs (1977) extended this to the use of $C_{31}$ (e.g.\ Kent 1985).
The linear concentration-magnitude relation for dwarf and ordinary ETGs has
been known since at least Binggeli et al.\ (1984, their Figure~10) and
Ichikawa et al.\ (1986, see their Figure~11).  Using the $B$-band absolute
magnitude $\mathfrak{M}_B$, the left panel of Figure~\ref{fig0} shows the
$\mathfrak{M}_B$--$\log n$ (hereafter $\mathfrak{M}_B$--$n$ for
brevity) diagram, taken from Graham \& Guzm\'an (2003, their Figure~10).  The S\'ersic
index is a measure of the radial concentration of galaxy light (King 1966, see
the end of his section IV; Trujillo et al.\ 2001, their section 3).  Other
examples of the $\mathfrak{M}_B$--$n$ diagram can be seen in Caon et
al.\ (1993), James (1994), Young \& Currie (1994, 1995), Graham et
al.\ (1996), Jerjen et al.\ (2000, their Figure~6); Ferrarese et al.\ (2006)
and K09.

The right panel of Figure~\ref{fig0} reproduces 
the $\mathfrak{M}_B$--(central surface
brightness, $\mu_{0,B}$) diagram from Graham \& Guzm\'an (2003, their Figure~9). 
The two relations in Figure~\ref{fig0} are such that 
\begin{eqnarray}
\mathfrak{M}_B &=& -9.4\log(n) - 14.3, {\rm and} \label{Eq_Mn} \\
\mathfrak{M}_B &=& (2/3)\mu_{0,B} - 29.5. \label{Eq_Mmu}
\end{eqnarray}  
All parameters are measured in the $B$-band, on the Vega magnitude system. 
To avoid confusion, no subscript $B$ is assigned to the S\'ersic index
$n$ --- nor will such a subscript be assigned to any scale radii in this paper --- although these parameters {\it are} slightly 
dependent on the filter used (e.g.\ Kelvin et al.\ 2012; H{\"a}u{\ss}ler et
al.\ 2013; Kennedy et al.\ 2016a,b).

There is no bend at $\mathfrak{M}_B \approx -18$~mag in either of the above
two relations (equations~\ref{Eq_Mn} and \ref{Eq_Mmu}), with the exception
that luminous ($\mathfrak{M}_B \lesssim -20.5$ mag) galaxies, with cores that
are depleted of stars, have central surface brightnesses that deviate from the
$\mathfrak{M}_B$--$\mu_0$ relation.  Such galaxies were discussed half a
century ago by King \& Minkowski (1966, 1972) and King (1978), and were known
to produce a departure from the otherwise linear $\mathfrak{M}_B$--$\mu_0$
relation (e.g.\ Gudehus 1973, his Figure~6; see Oemler 1973 for further
discussion).  The cores of these ``core-S\'ersic'' 
galaxies are nowadays thought to be depleted
by the coalescence of massive black holes, which kick (up to a few percent of)
the galaxy's inner stars to higher orbits, even ejecting some as hypervelocity
stars (Hills 1988) from the galaxy (Begelman et al.\ 1980; Thomas et
al.\ 2014).  Binggeli et al.\ (1984, their Figure~11; see also Binggeli \&
Cameron 1991, their Figures~9 and 18) showed that if they used the central
surface brightness coming from the inward extrapolation of King models, fit
outside of the depleted core region, then they recovered a near linear
$\mathfrak{M}_B$--$\mu_0$ relation.  Jerjen \& Binggeli (1997) and Jerjen,
Binggeli, \& Freeman (2000, their Figure~5) subsequently noted that bright
elliptical galaxies with depleted cores follow a linear
$\mathfrak{M}_B$--$\mu_0$ relation if one uses the central surface brightness
of the best-fitting S\'ersic model fit outside of the core region.  The
continuity between the ``dwarf'' and ``ordinary'' ETGs, that Binggeli had
repeatedly demonstrated, supported a single population of ETGs, from faint to
bright, until the modification of galaxy cores at $\mathfrak{M}_B \approx
-20.5$~mag. (see also Graham \& Guzm\'an 2003 and Ferrarese et al.\ 2006,
their Figure~116).

\begin{figure*}[ht]
\begin{center}
\includegraphics[angle=270, trim=6.7cm 2.cm 1.3cm 2.0cm, width=0.8\textwidth]{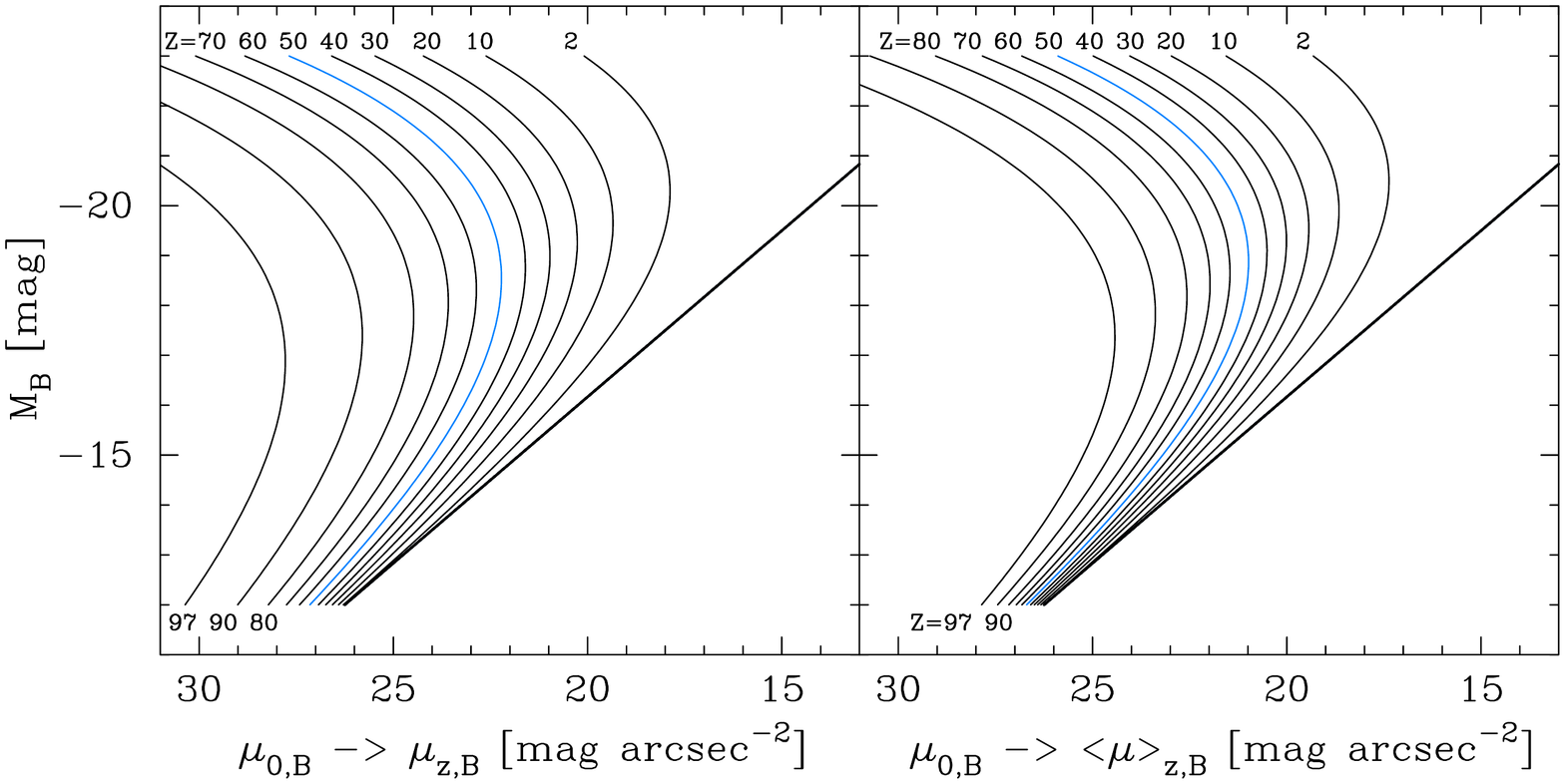}
\caption{ETG scaling relations between absolute $B$-band magnitude and the
  $B$-band surface brightness at projected radii containing different
  percentages ($Z=$2, 10, 20... 80, 90, 97) of the total light (left panel)
  and the mean surface brightness within these radii (right panel).  The thick
  straight line is the relation from Figure~\ref{fig0} involving the central
  surface brightness $\mu_{0,B}$.  The curved lines corresponding to $R_{\rm
    e}$, i.e.\ the radius enclosing 50 percent ($Z=50\%$, $z=0.5$) of the
  total light, shows the behaviour of both the effective surface brightness
  $\mu_{\rm e}$ and the mean effective surface brightness $\langle \mu
  \rangle_{\rm e}$.  As revealed in Graham (2013, his Figure~2-8), the ETGs in
  Figure~\ref{fig0} follow the $Z=50$ curves shown here.  The
  different absolute magnitude associated with the apparent midpoint or bend
  in the curves with different values of $Z$ is not due to different formation
  physics at brighter or fainter magnitudes.  }
\label{Fig2}
\end{center}
\end{figure*}

There are many computer simulations attempting to mimic, and thereby provide
insight into, the evolution of real galaxies in the Universe, such as 
the {\it Illustris} simulation (e.g.\ Genel et
al.\ 2014; Vogelsberger et al.\ 2014; Mutlu-Pakdil et al.\ 2018), IllustrisTNG
(Weinberger et al.\ 2018; Wang 2019), the EAGLE simulation (Schaye et
al.\ 2015; Trayford \& Schaye 2018), the Magneticum simulation (Remus et
al.\ 2015; Schulze et al.\ 2018), plus others (e.g.\ Ragone-Figueroa et
al.\ 2013; Barai et al.\ 2014; Gabor \& Bournaud 2014; Taylor \& Kobayashi
2014; Steinborn et al.\ 2015; Angl\'es-Alc\'azar et al.\ 2017; Taylor et
al.\ 2017).  In order to check if they are realistic, they must be able to
reproduce the $\mathfrak{M}_B$--$n$ and $\mathfrak{M}_B$--$\mu_0$ relations
for ETGs.  As we will see, these two relations additionally define the
$\mathfrak{M}_B$--$R_{\rm e}$ luminosity-size relation (which is used to
calibrate some of the simulations, such as the EAGLE project) plus the
$\mathfrak{M}_B$--$\mu_{\rm e}$ relation and the $R_{\rm e}$--$\mu_{\rm e}$
relation.  It is recognised that constraints on the spatial resolution of
simulations may inhibit the direct observation of $\mu_0$, but it should be
recoverable by fitting $R^{1/n}$ models to their light distributions.

\subsection{A representative set of ETG light profiles}

Given the S\'ersic function and luminosity (equations~\ref{Eq_Ser} and
\ref{Eq_LnT}), and armed with the two empirical equations~\ref{Eq_Mn} and
\ref{Eq_Mmu}, one can readily determine not only the typical S\'ersic index
and central surface brightness for a given ($B$-band) absolute magnitude, but
also the typical effective surface brightness at $R_{\rm e}$, the mean
effective surface brightness within $R_{\rm e}$, and the effective half light
radius in kpc.  This information has been used here to construct a representative
set of surface brightness profiles for ETGs having five different absolute
magnitudes, or rather, five different S\'ersic indices (Figure~\ref{Fig1},
upper panel).  The associated set
of mean surface brightness profiles, which display the average surface
brightness enclosed within the radius $R$, are also shown in the lower panel
of Figure~\ref{Fig1}.

\section{Projected parameters}

\subsection{Relations involving effective surface brightnesses and effective radii}\label{Sec_eff}

This section reveals how the absolute magnitude associated with
the bend in diagrams using effective radii, and effective surface
brightnesses, changes depending on
the percentage of light that these radii enclose. That is, it shows
that the absolute magnitude associated with the bend does not
relate to different formation processes, but rather relates to the
arbitrary definition of galaxy size.

\subsubsection{Luminosity-(effective surface brightness) diagram}

As was noted, given the absolute magnitude of an ETG, equations~\ref{Eq_Mn} and
\ref{Eq_Mmu} inform one of the typical S\'ersic index and central surface
brightness $\mu_0$ associated with this magnitude. This is enough information
to determine the surface brightness $\mu_z$, at a radius $R_z$, containing any
fraction $z$ (between 0 and 1, or percentage $Z$) of the ETG's total light.  
Using  $\mu(R)=-2.5\log I(R)$, at $R=R_z$, the S\'ersic model 
(equation~\ref{Eq_Ser}) gives 
\begin{equation} 
\mu_z = \mu_0 + 2.5b_{n,z}/\ln(10),
\end{equation}
and it can be shown that the mean surface brightness is such that 
\begin{equation}
\langle \mu \rangle_z = \mu_z - 2.5 \log[f(n)],
\end{equation}
where 
\begin{equation}
f(n)=\frac{z\,2n {\rm e}^{b_{n,z}}}{(b_{n,z})^{2n}}\Gamma(2n). 
\label{Eq_fn}
\end{equation}
To date, $z$ has invariably been set equal to 0.5, giving $R_{\rm e}$,
$\mu_{\rm e}$ and $\langle \mu \rangle_{\rm e}$.  
The quantity $b_{n,z}$ seen above is a function of both the S\'ersic index $n$ 
and $z$, and is obtained by solving 
\begin{equation}
\gamma(2n,b_{n,z})= z\, \Gamma(2n)
\label{bsubnz}
\end{equation}
(cf.\ equation~\ref{bsubn}).  Knowing $b_{n,z}$, one can additionally
calculate the radius $R_z$ containing Z percent of the total light, in terms of the
effective half light radius $R_{\rm e}$ containing 50 percent of the total light: 
\begin{equation}
R_z = \left( \frac{b_{n,z}}{b_n} \right)^n R_{\rm e},
\end{equation}
where $b_n$ is given by equation~\ref{bsubn}.

Figure~\ref{Fig2} reveals the difference between the central surface
brightness, $\mu_0$, and both the surface brightness $\mu_z$ at the scale
radius $R_z$ (left panel), and the mean
surface brightness $\langle \mu \rangle_z$ within this radius (right panel).  
%
%
The orthogonal behaviour (at faint and bright magnitudes) seen here for any $z$ 
is a consequence of the S\'ersic index 
changing systematically and monotonically with absolute magnitude, i.e.\ 
``structural non-homology''.  

While the ETG population are unified by the linear $\mathfrak{M}$--$\mu_0$ and
$\mathfrak{M}$--$\log(n)$ relations --- with no evidence for a divide at
$\mathfrak{M}_B \approx -18$ mag --- the peak in the bend of the ($z=0.5$) 
$\mathfrak{M}$--$\mu_{\rm e}$ and $\mathfrak{M}$--$\langle \mu \rangle_{\rm
  e}$ distribution occurs at $\mathfrak{M}_B \approx -18$ mag.  This has
contributed to decades of belief that different physical processes have shaped
the ETGs brighter and fainter than $\mathfrak{M}_B \approx -18$ mag.  However,
Figure~\ref{Fig2} reveals that had de Vaucouleurs used a radius containing 97
percent of the total light, then some might today be claiming that
the divide between dwarf and ordinary ETGs occurs at $\mathfrak{M}_B=-17$ mag; 
or had de~Vaucouleurs used a radius containing 2 percent of the galaxy's total
light, then they might be advocating for a divide at $\mathfrak{M}_B=-20.5$ mag.

The crucial point is that one should not assign a physical interpretation to
the bend.  Graham \& Guzm\'an (2003), and Graham (2013), tried to make this
point using only the $Z=50$ curves in Figure~\ref{Fig2} and explaining that the
bend is due to the light profile shape changing smoothly as the absolute
magnitude changes.  That is, it is not due to different physical processes
operating at absolute magnitudes fainter and brighter than $-18$~mag (or
$-17$~mag, or $-20.5$~mag).

Despite the above, there has been a remarkable number of claims of supporting
evidence for the false divide at $\mathfrak{M}_B \approx -18$ mag.  This often
pertains to observations that some quantity (for example, S\'ersic index, or
colour, or dynamical mass-to-light ratio) is, on average, different between
ETGs brighter and fainter than $\mathfrak{M}_B \approx -18$ mag.  This paper
has endeavoured to more fully explain the nature of ETGs by including the
additional curves in Figure~\ref{Fig2}, and by revealing in the coming
sections what the distribution of ETGs looks like in related diagrams
involving effective radii and other measures of radii.  There is much that
needs addressing given the decades of literature on this subject, the
engrained nature of assigning a divide between dwarf and ordinary ETGs at
$\mathfrak{M}_B= -18$ mag, and the many (yet to be widely recognised and utilised) 
insights from understanding these curved scaling relations.

\begin{figure*}[ht]
\begin{center}
\includegraphics[angle=270, trim=9.0cm 2.5cm 2.5cm 2.5cm, width=0.8\textwidth]{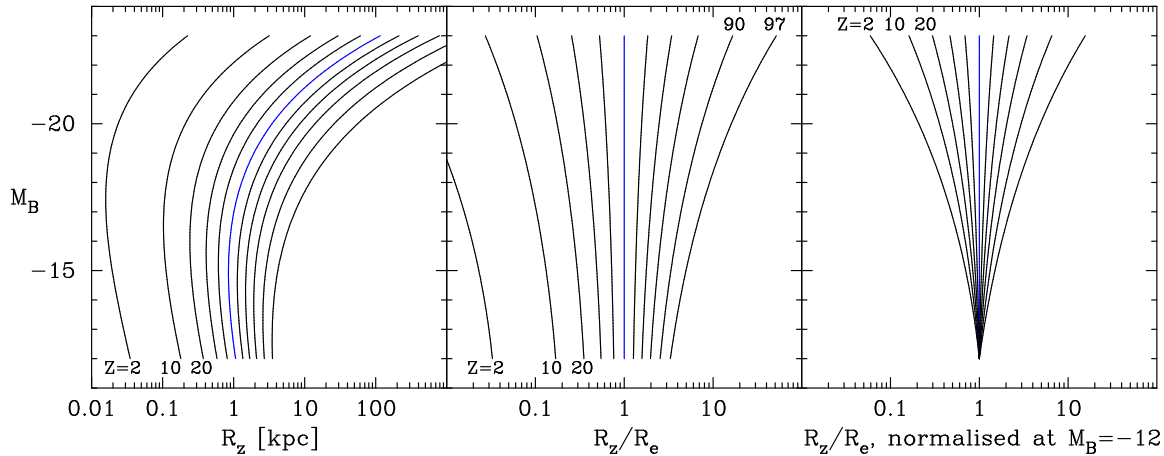}
\caption{Left panel: relations describing the distribution of ETG $B$-band
  absolute magnitude versus the projected radii enclosing various percentages
  ($Z=$2, 10, 20... 80, 90, 97) of their total flux (equation~\ref{Eq_Mag_R}).  
 The relation involving the effective half light radius corresponds to the $Z=50$ curve.
  Middle and right panels: It can be seen how much the scale radii vary depending on
  the arbitrary percentage of light used to define them.}
\label{Fig3}
\end{center}
\end{figure*}

\subsubsection{Luminosity-(effective radius) diagram}\label{Sec_LR}

Due to how the light profile smoothly and systematically changes shape with
absolute magnitude (e.g.\ Fisher \& Drory 2010, their Figure~13), when using
effective half light radii ($z=0.5$), it results in a distribution of ETGs ---
and bulges --- which is curved (e.g.\ Lange et al.\ 2015, and references
therein).  Here, Graham et al.\ (2006, their Figure~1) and Graham \& Worley
(2008, their Figure~11) is expanded upon by additionally showing what the
size-luminosity relation looks like when using scale radii that effectively
enclose different fractions of the total galaxy light.  This also reveals how
the absolute magnitude associated with the alleged dichotomy between dwarf
($\mathfrak{M}_B > -18$ mag) and ordinary ($\mathfrak{M}_B < -18$ mag) ETGs is
ficticious, purely dependent on the arbitrary fraction $z$ rather than
different physical formation processes.

Building upon equation~12 from Graham \& Driver (2005), which 
used $R_{\rm e}$ and thus $z=0.5$, the generalised expression for the total
absolute magnitude, in terms of the radius $R_z$ containing the fraction $z$ of the total
light, is given by 
\begin{equation}
\mathfrak{M}_{\rm tot,B} = \langle\mu\rangle_{\rm z,B} - 2.5\log(\pi R_{\rm z,kpc}^2/z)
- 36.57, 
\end{equation}
where $\langle\mu\rangle_{\rm z,B}$ is the mean surface
brightness within $R_z$. 
This can be rearranged to give the expression
\begin{eqnarray}
\log R_{\rm z,kpc} &=& \frac{\mu_0 - \mathfrak{M}_{\rm tot}}{5} + \frac{\log z - \log[f(n)]}{2}  
\nonumber \\
        &+& \frac{b_{n,z}}{2\ln(10)} - 7.065, 
\label{Eq_Magic}
\end{eqnarray} 
where $f(n)$ is given in equation~\ref{Eq_fn}. 
Using equation~\ref{Eq_Mmu} to replace $\mu_0$ with $\mathfrak{M}_{\rm tot}$,
this expression becomes 
\begin{eqnarray}
\log R_{\rm z,kpc} &=& \frac{\mathfrak{M}_{\rm tot}}{10} + \frac{\log z - \log[f(n)]}{2}
\nonumber \\
&+& 0.217b_{n,z} + 1.2874.
\label{Eq_Mag_R}
\end{eqnarray} 
The latter term in equation~\ref{Eq_Mag_R}, 
involving $z$, cancels with the same term in $f(n)$, and thus the
dependence of $R_z$ on $z$ occurs via the $b_{n,z}$ term (equation~\ref{bsubnz}).

Figure~\ref{Fig3} presents the ETG luminosity-size relations for a range of
fractions $z$, expressed there as percentages $Z$.  
The curved behaviour is, once again, due to the ETG population
smoothly changing its light profile shape --- as quantified by the S\'ersic
index --- with absolute magnitude. 
It can readily be appreciated that adopting
some fixed fraction $z$, such as 0.5, and then claiming that different
physical processes have shaped the luminosity-size relation on either  side of
the apparent bend-point, would be a misleading endeavour (e.g.\ 
Fisher \& Drory 2010, 2016). 

As can be seen in the left-hand panel of Figure~\ref{Fig3}, 
the bright-arm of the curved $\mathfrak{M}_B$--$\log R_{\rm e}$ relation for ETGs is
approximately linear.  This was noted by Fish (1963, who reported $\log L
\propto (1\,{\rm to}\,1.5)\log R_{\rm e}$, or equivalently $\mathfrak{M}_p \propto
-(2.5\,{\rm to}\,3.75)\log R_{\rm e}$, and can be seen in Figure~1 of S\'ersic (1968b,
using the data from Fish 1964; see also Brookes \& Rood 1971; Gudehus \& Hegyi
1991; Shen et al.\ 2003; Graham \& Worley 2008; Lange et
al.\ 2015)\footnote{S\'ersic (1968b) used $R_{\rm e}$ values from the
  $R^{1/4}$ model because he thought that the light profiles of elliptical
  galaxies, and the bulges of spiral galaxies, followed the $R^{1/4}$ model.
  His $R^{1/n}$ model (S\'ersic 1968a) was intended to capture varying
  combinations of $R^{1/4}$ bulge plus exponential disc.}$^,$\footnote{Excluding
  four erroneous references to S\'ersic (1968b), instead of S\'ersic (1968a)
  which presents the $R^{1/n}$ model, S\'ersic (1968b) has only
  been cited once since 1973.  S\'ersic (1968b), and Fish (1963), are indeed 
    over-looked papers.}.   
S\'ersic (1968b) was perhaps the first to remark upon the offset nature of the
faint ETGs from the bright ETGs in the $\mathfrak{M}_B$--$\log R_{\rm e}$ diagram. Not
understanding the bend in this diagram --- referred to as the ``transition
region'' by S\'ersic (1968b) --- coupled with the inclusion
of three unusually
small galaxies, S\'ersic attributed the bend to two populations of (dwarf and
giant) elliptical galaxies, rather than one population with smoothly varying
properties. 

Confounding the situation further, S\'ersic added late-type
galaxies into his $\mathfrak{M}_p$--$\log R_{\rm e}$ diagram (see his Figure~2;
cf.\ Figures~9 and 14 from Cappellari et al.\ 2013b).  Involving $R_{\rm e}$
measures from both two-dimensional spirals and three-dimensional 
ellipticals, S\'ersic (1968b) observed a slight
overlap, and wrote that ``it seems difficult to deny the existence of the
sequence of irregulars and spirals joining that of the ellipticals in the
transition region''. 
Kormendy (1985) adopted this same practice.

\begin{figure*}[ht]
\begin{center}
\includegraphics[angle=270, trim=6.5cm 1.7cm 1.5cm 2.0cm, width=0.8\textwidth]{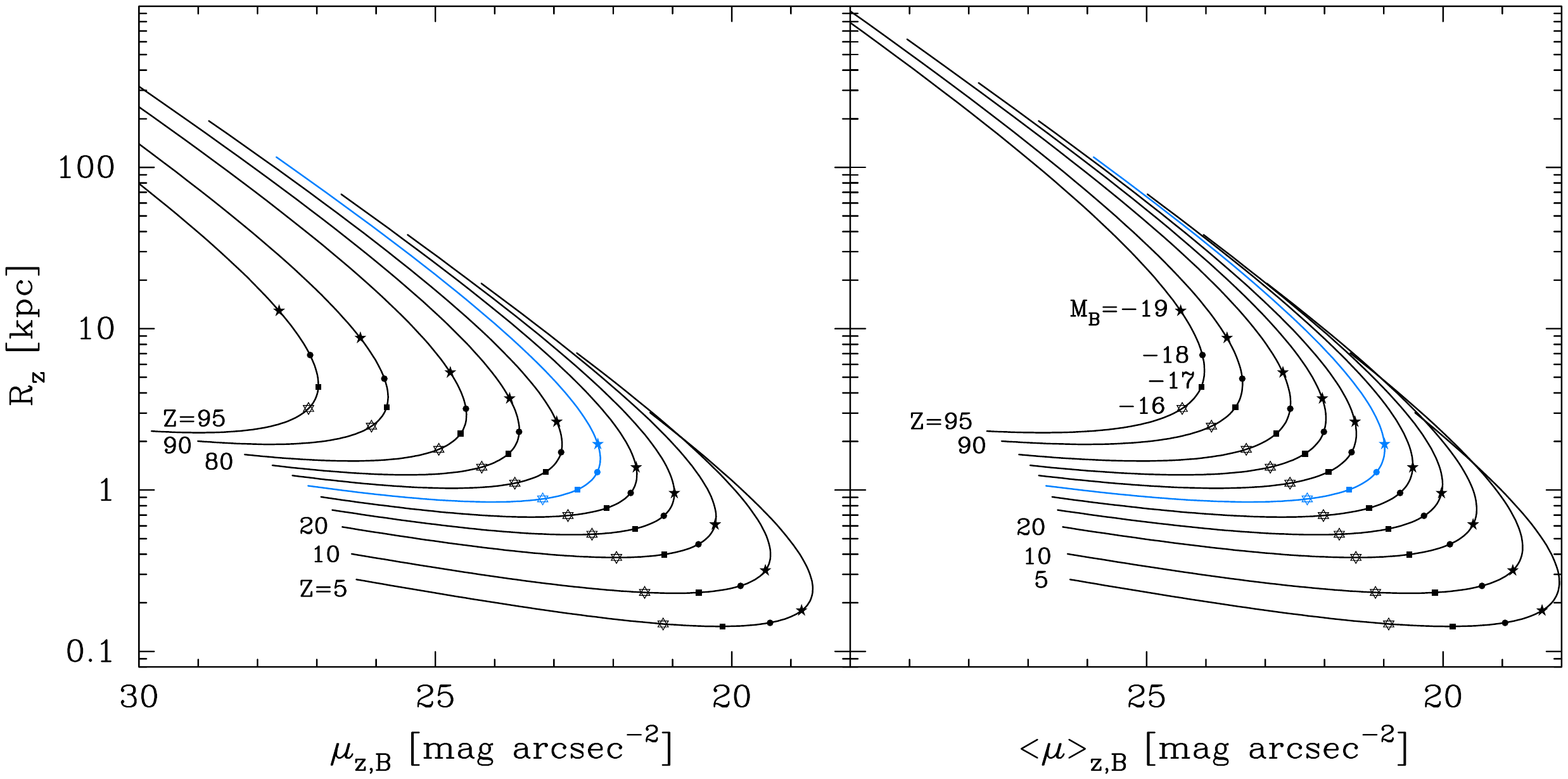}
\caption{Size-(surface brightness) relations --- describing the distribution
  of galaxies having a range of $B$-band absolute magnitudes (from $-12$ to
  $-23$ mag) --- based on projected radii, $R_z$, enclosing different
  percentages ($Z=$5, 10, 20... 80, 90, 95) of the galaxy light.  The
  innermost curve is associated with the 50 percent radius, known as the
  effective half light radius $R_{\rm e}$, the effective surface brightness
  $\mu_{\rm e}$ at this radius (left panel), and the average surface
  brightness $\langle \mu \rangle_{\rm e}$ within this radius (right panel).
  The upper envelope in the left-hand panel has a slope of $\sim$3.5, while
  the upper envelope in the right-hand panel has a slope of $\sim$2.8.}
\label{Fig4}
\end{center}
\end{figure*}

For $R_{\rm e} \gtrsim$ 1--2~kpc and a photographic absolute magnitude
$\mathfrak{M}_p$ brighter than $-19.5$~mag\footnote{Based on the Hubble-Lema\^itre constant of 100 km
s$^{-1}$ Mpc$^{-1}$ used by Fish 1964; or $R_{\rm e} \gtrsim$ 
1.5--3.0~kpc and $\mathfrak{M}_p \approx -20.3$ mag using
$H_0 = 67.8$ km s$^{-1}$ Mpc$^{-1}$.  
This therefore matches the S\'ersic /
core-S\'ersic divide seen in Figure~\ref{fig0}.}, 
S\'ersic (1968b) fit a line with a slope of unity to the distribution of giant
elliptical galaxies in his ($\log {\rm Mass}$)--($\log R_{\rm e}$) diagram.  
This distribution resembled that in his
$\mathfrak{M}$--$\log R_{\rm e}$ diagram because he claims to have used a constant
mass-to-light ratio of 30.  As such, S\'ersic (1968b) reported a distribution in which the
absolute magnitude scaled as $-2.5\log R_{\rm e}$.  Given that the magnitude of a
galaxy is proportional to $\langle \mu \rangle_{\rm e} - 5\log R_{\rm e}$
(e.g.\ de Vaucouleurs \& Page 1962, their equation~6), one immediately has the
relation $\langle \mu \rangle_{\rm e} \propto 2.5\log R_{\rm e}$ for
the distribution of giant elliptical galaxies. Furthermore, given that
$\mu_{\rm e} - \langle \mu \rangle_{\rm e} = 1.393$ for the $R^{1/4}$ model
that S\'ersic (1968b) was using, 
one also immediately has that $\mu_{\rm e} \propto 2.5\log R_{\rm e}$.  This
can be compared with  
Kormendy (1977)\footnote{Kormendy (1977) additionally noted that the magnitude
  of each galaxy is equal to $\mu_{\rm e} - 5\log R_{\rm e} -1.9995$ based on
  the $R^{1/4}$ model, and thus one has $\mathfrak{M}_B \propto -1.99 \log
  R_{\rm e}$, cf.\ S\'ersic (1968b) who reported $\mathfrak{M}_B \propto -2.5
  \log R_{\rm e}$ for the bright ETGs.}  who reported $\mu_{\rm e} \propto
3.02\log R_{\rm e}$.

Somerville \& Dav\'e (2015, their section~1.1.4)
refer to the ($\log {\rm Mass}$)--($\log R_{\rm e}$) relation as the Kormendy
relation (see also Cappellari 2016, his section~4.1.1), but it would be more
appropriate if that title was assigned to the linear relation which Kormendy
fit to the bright arm of what we now know is the curved $\mu_{\rm e}$--$\log R_{\rm e}$ relation,
and to instead refer to the linear ($\log {\rm Mass}$)--($\log R_{\rm e}$) and
$\mathfrak{M}$--($\log R_{\rm e}$) relation used to describe the distribution
of bright elliptical galaxies as the Fish\footnote{Using the $R^{1/4}$ model,
  Fish (1963) had previously claimed $-2.5\log L \propto -$(2.5 to 3.75)$\log
  R_{\rm e}$ for ETGs.}  or Fish-S\'ersic relation.  The curved $\mu_{\rm
  e}$--$\log R_{\rm e}$ relation is explored in Section~\ref{Sec_mure}.

Three additional insights from Figure~\ref{Fig3} can readily be made. 
The first has implications 
for dark matter (Kent et al.\ 1990, and references therein) if using
$\sigma^2R_{\rm e}$ (e.g.\ Poincare \& Vergne 1911; Poveda 1958, 1961) as a
proxy for mass\footnote{As derived from the virial theorem (Clausius 1870).} in a population of ETGs with a range of absolute magnitudes and
thus a range of light profile shapes. 
Considering how the ratio of radii ($R_{z_1}/R_{z_2}$) at fixed absolute
magnitude change, for different values of $z_1$ and $z_2$ in
Figure~\ref{Fig3}, one should pause for thought when using effective half
light radii ($z=0.5$) to measure dynamical (stellar plus dark matter) masses
via the proportionality $\sigma^2 R_{\rm e}$ (e.g.\ Drory et al.\ 2004, their
Figure~3; Cappellari et al.\ 2006 who use the luminosity-weighted $\sigma$
values within half light radii determined from $R^{1/4}$ models; Cappellari et
al.\ 2013a). Using 
$\sigma^2R_z$, with $z\ne0.5$, will produce a different trend because the
ratio $R_{\rm e}/R_z$ is not constant for different $\mathfrak{M}_B$ (see also
the telling merger simulations by Farouki et al.\ 1983 revealing how
$R_{10}/R_{\rm e}$ changes with mass, and the work by Campbell et al.\ 2017
and Lyskova et al.\ 2015).  Therefore, it may not be appropriate to solely invoke
varying fractions of dark matter to explain the 
systematic differences, as a function of varying absolute magnitude, 
between (i) this dynamical mass estimate (based on the arbitrary radius
$R_{\rm e}$) and (ii) the stellar mass estimate (obtained from the absolute
magnitude.   
This will be broached in a subsequent study, covering the ``Fundamental
Plane'' (Djorgovski \& Davis 1987), improved planes, and implications for both dark
matter estimates (e.g.\ Graves \& Faber 2010) and ETG formation (see
Cappellari 2016, his section 4, for an overview).

Second, it is noted that the acceleration at some radius $R$, inside a
symmetrical pressure supported system with velocity dispersion $\sigma$, is
proportional to $GM/R^2$, or $\sigma^2/R$. Due to the structural non-homology
of ETGs, this ratio will vary with $M$ in different ways depending on what
fraction $z$ has been used to measure $R$.  This has relevance to the critical
acceleration parameter $a_0$, or characteristic surface density $M/R^2$, in
Modified Newtonian dynamics (MOND: Milgrom 1983; Sanders \& McGaugh 2002, see
their Figure~7; Milgrom \& Sanders 2003; Kroupa et al.\ 2010, their Figure~7;
Misgeld \& Hilker 2011, their Figure~7; Famaey \& McGaugh 2012).  

It may also be insightful to explore the near-constant $R_{\rm e,bulge}/h_{\rm
  disc}$ ratio of $\sim$0.2 observed in spiral galaxies (e.g.\ Courteau et
al.\ 1996; Graham \& Worley 2008, and references therein), which appears
irrespective of whether the bulge is considered to be a ``classical'' bulge or
a ``pseudobulge''.  For instance, the use of $z=0.1$ or $z=0.9$, rather than
$z=0.5$, is expected to result in this ratio systematically changing, with
magnitude, by a factor of $\sim$3 for spiral galaxies.

\subsubsection{(Effective radius)-(effective surface brightness) diagram}\label{Sec_mure}

In addition to the $R_{\rm e}$--$\mu_{\rm e}$ and $R_{\rm e}$--$\langle \mu
\rangle_{\rm e}$ distributions (derived using $z=0.5$), it is instructive to
show the size-(surface brightness) distributions $R_z$--$\mu_z$ and
$R_z$--$\langle \mu \rangle_z$ that one would obtain for different values of
$z$, corresponding to the fraction of light contained within $R_z$.
Figure~\ref{Fig4} reveals a number of things, three of which are worth
explicitly pointing out here, while many other important but less recognised
aspects will be saved for a follow-up paper pertaining to both understanding
the ``Fundamental Plane'' and constructing an improved plane/surface.

First, had the community been using radii enclosing 95 percent, or 5 percent,
of the total light, then those interpreting the bend in the corresponding
size-(surface brightness) diagram may likely be claiming evidence of
distinctly different formation physics for galaxies brighter and fainter than
$\sim -16.5$ mag or $\sim -19.5$ mag, respectively.

\begin{figure*}[ht]
\begin{center}
\includegraphics[angle=270, trim=8.0cm 2.5cm 1.5cm 2.5cm, width=0.8\textwidth]{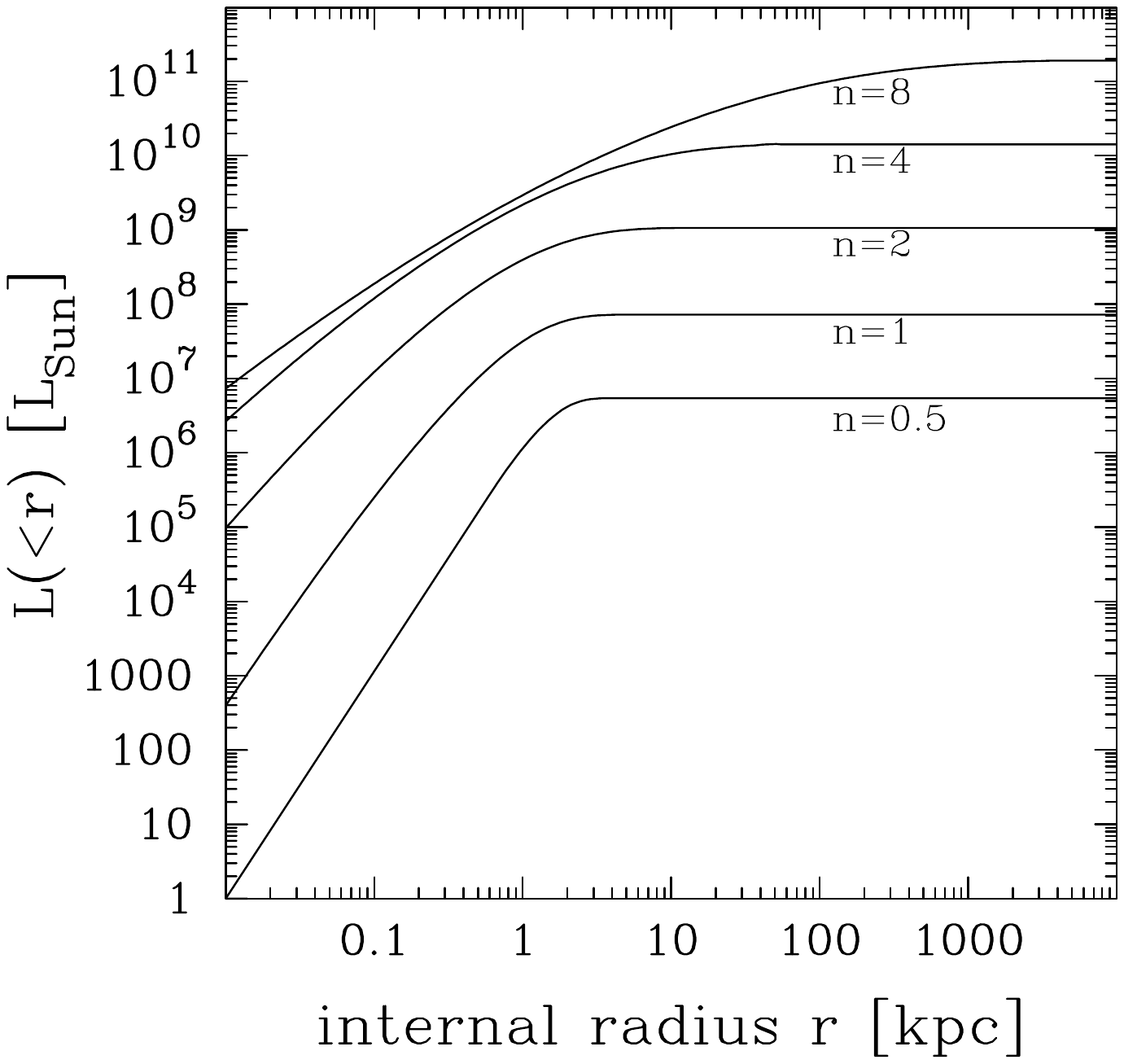}
\caption{Left panel: Relations between the absolute $B$-band magnitude,
  $\mathfrak{M}_B$, and the radius where the associated light profile's
  surface brightness has dropped by a fixed amount from the central $R=0$
  value $\mu_0$ (equation~\ref{Eq_Mmu}) of the S\'ersic model having a
  S\'ersic index $n$ dictated by the value of $\mathfrak{M}_B$
  (equation~\ref{Eq_Mn}).  The leftmost curve in each panel shows the result when using
  $\Delta\mu=8.327$ mag arcsec$^{-2}$, which is the difference in surface
  brightness between $\mu_0$ and $\mu_{\rm e}$ of de Vaucouleurs' $R^{1/4}$
  model.  Middle panel: Similar, except that the radii shown here denote where
  the surface brightness profile has dropped by the same set of constant
  values used in the left panel, but now starting from $R=0.01$~kpc rather
  than from $R=0$.  This helps to bypass the rapidly rising inner light
  profile of systems with high values of $n$, but which typically contain
  depleted cores.  Right panel: Similar to the middle panel but starting from
  $R=0.1$~kpc.  }
\label{fig5}
\end{center}
\end{figure*}

Second, the bunching up of tracks in the top right of Figure~\ref{Fig4}
reveals why the $R_{\rm e}$--$\mu_{\rm e}$ and $R_{\rm e}$--$\langle \mu
\rangle_{\rm e}$ relations have a low level of scatter for ETGs with $B$-band
absolute magnitudes brighter than $\approx -19$ mag.  If one mis-measures the
half light radius, and instead captures the radius enclosing 20, 30, 40, 60,
70, or 80 percent of the total light, the surface brightness terms associated
with these radii are such that the galaxy's location in the $R_{\rm
  e}$--$\mu_{\rm e}$ diagram moves along the upper envelope seen in
Figure~\ref{Fig4} and thereby maintains a tight $R_{\rm e}$--$\mu_{\rm e}$
relation.

Third, the inclusion of ETGs fainter than $\mathfrak{M}_B \approx -19$ mag
results in a thickening of the distribution in the $R_{\rm e}$--$\mu_{\rm e}$
diagram (e.g.\ Kodaira et al.\ 1983; Capaccioli et al.\ 1994) as mis-measures
of the half light radius will shift galaxies perpendicular to the curved
$z=0.5$ relation at faint absolute magnitudes.

Bildfell et al.\ (2008) report that `{\it The Kormendy relation of our BCGs is
  steeper than that of the [less luminous] local ellipticals, suggesting
  differences in the assembly history of these types of systems}'.  Although
the literature is full of similar claims, such interpretations are not
appropriate given the curved $R_{\rm e}$--$\mu_{\rm e}$ relation's dependence
on the arbitrary value $z=0.5$.  Countless studies which have attached a
physical significance to slopes, and bends, in scaling diagrams involving the
logarithm of $R_{\rm e}$, $\mu_{\rm e}$, and/or $\langle \mu \rangle _{\rm e}$
should be questioned.  As already noted in Graham \& Guzm\'an (2004) and
Graham (2005), this remark extends to studies of the `Fundamental Plane'
(Guzm\'an et al.\ 2019, in preparation).

\vspace{5mm}

A range of other measures for galaxy size is explored in the remainder of this
section.

\subsection{An alternative scheme for defining projected radii}\label{Sec_Alt}

The previous text focussed on projected radii that enclosed an arbitrary
fraction of light relative to the light enclosed within a radius of infinity,
i.e.\ the total light.  One can, alternatively, define a radius where the
intensity is an arbitrary fraction of the intensity at $R=0$.  In the case of
the exponential galaxy light profile model, the parameter $h$ denotes the
scalelength where the intensity has dropped by a factor of ${\rm
  e}\approx2.718$.  This subsection explores radii where the intensity of the
$R^{1/n}$ model has dropped by fixed amounts, effectively replacing the
variable ${\rm e}^{b_n}$ term in equation~\ref{Eq_Ser} with a constant.

For de Vaucouleurs' $R^{1/4}$ model
(equation~\ref{Eq_deV}), it was noted that the intensity at $R_{\rm e}$ 
is e$^{7.669} \approx 2141$ times fainter than the
intensity at $R=0$.  This corresponds to a surface brightness which is 8.327
mag arcsec$^{-2}$ fainter than the central surface brightness. It is
informative to explore what the size-luminosity diagram looks like when using
this alternative, but equally valid, measure of ETG size, i.e.\ the radius
where the surface brightness has dropped by a constant 8.327 mag
arcsec$^{-2}$. This is done in Figure~\ref{fig5}, where a few
other constant values are also used.

The left panel of Figure~\ref{fig5} reveals little evidence for a divide at
$\mathfrak{M}_B \approx -18$ mag between the so-called dwarf and ordinary ETGs.  Had
astronomers used the above system of radii, calibrated to the S\'ersic model's
central surface brightness (at $R=0$), rather than calibrated to the S\'ersic
model's total luminosity (at $R=\infty$), then they might well have concluded that
there is a dichotomy between bright and faint ETGs at $\mathfrak{M}_B \approx
-20$ mag, and speculated that different physical processes must be responsible
for the formation of ETGs fainter and brighter than this absolute magnitude.
Some astronomers may have even heralded the observation of partially depleted
cores in ETGs more luminous than $\mathfrak{M}_B \sim -20.5\pm1$ mag ---
thought to have formed their spheroids from major dry merger events --- as the
explanation for the bend seen in this alternative luminosity-size diagram.

In case some readers might be entertaining the 
$\Delta\mu=8.327$ mag arcsec$^{-2}$ 
 curve in the
left-hand panel of Figure~\ref{fig5} as evidence for a division at
$\mathfrak{M}_B \approx -20$ mag, additional measures of radii
based on larger differences in surface brightness from the central surface
brightness have been included.  One can see that the location of the bend in the scaling
relations shifts from a $B$-band magnitude of roughly $-20$ to $-22$ mag as
one samples more of the galaxy light.  Once again, this demonstrates that
these 
bends are not revealing the existence of different physical processes which
operate at magnitudes brighter and fainter than the location of the bend.  The
whip around to smaller radii seen at bright magnitudes in the left-hand panel
of Figure~\ref{fig5} is due to the rapidly rising (with decreasing radii),
inner light profile of systems with high S\'ersic indices.  One can devise
schemes to circumvent this (see the middle and right-hand panel), 
which may be desirable given the partially-depleted cores in these
galaxies which prevent such bright $\mu_0$ values actually being realised.
The monotonic size-luminosity relations in Figure~\ref{fig5}, 
which do not use radii where the intensity
has dropped by systematically different amounts as a function of luminosity
(as occurs with $R_{\rm e}$ and $R_z$) reveal no grounds for segregating
dwarf and giant ETGs at $\mathfrak{M}_B \approx -18$ mag.  

The middle and right-hand panels of Figure~\ref{fig5} show the distribution of
ETG sizes where their surface
brightness profiles have dropped by the same values as those
used in the left-hand panel, but starting the drop from a 
radius of 0.01 and 0.1 kpc, rather than from the central value.

\subsection{Isophotal radii}\label{Sec_iso}

Based on isophotal radii, the luminosity-size relation for ETGs was initially
considered to be log-linear, i.e.\ linear in log space, 
unifying dwarf and giant ETGs (e.g.\ Heidmann 
1967, 1969; Holmberg 1969; Oemler 1976; Strom \& Strom 1978), and it largely
still is (e.g.\ Forbes et al.\ 2008, their Figure~3; van den Bergh 2008; Nair et
al.\ 2011).  This section would therefore be somewhat incomplete if it did not
include isophotal radii.

Using a photographic (Pg, i.e.\ blue filter, Vega mag system) surface
brightness of 26.5 mag arcsec$^{-2}$ to define galaxy diameters, Holmberg
(1969, his Figure~9) reported a linear relation, with a slope of $-6$, between
the absolute magnitude and the logarithm of the isophotal major axis
diameter\footnote{Using ``cut-off'' radii, Oemler (1976) also reported a
  log-linear luminosity-size relation unifying dwarf and giant ETGs.}.  Using
the major axis diameter of the isophote corresponding to a photographic
surface brightness of 25 (Vega) mag arcsec$^{-2}$, Heidmann (1967, 1969: see
also Fraser 1977 and Bigay \& Paturel 1980) obtained a less steep slope of
$-4.75$ for ETGs\footnote{Heidmann (1967, 1969) reported a slope of 2.8 for
  spiral galaxies.}  in the $\mathfrak{M}$--($\log R_{\rm iso}$) diagram, which he
reported as a slope of 1.9 in the $\log L$--$\log R_{\rm iso}$ diagram.

Some half a century later, using the semimajor axis radius of the 3.6 $\mu$m
isophote whose surface brightness equals 25.5 (AB) mag arcsec$^{-2}$, 
Mu\~noz-Mateos et al.\ (2015, their Figure~14) presented a log-linear
radius-(stellar mass) relation for different morphological types.  The bright
ETGs have the same slope as reported by Heidmann, with $\log M \propto
(1.9\pm0.1)\log R_{3.6\mu {\rm m}=25.5}$.  Approximating the low-luminosity end of
the moderately curved $L$--$R_{\rm iso}$ relation with a power-law, the faint
ETGs in Mu\~noz-Mateos et al.\ roughly follow a relation with a slope of
$2.7\pm0.2$.  Mu\~noz-Mateos et al.\ additionally show, in their Figure~15,
that the use of $R_{\rm e}$, rather than isophotal radii, results in the
strongly curved size-luminosity relation seen in Figure~\ref{Fig3}. 

\begin{figure}[ht]
\begin{center}
\includegraphics[angle=270, trim=7cm 3.5cm 1.4cm 12.0cm, width=\columnwidth]{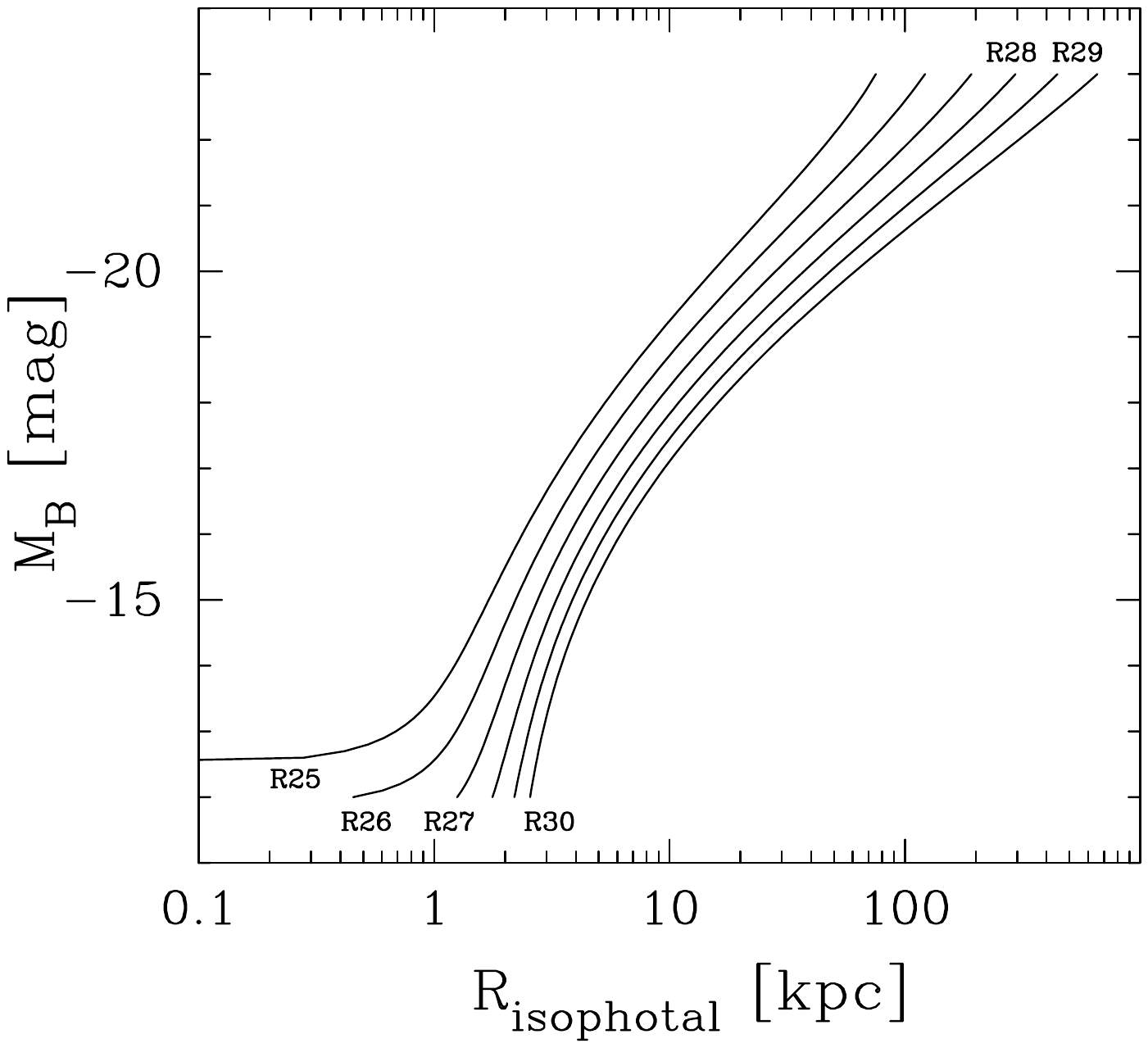}
\caption{Six different isophotal radii are shown as a function of the $B$-band
  absolute magnitude (for which the typical S\'ersic profile and S\'ersic parameters are known
  from equations~\ref{Eq_Mn}, \ref{Eq_Mmu}, and \ref{Eq_Mag_R}).}
\label{Fig6}
\end{center}
\end{figure}

Figure~\ref{Fig6} reveals what the size-luminosity relation for ETGs looks
like when using six different isophotal radii (specifically, those radii
where the $B$-band surface brightness equals 25, 26, ...\ 30 mag
arcsec$^{-2}$) and using the total $B$-band magnitude $\mathfrak{M}_B$ within
a radius of infinite aperture.  The smoothly changing slope is consistent with
the slight curve observed for 50 years in magnitude-(isophotal radii)
diagrams.  For example, as noted above, a moderate change in slope is seen
among the ETGs in the $\log \mathfrak{M}$--$\log R_{3.6\mu {\rm m}=25.5}$
diagram of Mu\~noz-Mateos et al.\ (2015) at 4 to 6 kpc.  This can be
understood in terms of the S\'ersic index varying with absolute magnitude,
which gives rise to the curves in Figure~\ref{Fig6}.  Although, it should be
noted that the mapping between Figure~14 in Mu\~noz-Mateos et al.\ (2015) and
Figure~\ref{Fig6} shown here is not linear because of the colour-magnitude
relation for ETGs (e.g.\ Ferrarese et al.\ 2006, their Figure~123), in which
fainter ETGs are bluer than luminous ETGs. To help anyone who may wish to
explore this further, it is quickly noted that given that luminous ETGs have a
($B-3.6$) colour of 4 to 5, the radius where the 3.6~$\mu$m surface brightness
equals 25.5 mag arcsec$^{-2}$ will roughly correspond to the $R_{B=30}$
isophotal radii seen in Figure~\ref{Fig6}, while ETGs with
$\mathfrak{M}_B=-16$ mag have a ($B-3.6$) colour of $\approx$2.5.

Past studies which did not include ETGs fainter than $\mathfrak{M}_B\approx
-16$ mag could have missed the slight curvature in the $L$--$R_{\rm iso}$
diagram.  The horizontal flattening of the curves associated with the brighter
isophotal levels, seen at small radii in Figure~\ref{Fig6}, reflects that the
central surface brightnesses in galaxies with these low absolute magnitudes is close to the
isophotal value.  Given that the ETGs in Figure~\ref{fig0} have
$\mathfrak{M}_B < -13$ mag, the curves seen in Figure~\ref{Fig6}, and
elsewhere, may not be reliable at $\mathfrak{M}_B > -13$ mag.  At these low
magnitudes, one encounters galaxies which may be a
different, more heterogeneous class of galaxy with a broad range of colours
(e.g.\ Jerjen et al.\ 2000; Hilker et al.\ 2003; Penny \& Conselice 2008).

\section{Internal parameters}

Projected quantities, such as those seen in the previous section, pertain to
the line-of-sight column densities through a galaxy.  To stave off criticisms
that the analysis presented thus far needs to be performed using internal
quantities, rather than projected fluxes and densities, this is now done.

The internal luminosity density profile, $\nu (r)$, 
can be computed from the observed, i.e.\ the projected, 
intensity profile $I(R)$.   Under the assumption of sphericity\footnote{This 
  may not be particularly attractive given the bulge/disc nature of ETGs,
  but it offers some insight.
  Moreover, this approach should be roughly applicable to bulges, which are 
  nowadays similarly misdiagnosed in structural scaling diagrams.  It is
  therefore helpful to include this section.} 
$\nu (r)$ can be obtained by solving the following Abel 
integral (e.g.\ Binney \& Tremaine 1987) 
\begin{equation} 
\nu (r)=\frac{-1}{\pi}\int
_{r}^{\infty}\frac{dI(R)}{dR}\frac{dR}{\sqrt{R^{2}-r^{2}}}. 
\end{equation}
For the S\'ersic $R^{1/n}$ profile, $I(R)$, one has that 
\begin{equation} 
\nu (s) = \frac{I_{\rm e}{\rm e}^{b}}{R_{\rm e}} \frac{b^{n}x^{1-n}}{\pi }\int
_{0}^{1}\frac{1}{t^{2}}\frac{\exp(-x/t)}{\sqrt{t^{-2n}-1}}dt,
\label{Eq_dens}
\end{equation}
where $s=r/R_{\rm e}$, $x=bs^{1/n}$, and $t$ is the dummy variable 
(Ciotti 1991; Graham \& Colless 1997).  
In passing, it is noted that Prugniel \& Simien (1997) provide a useful
parameterized model which approximates this, while Terzi\'c \& Graham (2005,
see also Terzi\'c \& Sprague 2008 for triaxial models) 
provide a modified expression with a power-law core. 

\begin{figure}[ht]
\begin{center}
\includegraphics[angle=270, trim=5.8cm 5.5cm 2.3cm 10.0cm, width=\columnwidth]{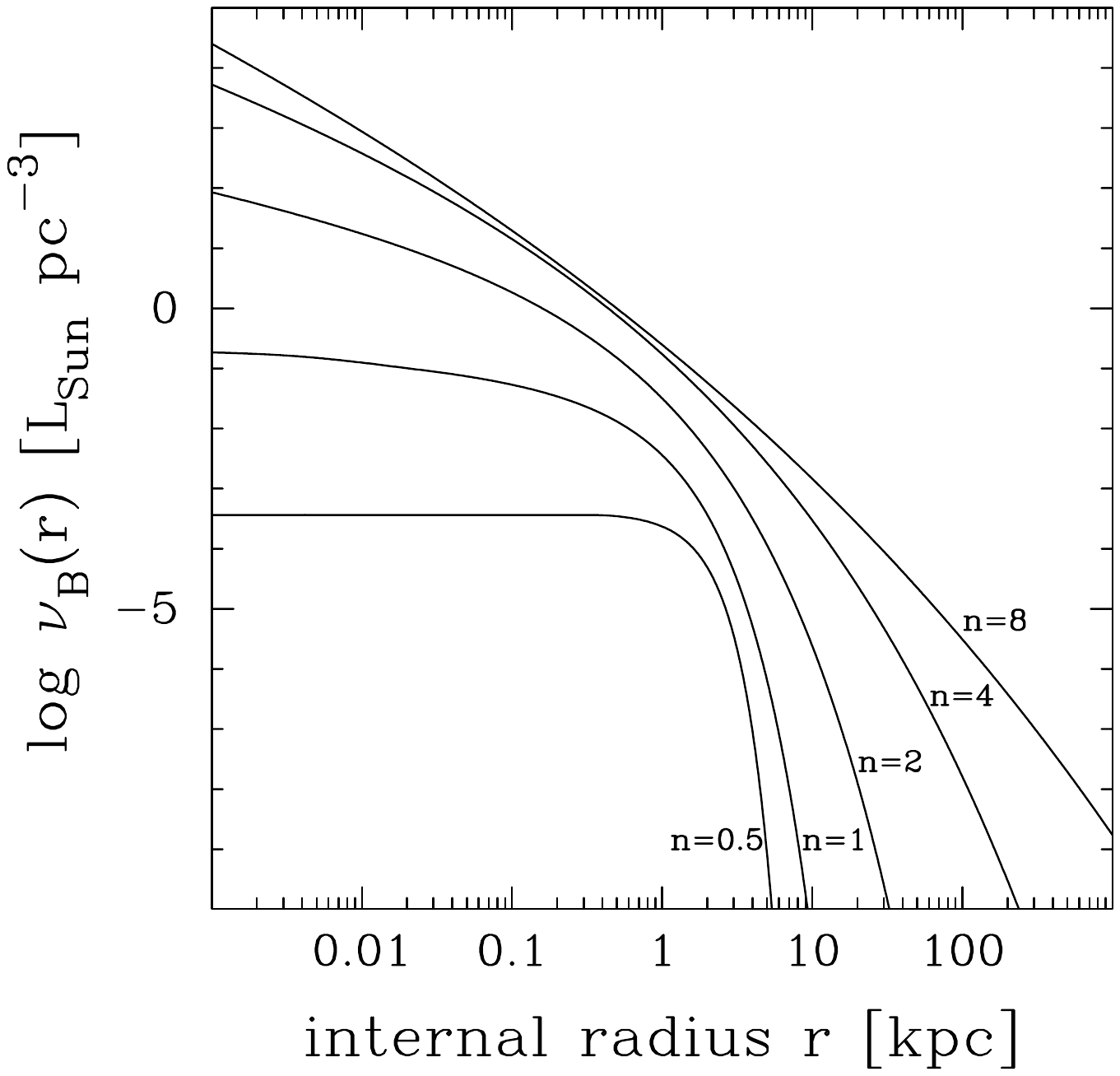}
\caption{Internal, $B$-band, luminosity density profiles associated with the
  projected (surface luminosity density, i.e.\ surface brightness) profiles
  seen in Figure~\ref{Fig1}. }
\label{Fig7}
\end{center}
\end{figure}

Figure~\ref{Fig7} shows the internal luminosity density profiles pertaining to
the (projected) S\'ersic light profiles displayed in Figure~\ref{Fig1}.  A
solar absolute magnitude $\mathfrak{M}_{\odot,B} = 5.44$ mag was used (Mann \&
von Braun 2015).  For S\'ersic indices $n<0.5$, the internal density profile
is known to have a central dip.

\subsection{Internal effective radii}

\begin{figure*}[ht]
\begin{center}
\includegraphics[angle=270, trim=10.7cm 2.2cm 1.9cm 3.2cm, width=\textwidth]{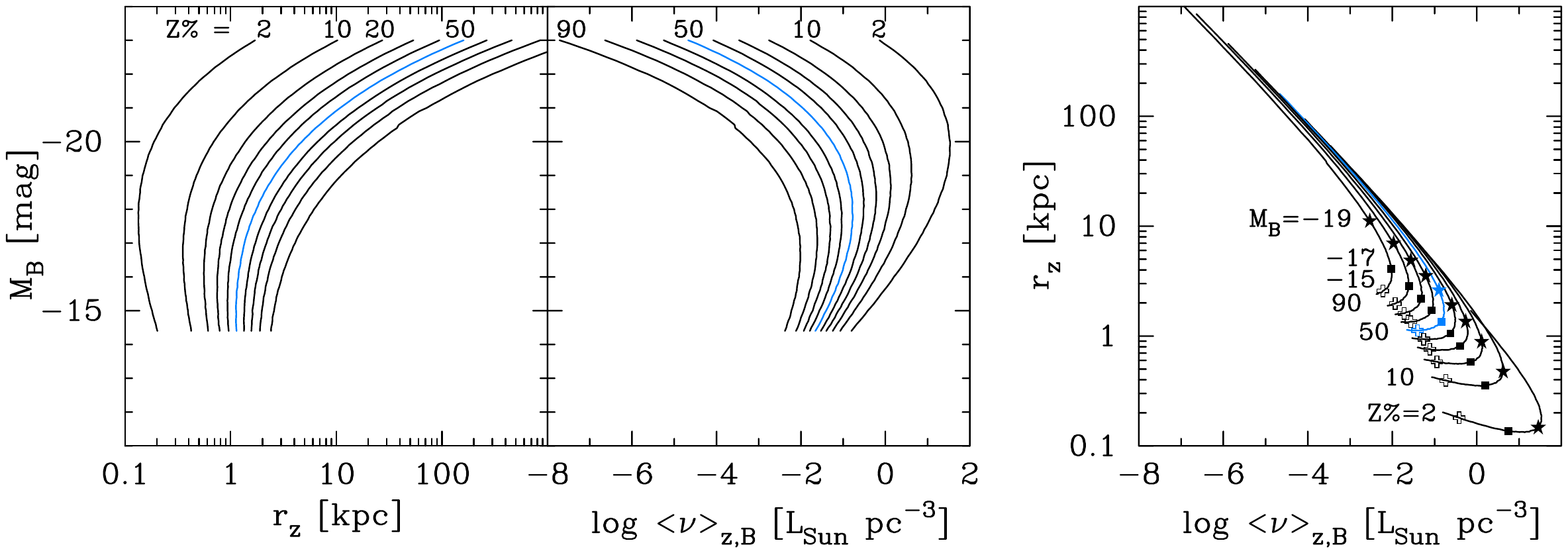}
\caption{For a range of absolute $B$-band magnitudes with S\'ersic indices
  $n>1$, the internal radius $r_z$ enclosing a sphere with $Z$ percent of the
  total light, is shown (left panel), as is the mean luminosity density
  $\langle \nu \rangle_z$ within this radius (middle panel: a somewhat similar
  pattern exists when using the internal luminosity density $\nu_z$ at $r_z$).
  The right-hand panel shows the $\log r_z$--$\log \langle \nu \rangle_z$
  relations for fractions $Z = 2, 10, 20, 30,... 90$ percent. The over-lapping
  nature of the relations for the brighter galaxies is the reason behind the
  tight $\log r_z$--$\log \nu_z$ relation reported by Graham et al.\ (2006).}
\label{Fig8}
\end{center}
\end{figure*}

Subsection~\ref{Sec_eff} explored parameters arising from projected radii, $R$,
that effectively enclosed different percentages, $Z$, of the total galaxy
light.  Here we explore parameters arising from internal radii, $r$, defining
spheres which effectively enclose different percentages of the total galaxy
light.  Trends with these internal radii $r_z$, the average
luminosity densities $\langle \nu \rangle_z$ contained within the spheres
defined by these radii\footnote{A similar pattern (not shown in order to
  minimise repetitive figures) occurs when using the internal luminosity
  density, $\nu_z$, at the internal radius, $r_z$.}, and the absolute
magnitude, are investigated.  The results can be seen in Figure~\ref{Fig8}.

As with the projected effective parameters, the internal effective parameters
display a similar behaviour of strongly curved relations, in which the midpoint of
each curve depends on $Z$ and therefore obviously does not reflect a
separation based on physically different formation processes.  
The midpoint of the 
bend shifts from roughly $-$19 to $-$16 mag as $Z$ changes from 2 to 90
percent.  A value of $Z=50$ corresponds to the internal half light radius
$r_{\rm e}$ and the mean luminosity density $\langle \nu \rangle_{\rm e}$, traced
by the $Z=50$ curves in Figure~\ref{Fig8}. The similarity between the $Z=50$
curve in the left-hand panel of Figure~\ref{Fig8} and the $Z=50$
curve in the left-hand panel of Figure~\ref{Fig3} was expected, given that 
$r_{\rm e} \approx 4/3 R_{\rm e}$ (Ciotti 1991). In addition, the similar
patterns seen in both panels means that the different 
ratio of radii ($r_{z_1}/r_{z_2}$) at fixed absolute
magnitude, for different percentages $z_1$ and $z_2$, will result in 
$\sigma^2 r/G$ mass estimates that depend on the percentage used to define
$r$.  As was seen in the middle panel of Figure~\ref{Fig3}, the ratio of radii again
increasingly vary as the luminosity increases.  This also coincides with an
increased steepening of the velocity dispersion profile, impacting estimates
of the dynamical mass (e.g.\ Wolf et al.\ 2010, 2011; Forbes et al.\ 2011, see
their
section~9.2) and further undermining the use of $\sigma^2 r_{\rm e}/G$ in the
brighter, non-dwarf, ETGs. 

The right-hand panel of Figure~\ref{Fig8} reveals that 
one can expect a strong $\log r_{\rm e}$--$\log \langle \nu \rangle_{\rm e}$
relation for bright ETGs.  This is because if one mis-measures the internal
radius enclosing 50 percent of the light, and obtains a radius containing say
20 or 80 percent of the total light, the associated mean luminosity density
that one measures will largely shift one along the $\log r_{\rm e}$--$\log
\langle \nu \rangle_{\rm e}$ relation for bright ETGs (see also Trujillo et
al.\ 2001, their section~4). 
At low luminosities, faint of the midpoint of the bend in these
curved relations, the same such mis-measurement will move one away from the
curved $\log r_{\rm e}$--$\log \langle \nu \rangle_{\rm e}$ relation.  This
behaviour can be seen in the $\log r_{\rm e}$--$\log \nu_{\rm e}$ diagram of
Graham et al.\ (2006, their Figure~2b).  

The collective broadening that can be
seen at faint absolute magnitudes in the right-hand panel of
Figure~\ref{Fig8} can be compared with the right-hand panel of
Figure~\ref{Fig4}. Note that Figure~\ref{Fig8} only shows data for
light profiles with $n>1$ ($\mathfrak{M}_B < -14.3$ mag). Also bear in mind
that the mean surface brightness (Figure~\ref{Fig4}) is 2.5 times the
logarithm of the mean intensity, hence the greater range along the $\langle
\mu \rangle_{\rm e}$ axis in Figure~\ref{Fig4} than compared to the $\log
\langle \nu \rangle_z$ axis in Figure~\ref{Fig8}.

\begin{figure*}[ht]
\begin{center}
\includegraphics[angle=270, trim=6.5cm 1.7cm 1.5cm 2.0cm, width=0.8\textwidth]{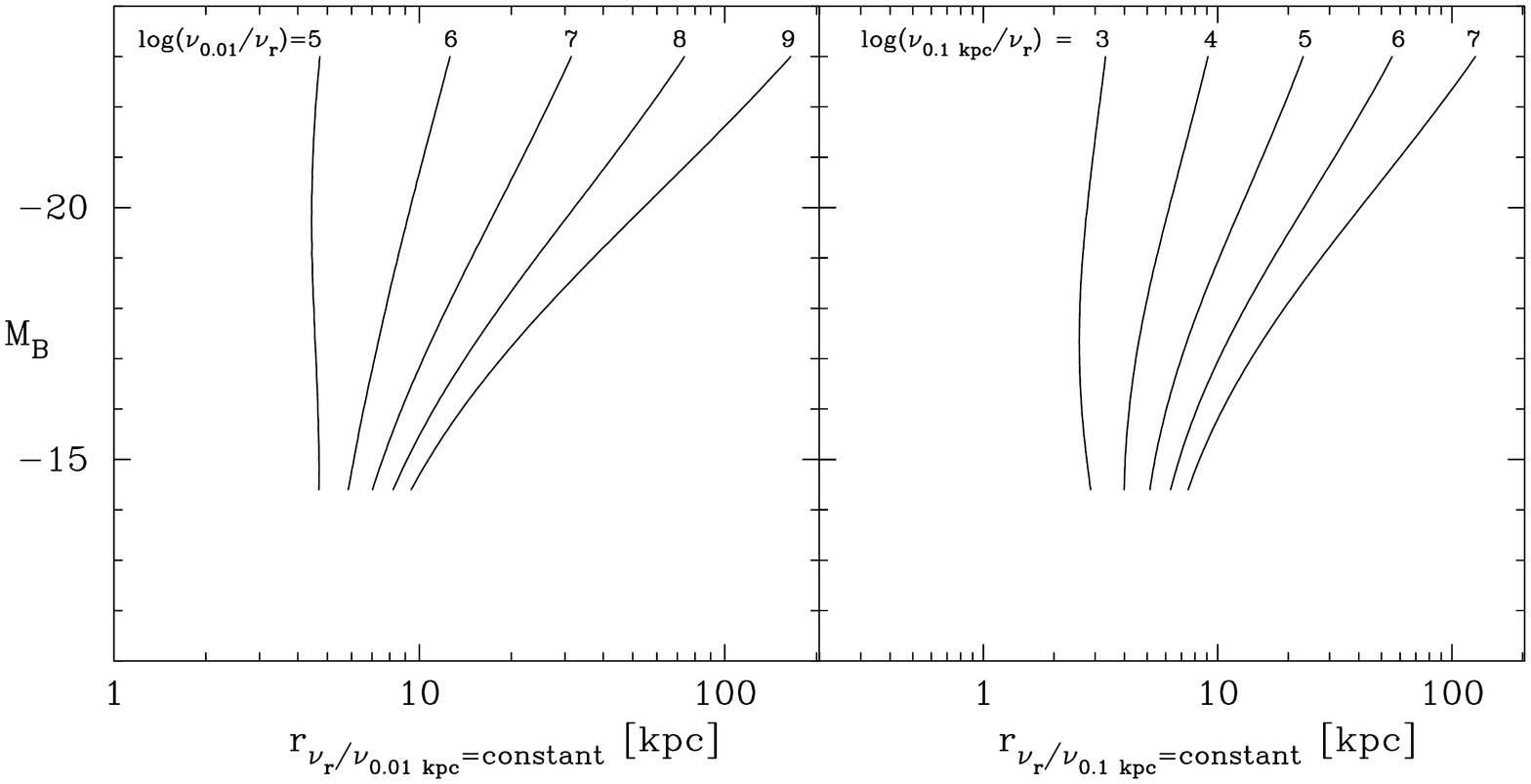}
\caption{For each absolute magnitude, $\mathfrak{M}_B$, one can see the internal radius
  where the logarithm of the luminosity density ($\nu$, in units of $L_{\odot,B}$ pc$^{-3}$)
has decreased by a fixed amount from its value at $r=0.01$~kpc (left panel)
and $r=0.1$~kpc (right panel).} 
\label{fig9}
\end{center}
\end{figure*}

\subsection{Alternative internal radii}

Similar to subsection~\ref{Sec_Alt}, we can explore the internal radii $r$ where
the internal density, $\nu$, has dropped by a fixed amount from the value at
some inner radius. Figure~\ref{Fig7} reveals that, for S\'ersic indices
$n>0.5$--1, the internal density profile rises steeply with decreasing radius.
In the middle and right-hand panels of Figure~\ref{fig5}, this rapid
brightening of the projected surface brightness was circumvented by starting
from the radius $R=0.01$ and 0.1~kpc.  Here, we start at $r=0.01$ and
0.1~kpc to compute the internal radii where the internal luminosity density
profile has dropped by a fixed amount from the density at these two inner radii.
Figure~\ref{fig9} shows these alternative scale radii as a function of the
absolute $B$-band magnitude (for magnitudes corresponding to $n>1$).  The
relations seen there do not support an ETG divide at $\mathfrak{M}_B \approx
-18$~mag.

\subsection{Isodensity radii}

Isodensity radii define a two-dimensional surface, such as a sphere, within a
three-dimensional space.  These radii are the internal analog to the projected
isophotal radii seen in subsection~\ref{Sec_iso}, and are naturally considered a
better measure to define the radii of three-dimensional stellar systems. This
is simply because isophotes can display an artificial (not physical) contour,
arising from the projected column density through a galaxy, rather than a real
boundary of equal density.

\begin{figure}[ht]
\begin{center}
\includegraphics[angle=270, trim=7cm 3.8cm 1cm 12.0cm, width=\columnwidth]{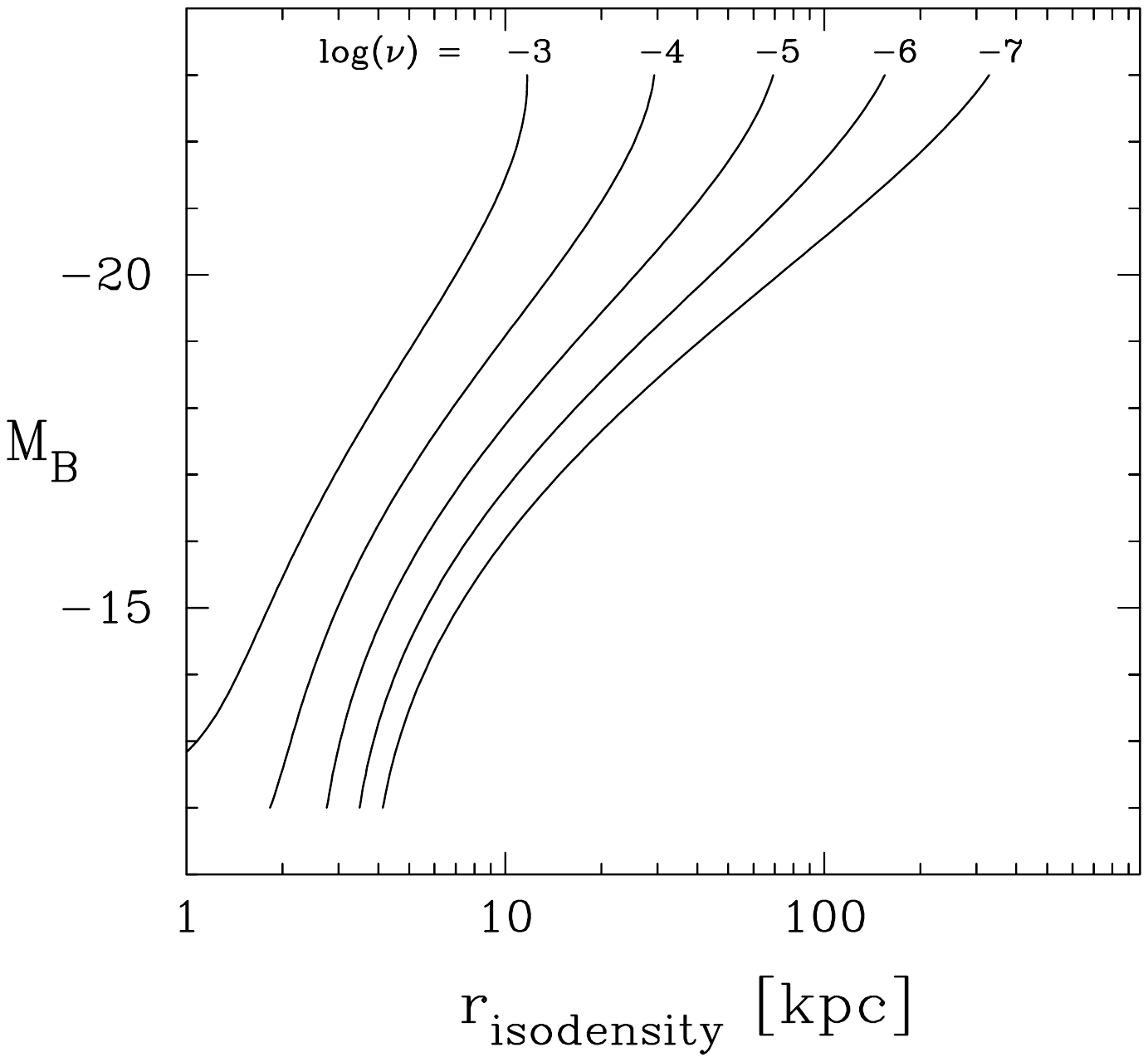}
\caption{For each absolute magnitude, $\mathfrak{M}_B$, one can see the internal radius
  where the logarithm of the luminosity density ($\nu$, in units of $L_{\odot,B}$ pc$^{-3}$)
equals one of five different values.} 
\label{Fig10}
\end{center}
\end{figure}

Figure~\ref{Fig10} displays the $\mathfrak{M}_B$--$R_{\rm isodensity}$ 
relations for five different luminosity densities. 
The trends reveal no evidence for a divide at $\mathfrak{M}_B \approx -18$ mag.

While this concludes the recapitulation of the previous section, but performed
using internal parameters, it would be somewhat incomplete to proceed without
having used the virial radii which are popular among theorists.  Therefore,
the following subsection presents this, along with an observer-inspired
variation.

\subsection{Virial radii}

\begin{figure}[ht]
\begin{center}
\includegraphics[angle=270, trim=4cm 3.8cm 2cm 10.0cm, width=\columnwidth]{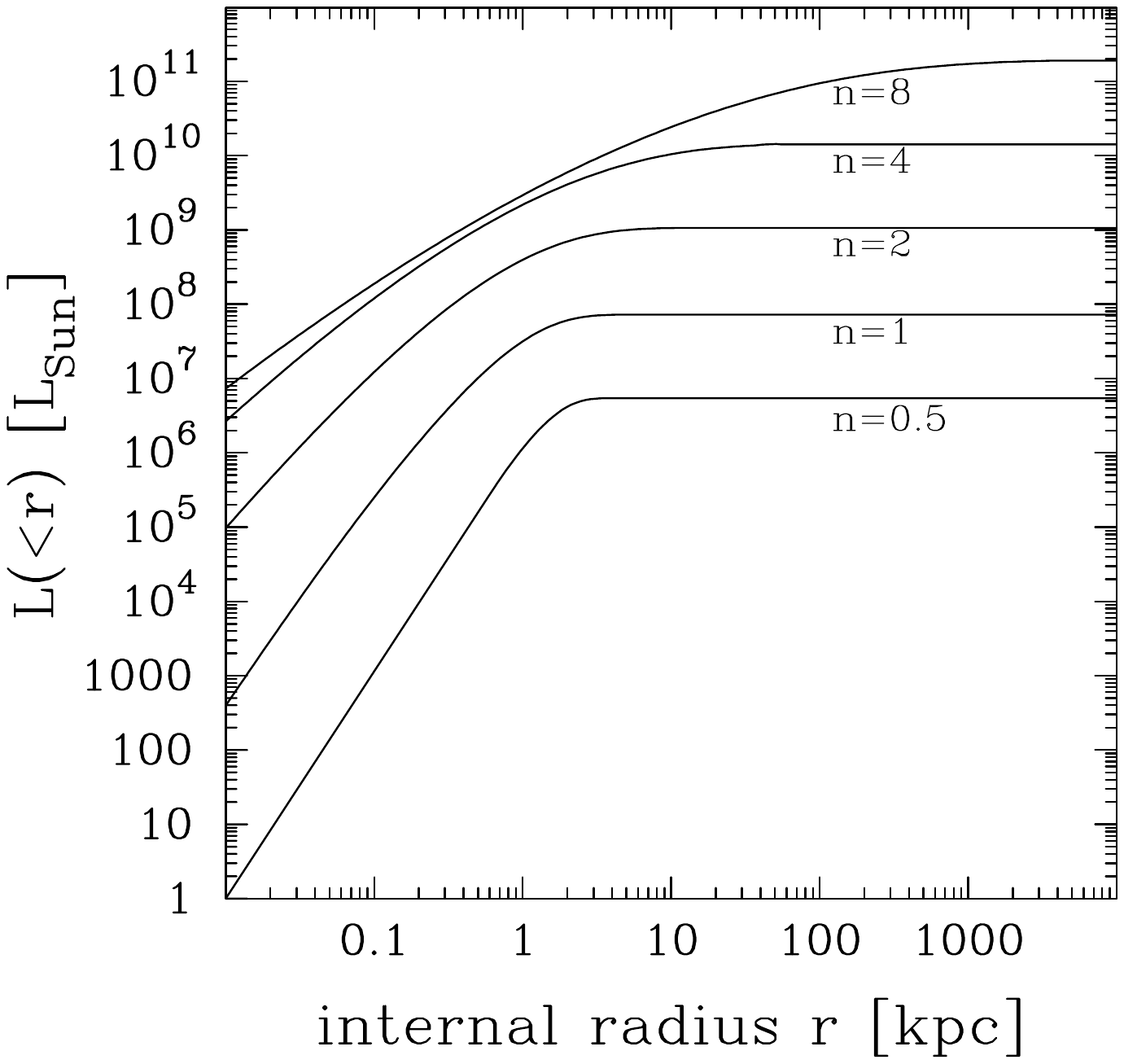}
\caption{Representative, cumulative luminosity profiles for different S\'ersic
  indices $n$, as matched to the light profiles shown in Figure~\ref{Fig1}.}
\label{Fig11}
\end{center}
\end{figure}

The integrated luminosity, within spheres centered on a galaxy, is given by 
\begin{equation}
L_{n}(s)=I_{\rm e}{\rm e}^{b}R_{\rm e}^{2}\, 4\pi \int _{0}^{s}\nu _{n}(s')s'^{2}ds',
\end{equation}
and is shown in Figure~\ref{Fig11} for our representative set of profiles
from Figure~\ref{Fig7}.

Multiplying by a stellar mass-to-light ratio gives the cumulative stellar mass
profiles. 
The luminosity density profiles $\nu(r)$ (equation~\ref{Eq_dens}) 
were converted into stellar mass density profiles $\rho(r)$, using 
a constant $B$-band stellar mass-to-light ratio of $M/L_B=8$. 

\begin{figure}[ht]
\begin{center}
 \includegraphics[angle=270, trim=7cm 3.8cm 1cm 12.0cm, width=\columnwidth]{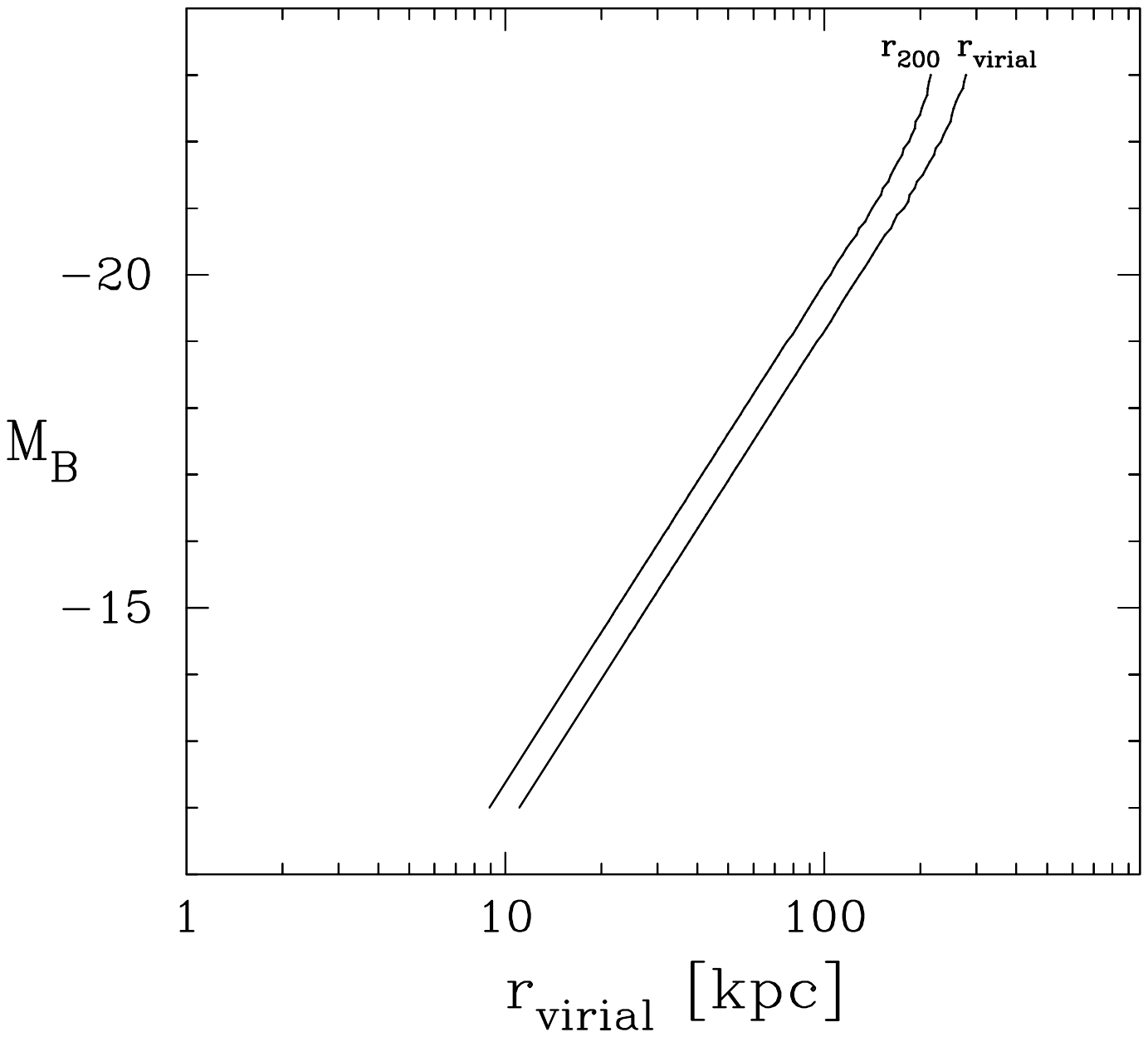}
\caption{The virial radius is shown as a function of the $B$-band absolute
  magnitude $\mathfrak{M}_B$.  The average stellar mass density within the virial radius
  equals $104.5\, \rho_{\rm critical}$.  Also shown is the radius $r_{200}$ 
 within which the average density equals $(200/\Omega_{\rm matter}) \rho{\rm
   critical}$.  The slope equals $-$3, except for the luminous galaxies with high S\'ersic
 indices, and thus long tails to their light profiles, with stars beyond the
 virial radius. 
}
\label{Fig12}
\end{center}
\end{figure}

Following Macci\`o et al.\ (2003, their equation~1.1; 
see also Bryan \& Norman 1998), 
a proxy\footnote{Zemp (2014) explains why this is a proxy for the
  virialised region.  Moreover, only the stellar mass for
  this alternate measure of galaxy size is considered here.} is used 
for the virial radius defined as the radius of the sphere within which the
average (stellar mass) density is equal to 
$18\, \Pi^2\, \Omega_{\rm matter}^{0.45}\, \rho_{\rm critical} 
\approx 177.7 \times 0.589 \times \rho_{\rm critical} 
\approx 104.6\, \rho_{\rm critical} 
\approx (339.5\, \Omega_{\rm matter})\, \rho_{\rm critical}$.  
The Planck 2015 results (Planck et al.\ 2016) give 
$\Omega_{\rm matter}=0.308\pm0.012$, and thus $339.5\, \Omega_{\rm matter} =
104.6$.  
They also report 
$H_0 = 67.8\pm0.9$ km s$^{-1}$ Mpc$^{-1}$, 
and thus $\rho_{\rm critical} \equiv 3H^2_0/(8\Pi G) = 0.864\times10^{-26}$ kg
m$^{-3}$, or $39.3\, M_{\odot}\, {\rm kpc}^{-3}$.   
The popular, and smaller, $r_{200}$ radius
(Carlberg et al.\ 1997), within which the average (stellar mass) density is
equal to $200 \rho_{\rm critical}$ is additionally calculated. 
The virial radius and the $r_{200}$ radius (associated with the stellar mass,
and thus ignoring any potential dark matter halo) is shown in
Figure~\ref{Fig12} as a function of the absolute 
magnitude\footnote{Although Kravtsov (2013, see also Huang et al.\ 2017) 
reports an approximately linear $r_{\rm
  e}$--$r_{200}$ relation, apparently at odds with the results in
Figure~\ref{Fig12} and the left panel of Figure~\ref{Fig8}, it needs to be
remembered that 
his relation based on ``abundance matching'' of simulated dark matter halos with
stellar mass estimates of real galaxies
uses an $r_{200}$ that pertains to the dark matter halos.  As such, it is therefore not
equivalent to the $r_{200}$ used here and is thus not at odds with the results
reported here. 
}.  

For $\mathfrak{M}_B \gtrsim -22$ mag, the slope of the
$\mathfrak{M}_B$--$r_{\rm virial}$ and $\mathfrak{M}_B$--$r_{200}$ relation is
7.5.  Converting the magnitude axis to $\log({\rm luminosity})$, and applying a
constant stellar mass-to-light ratio (as was assumed for calculating the
virial radii), would give a slope of 7.5/2.5=3 for the logarithmic mass-size relation.
Thus, for $\mathfrak{M}_B \gtrsim -22$ mag, the radii are large enough that they
enclose the bulk of the stellar mass $M$, and thus the pursuit of a constant,
average enclosed density (mass/volume) is simply giving radii that meet the
condition $M \propto r^3$.  As such, the masses may as well be point masses, as
the information in the density profile is effectively lost.

\begin{figure}[ht]
\begin{center}
 \includegraphics[angle=270, trim=6.5cm 3.5cm 1.5cm 7.0cm, width=\columnwidth]{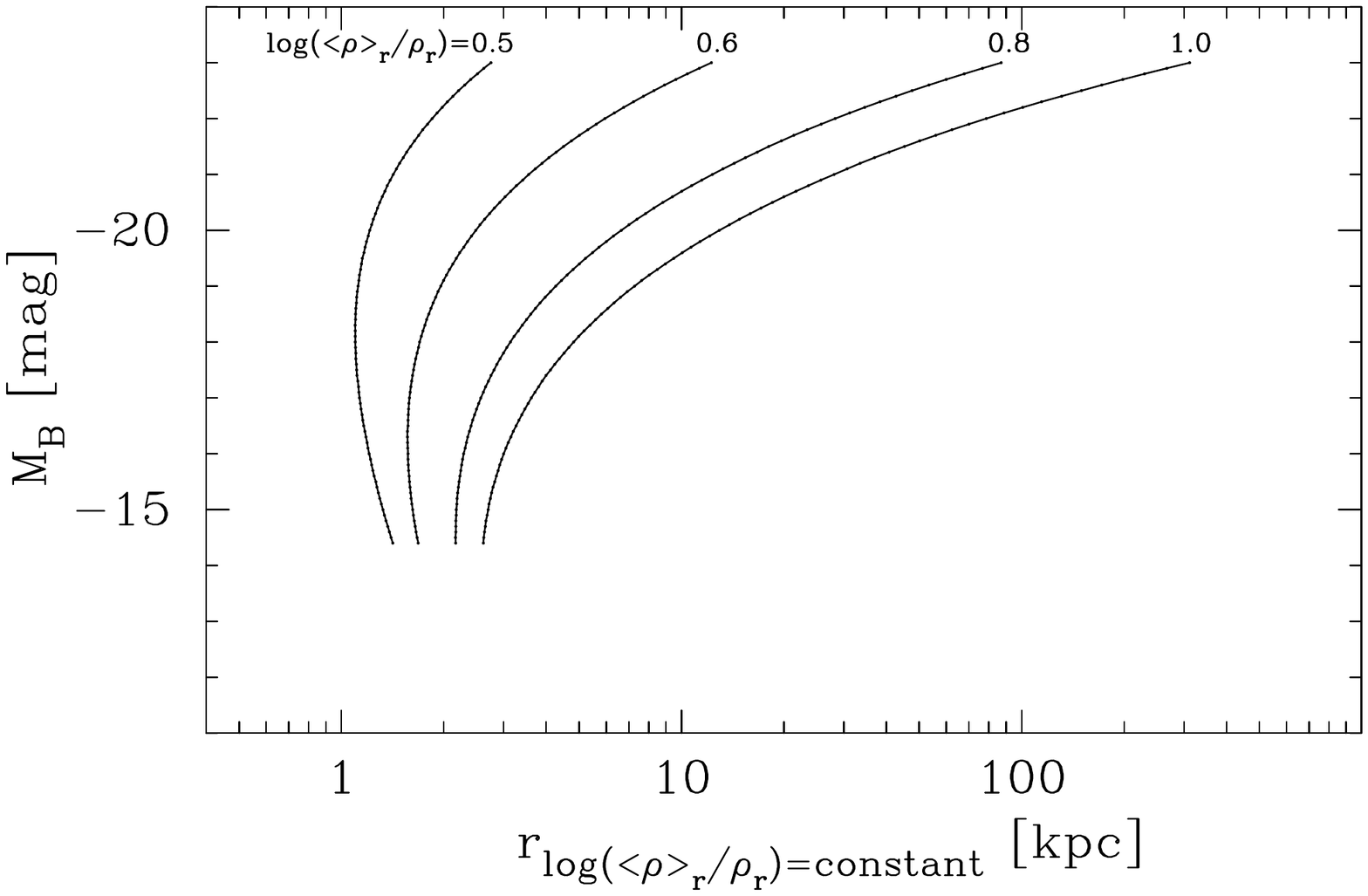}
\caption{Internal radius where the mean enclosed density equals some fraction of
  the density at that radius.
}
\label{Fig_Petro-like}
\end{center}
\end{figure}

\subsubsection{A variation}

Here, a new internal galaxy radius, 
$r_{\rm g}$, is introduced. It is such that the average density 
within this radius equals some fraction of the local density at that 
radius.  Mathematically, this can be thought of as a variation of the virial
radius, which can be expressed as 
\begin{equation} 
\log (104.6)=\log \langle \rho \rangle_{r_{\rm virial}} - \log \rho_{\rm
  critical}. 
\end{equation}
The variation introduced here can be written as 
\begin{equation}
\log ( \mathrm{H} )=\log \langle \rho \rangle_{r_{\rm g}} - \log \rho(r=r_{\rm g}). 
\label{Eqn-Eta}
\end{equation}

This radius is somewhat akin to Petrosian (1976) radii, used by observers,
which is such that the average intensity within some projected radius $R_{\rm
  P}$ divided by the intensity at that radius (denoted $\eta$) equals some
constant value, typically 5 (e.g., Bershady, Jangren, \& Conselice 2000;
Blanton et al.\ 2001). For Petrosian radii, one has the expression 
\begin{equation}
-2.5 \log [\eta=5]=\langle\mu\rangle_{R_{\rm P}}  -  \mu(R=R_{\rm P})
\label{eqn_eta} 
\end{equation}

Figure~\ref{Fig_Petro-like} presents these new galaxy radii
(equation~\ref{Eqn-Eta}) for values of $\log \mathrm{H} =$ 0.5, 0.6, 0.8 and 1.0.
One can see that they, unlike the virial radii, are no longer too large to be
unaffected by the galaxies' structure.  They behave in a fashion somewhat
similar to the internal radii containing different fractions of the galaxy
light.  Once again, no convincing evidence for a dichotomy at a fixed
magnitude is apparent.

\begin{figure*}[ht]
\begin{center}
\includegraphics[angle=270, trim=11.cm 5.0cm 0.5cm 0.0cm, width=0.75\textwidth]{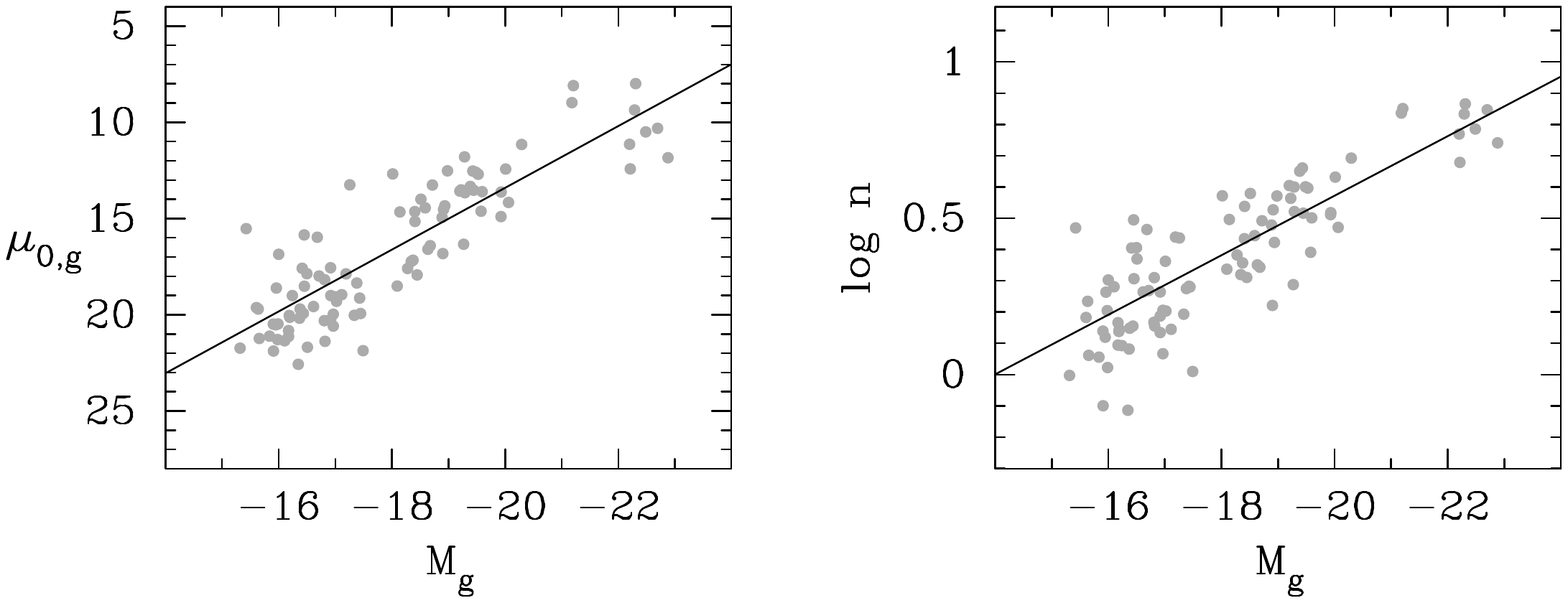}
 \includegraphics[angle=270, trim=7.5cm 3.0cm 11.0cm 1.0cm, width=0.75\textwidth]{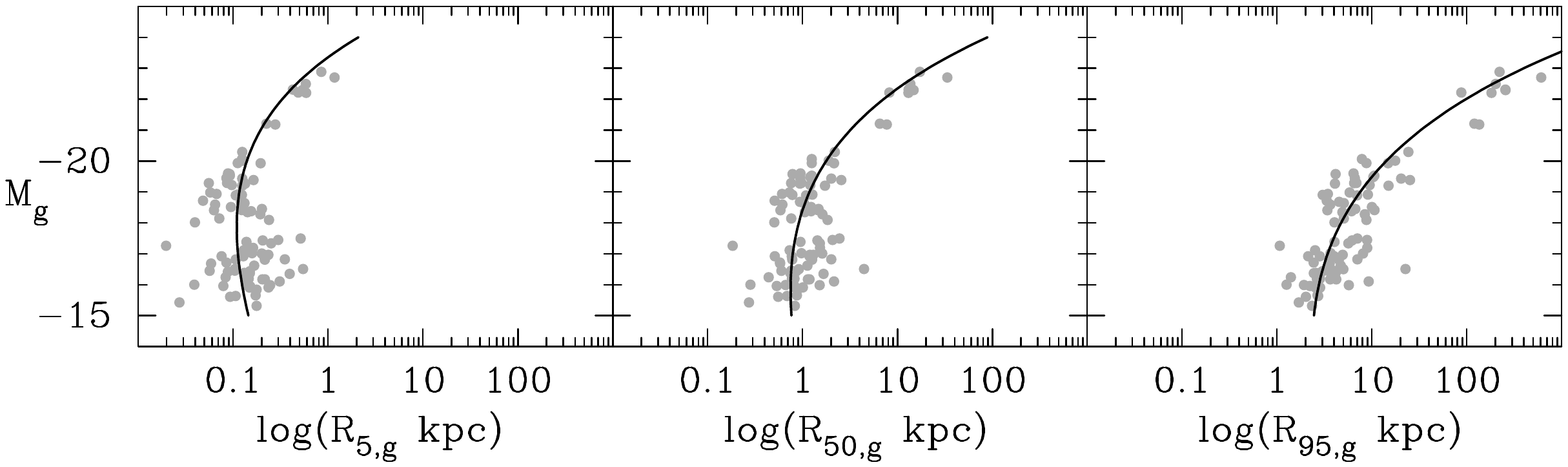}
 \includegraphics[angle=270, trim=2.0cm 3.0cm 2.0cm 1.0cm, width=0.75\textwidth]{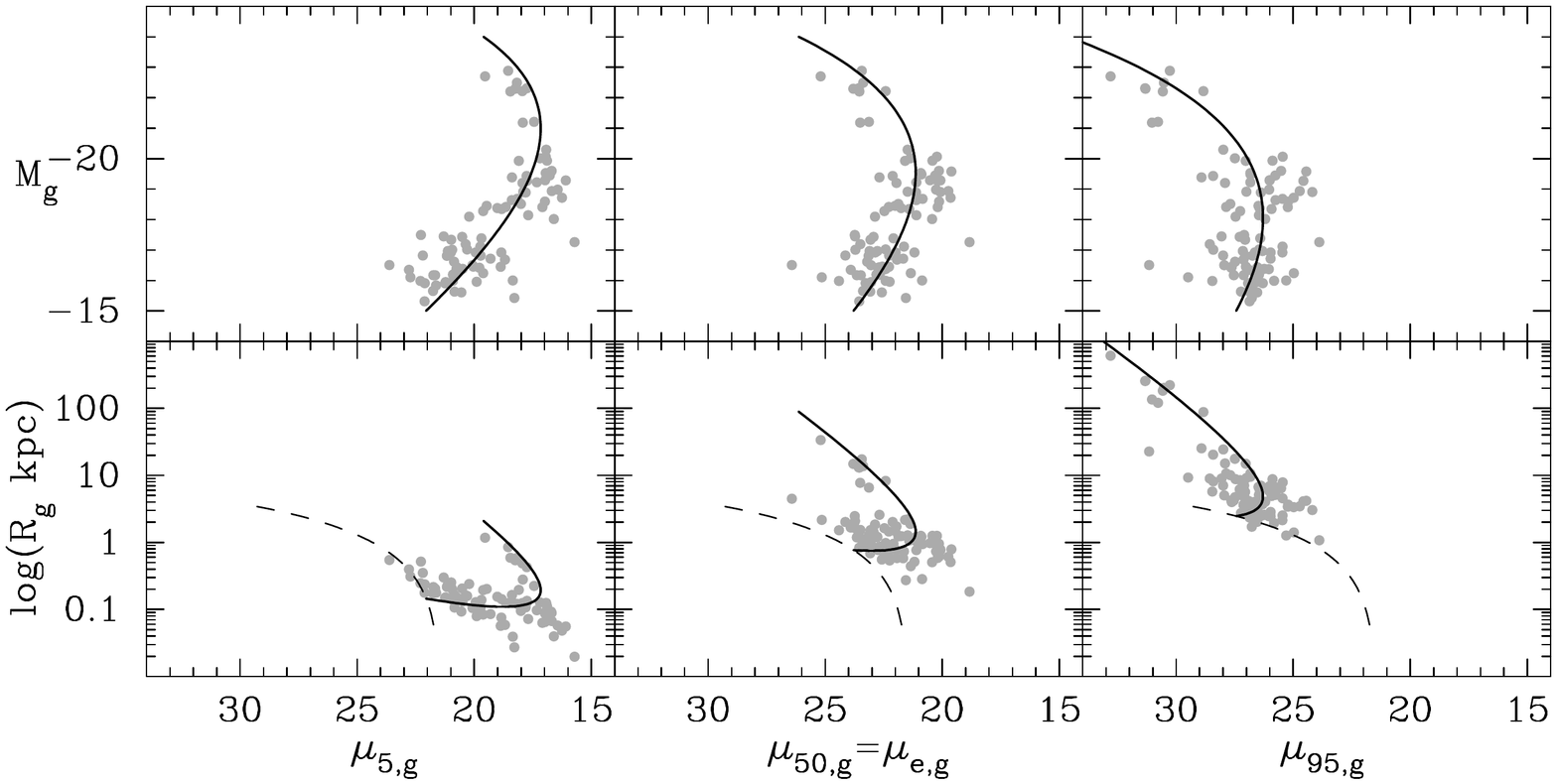}
\caption{ 
Sample of 94 Virgo cluster ETGs from Ferrarese et al.\ (2006).  Top
  row: The S\'ersic parameters, including $\mathfrak{M}_g$ and $\mu_{0,g}$,
  are from fits to the (geometric mean)-axis light profiles.  The lines are
  defined in equations~\ref{eqFerr1} and \ref{eqFerr2}.
Middle and bottom rows: 
The solid 
curves are predictions based on the linear fits in the top panels. 
The three equal dashed curves in the lower panels show 
the $\mathfrak{M}_g = -15$~mag boundary.  Looking at the two lower left-hand panels, one
might be inclined to call for a divide at $\mathfrak{M} = -20.5\pm0.5$~mag,
while looking at the two lower right-hand panels, one 
may instead be inclined to advocate for a divide at $\mathfrak{M} =
-18\pm1$~mag. 
}
\label{FigFerr2}
\end{center}
\end{figure*}

\section{Case Studies}\label{Sec_Case}

Graham \& Guzm\'an (2003) compiled data for $\sim$250 ETGs to demonstrate the
unified nature of dwarf and ordinary ETGs across the alleged divide at
$\mathfrak{M} \approx -18$~mag.  This data set included {\it Hubble Space
  Telescope} ({\it HST})-resolved dwarf ETGs plus {\it HST}-resolved,
ordinary, ETGs from Faber et al.\ (1997), among which those ETGs without
depleted cores followed the $\mathfrak{M}$--$\mu_0$ relation
(equation~\ref{Eq_Mmu}).  Since then, a few additional ETG data sets have come
forth, and it may be insightful to elaborate on alleged discrepancies or
differences using two of these.

\subsection{Case Study 1: Ferrarese et al.\ (2006)}

Ferrarese et al.\ (2006) imaged 100 Virgo cluster ETGs with {\it HST}  and the 
F475W filter (transformed to the AB photometric system's
$g$-band) as a part of the ``Advanced Camera for Surveys Virgo Cluster
Survey'' (ACSVCS; C\^ot\'e et al.\ 2004; Ferrarese et al.\ 2006).  
Ferrarese et al.\ (2006) fit seeing-convolved S\'ersic and core-S\'ersic
models (plus optional nuclear excesses) to the
(geometric mean)-axis\footnote{Geometric mean of the major- and minor-axis.}
surface brightness profiles.  As such, because they take the ellipticity
profile into account, their models can be readily integrated to obtain the
total 
galaxy magnitude.  The (surface brightness fluctuation)-based
distances from Mei et al.\ (2007) have been used to convert these model magnitudes into
absolute magnitudes, and they have been corrected for Galactic extinction using the
values from Schlafly \& Finkbeiner (2011), as tabulated in the NASA/IPAC
Extragalactic Database (NED)\footnote{http://nedwww.ipac.caltech.edu}.

Of these 100 galaxies, 2 (VCC~1535; VCC~1030) could not be modelled by
Ferrarese et al.\ (2006) due to dust, and 2 (VCC~1250, and VCC~1512) have
core-S\'ersic fits which Ferrarese et al.\ (2006, their section~4.2) discredit
--- and rightfully so given that the $R_{\rm e}$ values hit their limit of 490
arcseconds, as was also the case for VCC~575 ($\mathfrak{M}_B = -17.61$~mag).
In addition to these five galaxies,  the S0 galaxy VCC~1321
(NGC~4489, $\mathfrak{M}_B = -18.20$~mag) which was reported to have an
unusually high galaxy S\'ersic index of $\sim$6 (cf.\ $2.3\pm0.5$ from
Table~C1 of Krajnovi\'c et al.\ 2013) is also excluded. 

Here, we will see how the linear $\mathfrak{M}$--$\mu_0$ and
$\mathfrak{M}$--$n$ relations\footnote{Equations~27 and 30 in Ferrarese et
  al.\ (2006) are similar to equations~\ref{Eq_Mn} and \ref{Eq_Mmu}, but are a
  blend of $B$-band data and $g$-band data on the AB photometric system.}, 
spanning the Virgo ETG sample's full magnitude
range, explain the curved trends in diagrams involving effective radii and
effective surface brightnesses.  The following
two $g$-band equations approximate the distribution of 
data seen in the upper panels of Figure~\ref{FigFerr2}: 
\begin{eqnarray}
\mathfrak{M}_g &=& -10.5\log(n) - 14.0, {\rm and} \label{eqFerr1} \\ 
\mathfrak{M}_g &=& 0.63\mu_{0,g} -28.4, \label{eqFerr2}   
\end{eqnarray}  
The predicted $g$-band $\mathfrak{M}$--$\mu_z$ and $\mathfrak{M}$--$R_z$
distributions are presented in the middle and lower panels of
Figure~\ref{FigFerr2} for $z=0.05, 0.5$ and $0.95$.  Similar results are
obtained with the $\mathfrak{M}$--$\langle \mu \rangle$ diagram, and also when
using their data obtained through the F850LP filter.

The middle and lower panels of Figure~\ref{FigFerr2} should be compared with
Figure~117 in Ferrarese et al.\ (2006), which used quantities at $z=0.05$ and
0.5 ($Z=5$ percent and 50 percent), and compared with Figure~76 in K09 which
used quantities at 50 percent.  The series of linear equations~17 to 26 in
Ferrarese et al.\ (2006) --- used to approximate the bright and faint ends of
their $\mathfrak{M}$--$\mu_{\rm e}$, $\mathfrak{M}$--$R_{\rm e}$,
$\mathfrak{M}$--$\mu_5$ and $\mathfrak{M}$--$R_5$ distributions --- does not
adequately capture the curved nature of the scaling relations which unify the
faint and bright ETGs in these diagrams.  Their equations have been fit
separately to the core-S\'ersic and S\'ersic galaxies, implying a division
between these two galaxy types in these diagrams.  However, their set of
linear approximations are not only dependent upon the magnitude range included
in the fit, but they go against the premise of a continuity in these diagrams,
and against the understanding that the different slopes at bright and faint
magnitudes cannot be used to interpret signs of different galaxy types or
formation physics in diagrams involving ``effective'' parameters.  Ferrarese
et al.\ (2006) understood that there is a continuity at $\mathfrak{M} =
-18$~mag, and a divide at $\mathfrak{M} \approx -20.5$~mag --- as did Gavazzi
et al.\ (2005); C\^ot\'e et al.\ 2006, 2007; Misgeld et al.\ 2008, 2009; and
Chen et al.\ 2010 --- but diagrams involving effective radii and effective
surface brightnesses should not be used to make this diagnosis.  Similarly,
the colour-coding used by K09 (their Figure~76) is inappropriate and
misleading.

\subsection{Case Study 2: Kormendy et al.\ (2009)}

\begin{figure*}[ht]
\begin{center}
 \includegraphics[angle=270, trim=11.cm 5.0cm 0.5cm 0.0cm, width=0.7\textwidth]{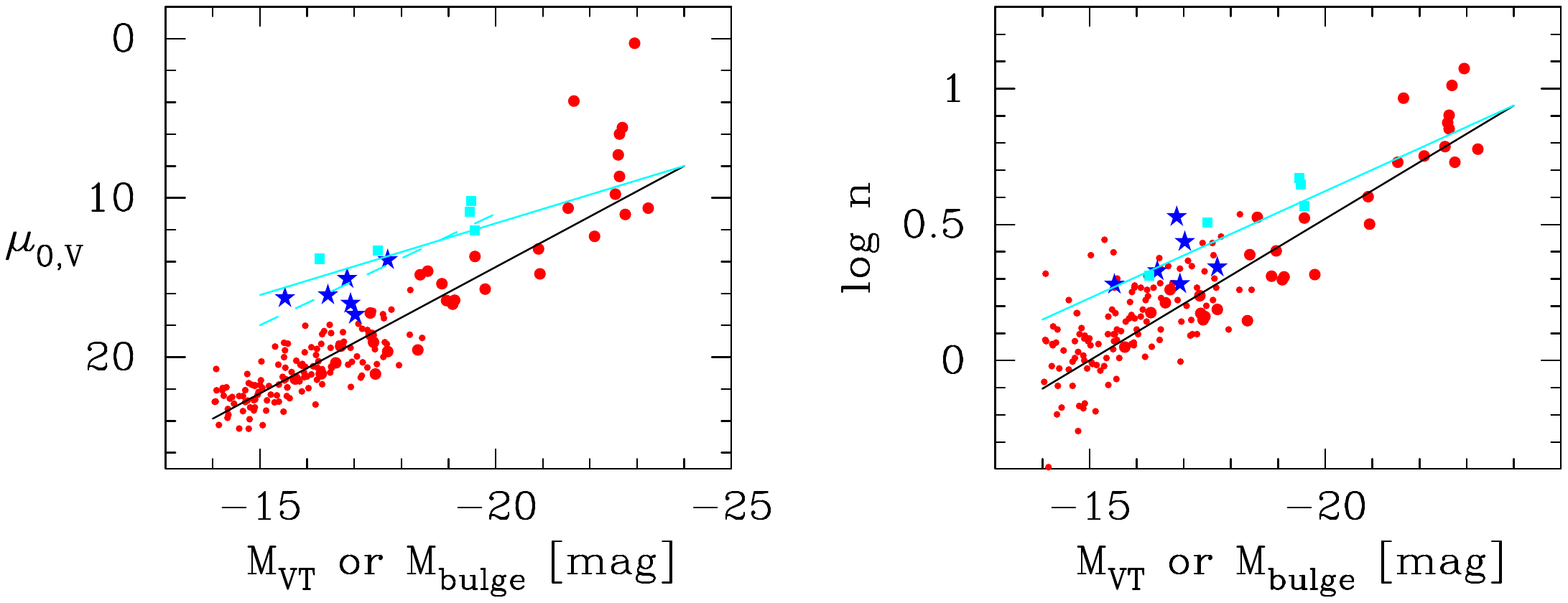}
 \includegraphics[angle=270, trim=7.5cm 3.0cm 11.0cm 1.0cm, width=0.7\textwidth]{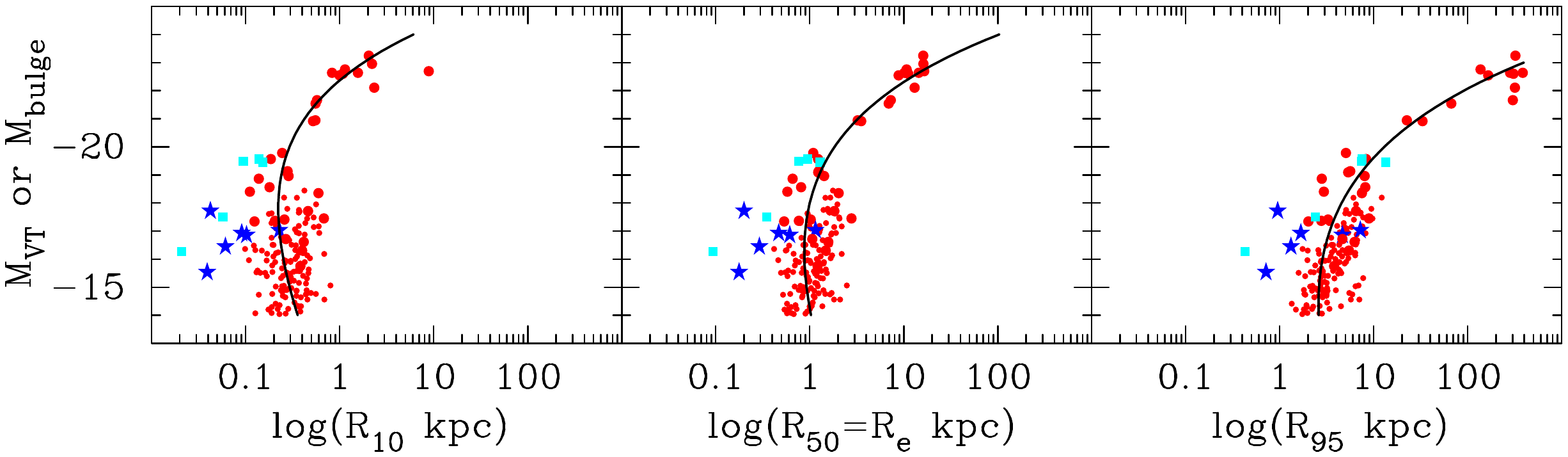}
 \includegraphics[angle=270, trim=2.0cm 3.0cm 2.0cm 1.0cm, width=0.7\textwidth]{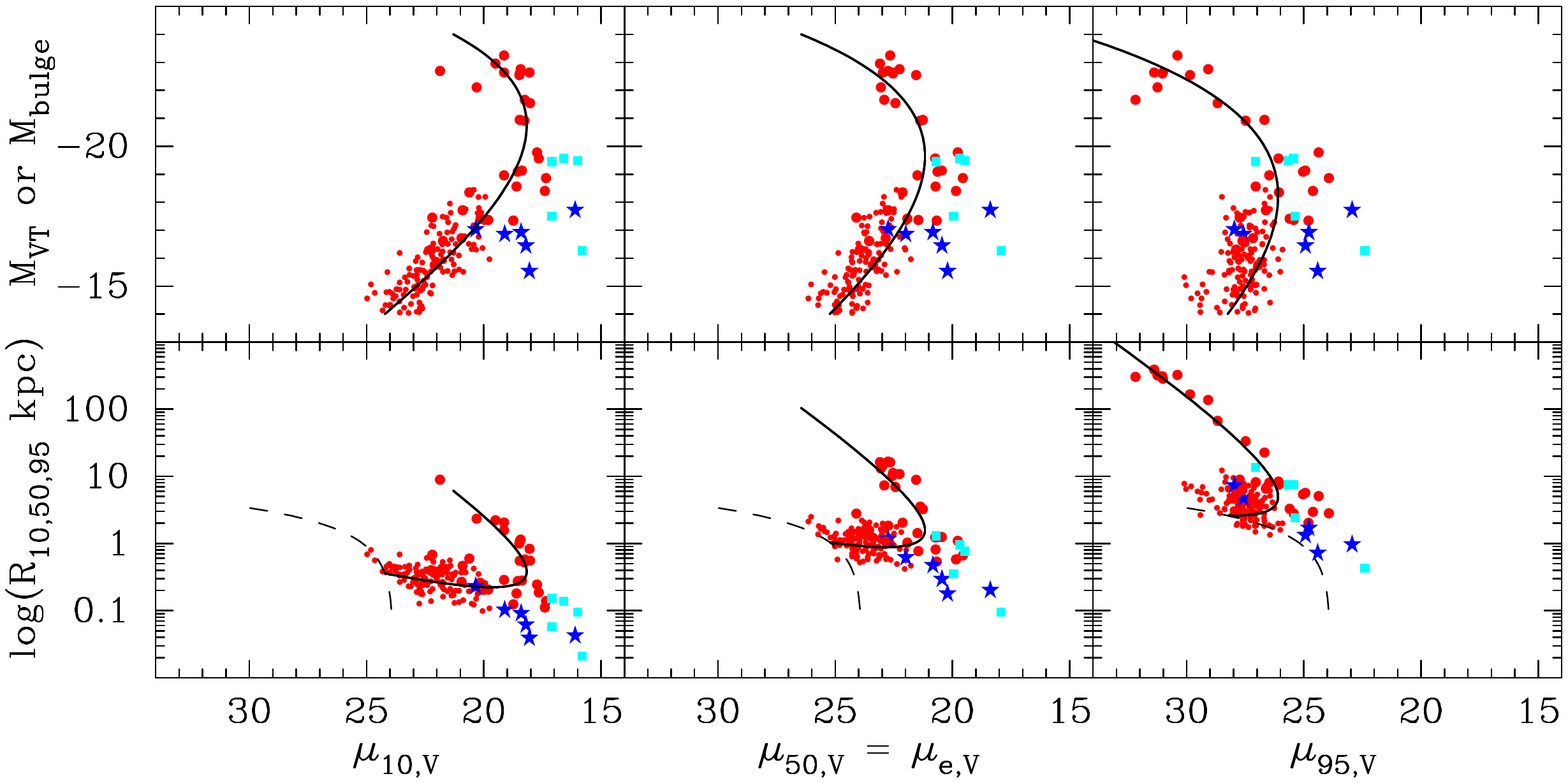}
\caption{
  Sample of 42 stellar systems in the Virgo cluster from K09,
  comprising ordinary ETGs (large red circles), 5 bulges of ETGs (light blue
  squares, NGC: 4570, 4660, 4564, 4489, 4318), 
  and 6 compact elliptical galaxies (dark blue stars, 
  VCC: 1297, 1192, 1440, 1627, 1199, 1545)  which are
  considered to be the remnant bulges of stripped disc galaxies.  
  An additional 128 Virgo cluster 
  ``dwarf'' ETGs from Binggeli \& Jerjen (1998) 
  show the extension to fainter magnitudes. A rough $B-V=0.8$ colour was
  applied uniformly to this latter sample of $B$-band data. 
  Top panel: The
  parameters are from Table~1 in K09, where: the absolute magnitudes
  $\mathfrak{M}_{V_T}$ were derived independently of the S\'ersic model for the
  ETGs and cE galaxies, and are from their column~11; the S\'ersic indices are
  major-axis values; and the central surface brightnesses are the $R=0$ values
  from their S\'ersic models fit either to the galaxy or, in 5 instances, the
  bulge component. Expressions for the red ETG lines are provided in
  equations~\ref{eqKorm1} and \ref{eqKorm2}.  
  Middle and lower rows: Similar to Figure~\ref{FigFerr2}. 
  The three equal dashed curves 
  show the $\mathfrak{M}_{V_T} = -14$~mag boundary.  Neither the absolute
  magnitudes, effective half light (50
  percent) radii, nor effective surface brightnesses in the lower middle panels
  are from the S\'ersic model but were instead obtained independently from 2D profile
  integration by K09.  The 10 percent and 95 percent radii
  and surface brightnesses were derived from the S\'ersic model.  }
\label{FigKorm2}
\end{center}
\end{figure*}

K09 accept\footnote{While K09 acknowledge
  the $\mathfrak{M}$--$n$ relation, they claim that it is insensitive to the
  formation physics of ETGs because there is no discontinuity at
  $\mathfrak{M}_B \approx -18$~mag.  The Abstract of K09 also claims that
  there is no $\mathfrak{M}$--$n$ correlation for the core-S\'ersic galaxies and almost
  no correlation for the S\'ersic galaxies.}  the $\mathfrak{M}$--$n$ relation
(see their Figure~33) but they deny the existence of a linear
$\mathfrak{M}$--$\mu_0$ relation unifying dwarf and ordinary ETGs (see their
Figure~1).  This follows on from Kormendy (1985, his Figure~3), which produced an
$\mathfrak{M}_B$--$\mu_0$ diagram with a sample selection that had an absence
of ETGs with magnitudes $-17 \gtrsim \mathfrak{M}_B \gtrsim -20.5$ mag
($H_0=50$ km s$^{-1}$ Mpc$^{-1}$), making it difficult to know where a
transition may occur.\footnote{Graham (2005) presented the Virgo cluster
  Luminosity Function for ETGs (with data provided courtesy of H.Jerjen,
  priv.\ comm.)  It showed no evidence of a division.  More recently,
  Ferrarese et al.\ (2016, their Figure~4) reviewed the luminosity function
  for all galaxy types in the core of the Virgo cluster.  A small dip is
  evident at $\mathfrak{M}_B=-17$~mag, which could disappear upon rebinning the
  histogram.}  Further complicating the interpretation was that the faint ETG
data in Kormendy (1985) did not produce the known 
$\mathfrak{M}_B$--$\mu_{0,B}$ trend (e.g.\ Binggeli et al.\ 1984, their
Figure~8)\footnote{Kormendy \& Djorgovski (1989) appear to dismiss this issue in a
  footnote, by claiming that seeing effects on dwarf galaxies contributed to
  the continuity seen by Binggeli et al.\ (1984) and others.}  whose
distribution points towards the {\it 
  faint}-end of the $\mathfrak{M}_B$--$\mu_{0,B}$ sequence for bright ETGs
with depleted cores.  That is, according to the data in Kormendy (1985), the ETGs fainter
than $\mathfrak{M}_B \approx -17$~mag follow a distribution with a steeper
slope in the $\mathfrak{M}_B$--$\mu_{0,B}$ diagram than shown in the
right-hand panel of Figure~\ref{fig0}, such that the distribution of faint
ETGs in Kormendy (1985) points to the {\it bright}-end of the distribution of
ETGs with depleted  
cores.  Despite the ongoing rejection by K09 for a unifying
$\mathfrak{M}$--$\mu_0$ relation across $\mathfrak{M}_B = -18$~mag
($\mathfrak{M}_{V_T} \approx -19$~mag), K09 did not actually show the
$\mathfrak{M}$--$\mu_0$ diagram for their data set, nor the
$\mathfrak{M}$--$\mu_5$ diagram used by Ferrarese et al.\ (2006).

K09 also excluded many of the ETGs in Ferrarese et al.\ (2006) because they wanted
to work with what they thought was a sample of predominantly one-component
galaxies, i.e.\ no lenticular galaxies.  That is, K09 effectively thinned-out
much of the population of ETGs from $-18 > \mathfrak{M}_B > -20.5$~mag. This
resulted in a sample of 42 ETGs, including 10 dwarf ETGs.  However, 5
of the 32 non-dwarf galaxies were subsequently identified as S0 galaxies by
K09.  Furthermore, another 5 were rare compact elliptical (cE) galaxies, taken
from Table~XIII of Binggeli et al.\ (1985; who note in Part~3 of their
Appendix that the M32-like galaxies are vastly outnumbered by dwarf ETGs of
similar magnitude: $-14 > \mathfrak{M}_B > -18$~mag).  The cE galaxies are
thought to be heavily stripped disc galaxies (e.g., Rood 1965; Bekki et
al.\ 2001; Graham 2002b; Chilingarian et al.\ 2009), while the isolated cE
galaxies may have either never acquired a significant disc, or may have been
ejected from a cluster after losing much of their disc (e.g.\ Chilingarian \&
Zolotukhin 2015).  The cE galaxies are two-component systems, likely dominated
by a remnant bulge, and are known to overlap with the bulges of spiral and
S0 galaxies in the scaling diagrams (e.g.\ Graham 2013, his Figure~1).  This
over-representation of cE galaxies, relative to normal galaxies, 
in K09 is inappropriate for two reasons. 
In terms of a sample providing a balanced representation of galaxies, there
should be $\sim$200 times (I.\ Chilingarian 2018, priv.\ comm.) fewer cE
galaxies than non-cE galaxies across their co-existing range in absolute 
magnitude. 
Second, the cE galaxies are more akin to bulges, and as such, they are better
compared with bulges than with parameters from single S\'ersic fits to ETGs
that typically contain bulges and discs.

As for the five galaxies identified by K09 as lenticular
galaxies, K09 performed a bulge/disc decomposition for these.  
As with their fitting of a single S\'ersic model, they did not convolve their
models with the central image's point spread function but excluded by eye the
region they considered to be affected by either nuclear excesses or a partially
depleted core. They then used 
the {\it bulge} parameters rather than the {\it galaxy} parameters for these 5
galaxies to compare with the {\it galaxy} parameters of the remaining galaxies
which they thought were pressure-supported, single-component
systems (with additional small nuclear excesses or cores).  However,
Emsellem et al.\ (2011, their Table~B1) report on the internal kinematics for
the brightest 19 of the supposed 27 ($=32-5$) ``elliptical'' galaxies in
K09.  They 
reveal that ten of these 19 are ``fast rotators'', and Krajnovi\'c et
al.\ (2013) provide bulge/disc decompositions for 7 of them.  Furthermore,
Toloba et al.\ (2015) contains internal kinematical information for 6 of the
10 ``dwarf spheroidal'' galaxies in K09, reporting that 4
of these 6 are ``fast rotators''.  K09 have therefore
plotted a mixture of bulge parameters (for 5 S0 galaxies) and galaxy
parameters (for at least 11, and likely more, S0 galaxies).  This blurs 
prospects for identifying connections in parameter scaling diagrams, and it
explains why K09 did not find the known $\mathfrak{M}$--$n$ or 
$\mathfrak{M}$--$\mu_0$ relations.  Given that bulges and ETGs follow a
different size-luminosity relation, they can not follow the same 
$\mathfrak{M}$--$n$ and $\mathfrak{M}$--$\mu_0$ relations (see
section~\ref{Sec_eff}). The 
$\mathfrak{M}$--$n$ diagram in K09 is thus a blurring of two distributions,
which have been separated here in Figure~\ref{FigKorm2}. 

The upper panels in Figure~\ref{FigKorm2} display the $\mathfrak{M}$--$\mu_0$
(Vega $V$-band mag) diagram using the data from K09.  One
can see that there is an  
$\mathfrak{M}$--$\mu_0$ relation for ETGs, although a couple of high$-n$ ETGs 
appear to have had their S\'ersic index over-estimated and their 
total magnitudes under-estimated\footnote{NGC~4552 (VCC~1632) is one of these high-$n$ galaxies, 
with a reported major-axis S\'ersic index 
equal to 9.22 (K09) and a (geometric mean)-axis S\'ersic index equal to 
7.6 (Ferrarese et al.\ 2006). 
Bonfini et al.\ (2018) reassign NGC~4552 from 
a core-S\'ersic galaxy type to a S\'ersic galaxy given that its `core' is due to
dust and, after accounting for interloping / undigested galaxy flux, 
Sahu et al.\ (2019) measure S\'ersic indices, from Spitzer 3.6~$\mu$m
images, equal to 5.42 and 5.36 for the major-axis and (geometric mean)-axis,
respectively. Krajnovi\'c et al.\ (2013) report a S\'ersic index of
6.2 for this galaxy.  Krajnovi\'c et al.\ also report indices of 5.5 and 2.9 for
NGC~4406 and NGC~4486, respectively (cf.\ 10.27 and 11.84 in K09, 
who offer reasons in their Section~7.5 
for why they may have obtained their unusually high S\'ersic index for
NGC~4406).  Their high S\'ersic index for NGC~4486 is likely due to 
the halo of intracluster light surrounding this cD galaxy (see Graham et
al.\ 1996 and Seigar et al.\ 2007 to appreciate this).
} by the (roughly) 
isophotal magnitudes advocated by K09 and used here
for comparative purposes.  
The following equations represent the lines for ETG shown in the upper row of Figure~\ref{FigKorm2}.
\begin{eqnarray}
\mathfrak{M}_{V_T} &=& -9.6\log(n) - 15.0, {\rm and} \label{eqKorm1} \\ 
\mathfrak{M}_{V_T} &=& 0.63\mu_{0,V} -29.0. \label{eqKorm2} 
\end{eqnarray}

\begin{figure}[ht]
\begin{center}
 \includegraphics[angle=270, trim=2.5cm 7.5cm 1.0cm 8.5cm, width=\columnwidth]{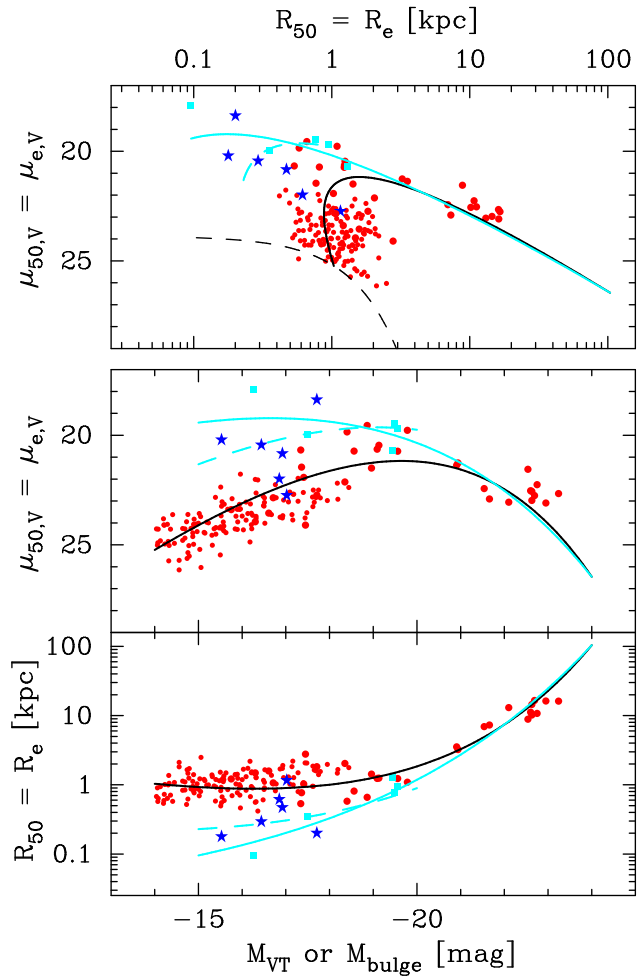}
\caption{Zoom in and summary of the effective half light parameters displayed
  in Figure~\ref{FigKorm2}.  Here, the predicted relations for bulges ---
  according to the two linear relations in Figure~\ref{FigKorm2} --- have
  also been included.  This current diagram facilitates comparison with, and
  understanding of, Figure~14 in Bender et al.\ (2015) --- where bright S0
  galaxies were often either excluded or their {\it bulge} parameters plotted, and where
  dwarf S0 galaxies always have their {\it galaxy} parameters plotted. 
}
\label{FigKorm3}
\end{center}
\end{figure}

As can be seen in Figure~\ref{FigKorm2}, the cE galaxies do not follow either
the $\mathfrak{M}$--$n$ nor the $\mathfrak{M}$--$\mu_0$ relations for ETGs.
The bulges of the S0 galaxies similarly do not follow these relations.  For a
given central surface brightness, the bulges have fainter absolute magnitudes
than the ETGs, which makes sense given that their disc light has been
excluded.  The offset to fainter absolute magnitudes in the
$\mathfrak{M}$--$n$ diagram is not as great, due to the reduced S\'ersic
indices of these bulges relative to their galaxy S\'ersic indices (which tend
to be higher due to the outer disc light).  K09 does not use
a different colour to denote the (i) cE galaxies, (ii) bulges, and (iii) ETGs 
This missing
information makes it difficult to appreciate what is going on in their scaling
diagrams.  

K09 elected to plot the
$\mathfrak{M}$--$\mu_{10}$, rather than the $\mathfrak{M}$--$\mu_{0}$ or
$\mathfrak{M}$--$\mu_{5}$, and the $\mathfrak{M}$--$\mu_{\rm e}$ diagrams,
which are shown here in Figure~\ref{FigKorm2}, along with the
$\mathfrak{M}$--$\mu_{95}$ diagram, with the bulge and galaxy-type
information included.  The exclusion of known S0 galaxies by K09, 
coupled with their use of bulge rather than galaxy parameters for
some S0 galaxies but not others, results in a thinning of the bridging
population of ETGs around $\mathfrak{M}_B = -18$ to $-20.5$~mag in their
diagrams.  This practice is particularly apparent throughout Kormendy \&
Bender (2012), Bender et al.\ 2015, and Kormendy (2016).  Furthermore, their
additional shift of focus to dynamically-hot\footnote{Due to the rotation of
  some classical bulges (e.g.\ Saha et al.\ 2012), one may hesitate to call this a
  ``dynamically-hot'' sequence for pure elliptical galaxies and bulges. 
  Sequences for bulges can be
  seen in Graham \& Worley (2008) and Graham (2013).} systems, i.e.\ the bulge
sequence (e.g.\ Balcells et al.\ 2007; Graham 2013, and references therein),
rather than the ETG sequence, came at the expense of realising the continuous
ETG sequence, i.e.\ the continuity between dwarf and ordinary ETGs, and
contributed to their ongoing belief in the artificial divide at $\mathfrak{M}
\approx -18$~mag.

Figure~\ref{FigKorm3} has been included to better help one evaluate the
colour-coding and information presented in the scaling diagrams of K09, Bender et
al.\ (2015), Kormendy (2016) and elsewhere.  The curved distribution for the
ETGs in these scaling diagrams involving the arbitrary ``effective''
parameters is not a sign of division but arises from the unity seen in the
$\mathfrak{M}$--$\mu_0$ and $\mathfrak{M}$--$n$ diagrams.  As will be broached
in Section~\ref{Sec_form}, numerous other scaling relations also display a
continuity across the alleged dwarf / ordinary ETG divide at $\mathfrak{M}_B =
-18$~mag.\footnote{Whether or not any type of sequence might unite
  three-dimensional ellipsoids, from nuclear star clusters and ultra-compact
  dwarf galaxies to bulges and pure-elliptical galaxies, does not undermine
  the continuity between dwarf and normal ETGs.}

\section{Discussion}\label{Sec_Dis}

\subsection{$\mathfrak{M}, \mu_{\rm e}$, and $R_{\rm e}$}

Although S\'ersic (1968b) plotted magnitude and mass versus $R_{\rm e}$
without yet fully appreciating the nature of ETGs\footnote{S\'ersic (1968b)
  used $R^{1/4}$-model $R_{\rm e}$ for the ETGs, most of which he
  thought were elliptical galaxies.}, i.e.\ how ETG light profiles
systematically change shape with absolute magnitude, S\'ersic (1968a) provided
the very tool, i.e.\ the $R^{1/n}$ model\footnote{S\'ersic introduced the
  $R^{1/n}$ model to describe galaxies consisting of what he thought was
  varying ratios of $R^{1/4}$-bulge and exponential-disc.}, that subsequently
resulted in the realisation that the curved distribution of ETGs in diagrams
using $R_{\rm e}$, or the associated $\mu_{\rm e}$ and $\langle \mu
\rangle_{\rm e}$, is a consequence of the continual change in the S\'ersic
index $n$ with absolute magnitude.

Both S\'ersic (1968b) and Fish (1963) deserve more recognition than received
to date.  Their ($R_{\rm e}$-based size)-luminosity (and size-mass) relations
are the linear approximation to the bright arm of the curved size-luminosity
relation for ETGs (see Graham et al.\ 2006) that is still fit by countless
papers today (e.g.\ Shen et al.\ 2003; Lange et al.\ 2015).  As noted earlier,
this linear approximation from half a century ago also simultaneously yields
the linear approximation to the bright arm of both the $\langle \mu
\rangle_{\rm e}$--$R_{\rm e}$ 
relation and the $\mu_{\rm e}$--$R_{\rm e}$
relation, which later became known as the Kormendy (1977) relation after
Kormendy was the first to plot $\mu_{\rm e}$ versus $R_{\rm e}$.

Like previous papers, Kormendy (1977) used de Vaucouleurs' $R^{1/4}$ model to
parametrize the galaxy sample, and wrote ``some experimentation suggests that
the most relevant correlation is between $B_{0V}$ [$\mu_{\rm e}$] and $r_0$
[$R_{\rm e}$]'' and that ``the interpretation of the $B_{0V}(\log\, r_0)$
[$\mu_{\rm e}$--$\log R_{\rm e}$] is unknown''.  As with Fish (1963), he had
  found that the $B$-band effective surface brightness was not constant among
  the ETG population.  For
  luminous ETGs, Kormendy (1977) 
  reported that $\mu_{\rm e} \propto 3\log R_{\rm e}$ over a
  range of 3.5~mag arcsec$^{-2}$.  Although the following papers did not fit a
  curved relation, the more complete curved distribution for ETGs can be seen
  in Kodaira et al.\ (1983), Ichikawa et al.\ (1986), and Capaccioli \& Caon
  (1991; see also Binggeli et al.\ 1984 in regard to the curved
  $\mathfrak{M}_B$--$\langle \mu \rangle _{\rm e}$ distribution).  As 
  explained here, the $\mu_{\rm e}$--$R_{\rm e}$ and $\langle \mu
  \rangle_{\rm e}$--$R_{\rm e}$ relations for bright ETGs are tight because
  ETGs with S\'ersic $n\gtrsim 3$ have similar (near over-lapping) light
  profiles.  As such, errors in $R_{\rm e}$, and thus $\mu_{\rm e}$ and
  $\langle \mu \rangle_{\rm e}$, shift bright ETGs of a given magnitude along
  these near over-lapping light profiles, maintaining a $\mu_{\rm e}$--$R_{\rm
    e}$ and $\langle \mu \rangle_{\rm e}$--$R_{\rm e}$ relation that resembles
  the $\mu$--$R$ and $\langle \mu \rangle$--$R$ light profiles (see
  Figure~\ref{Fig4}).
`<
We have seen why fainter ETGs, with progressively different light profiles, smoothly
depart from the $\mu_{\rm e}$--$R_{\rm e}$ and $\langle \mu \rangle_{\rm
  e}$--$R_{\rm e}$ relations.  Figure~\ref{Fig4} reveals why Ichikawa et al.\ (1986) were
mistaken when they interpreted the curved $\mu_{\rm e}$--$R_{\rm e}$
distribution, their Figure~12, as strong evidence for a separation of dwarf
and giant ETGs; and half of the answer as to why they were mistaken is evident
in their linear $\mathfrak{M}_B$--concentration relation (their 
Figure~11). However, they were in good company, following Michard (1979), and
their data quality and trends appear accurate. 

It should also be borne in mind that any continuum can be split into two, but
this does not mean that one has two distinct populations. For example, the
colours of ETGs become more blue as one progresses to fainter
magnitudes\footnote{As remarked in Graham \& Soria 2019, this is not the case
  for stripped galaxies, whose stellar mass has been reduced but whose colour
  remains red.} (e.g.\ Baum 1959; Visvanathan \& Sandage 1977; Driver et
al.\ 2006; Lee et al.\ 2006; Bamford et al.\ 2009; Deng et al.\ 2009;
Kannappan et al.\ 2009), and one could construct a red and blue bin with a
divide at say $\mathfrak{M}_B=-18$~mag.  Furthermore, a histogram would appear
bimodal if one excluded the bridging population from the sample, but this
would not be evidence against a single unified population but rather evidence
of sample selection.  Similarly, the observation (Figure~\ref{fig0}) that
ETGs fainter than $\mathfrak{M}_B=-18$ mag tend to have S\'ersic indices $n<2$
(exponential-like), while brighter ETGs tend to have $n>2$ ($R^{1/4}$-like),
is not evidence for two distinct populations.  Similar caution is required for
bulges.

\subsection{$\mathfrak{M}$--$\mu_0$}

Some further historical context surrounding the $\mathfrak{M}$--$\mu_0$
diagram should prove helpful and insightful. 

While de Vaucouleurs' (1948, 1953, 1959) $R^{1/4}$ model for describing bright
ETGs, and the bulges of disc galaxies, took hold, Hodge (1961a, 1961b) had
shown that de Vaucouleurs' model did not provide a good description of the
faint ETGs.  Hodge et al.\ (1965) noted, in the concluding sentence of their
article, that all of their dwarf ETGs displayed an exponential\footnote{It
  is noted that an exponential (S\'ersic $n=1$) light profile can apply to both a
  three-dimensional ellipsoid and a two-dimensional disc.}  light profile, as
was shown in their Figures~4a-4e.  The influential article by Hodge (1971)
also reported that the exponential model works well for the barred dwarf Im
galaxy IC~1613, see also de Vaucouleurs (1961a) for other Irregular galaxies.
This engendered suspicions of a connection between the discs at one end of the
Aitken-Jeans-Lundmark-Hubble
 sequence and faint ETGs at the other.  Faber \& Lin (1983) and Binggeli
et al.\ (1984) continued to show that the exponential model provides a
reasonable description of the light profiles of faint ETGs.  Kormendy (1985)
subsequently added two generic disc galaxies from Freeman (1970) into his
$\mathfrak{M}_B$--$\mu_0$ diagram, to represent dwarf Irregular and late-type
spiral galaxies near the bright-end of his dwarf galaxy distribution; see also
K09 (their Figure~1) and Tolstoy et al.\ (2009, their
Figure 1).  Kormendy (1985) went on to advocate that dwarf elliptical (dE) galaxies are better connected
with the dwarf spiral and dwarf irregular galaxies found at the other end of
the Aitken-Jeans-Lundmark-Hubble
 sequence, and do not connect with ordinary ETGs at
$\mathfrak{M}_B = -18$~mag.  This view differed from Sandage \& Binggeli (1984,
their Figure~1) who suggested a strong connection between dE and E galaxies,
and a weak connection between dE and dwarf Irregular/Magellanic-like galaxies. 
In addition, Wirth \& Gallagher (1984) had just suggested 
that compact elliptical (cE) galaxies (de Vaucouleurs 1961b; Zwicky \& Kowal
1968; Zwicky \& Zwicky 1971), rather than dwarf ETGs may be the 
low-luminosity analogs of ordinary ETGs. Indeed, Fish (1963) had also
noted that M32 defined the faint-end of his size-luminosity relation for
elliptical galaxies, and K09 maintains this view. 

With the availability of better spatially resolved data from the {\it HST},
Kormendy et al.\ (1994, see also Ferrarese et al.\ 1994 and Grillmair et
al.\ 1994) used the $R^{1/4}$ model to describe the global properties of
galaxies, and a double power-law model\footnote{See Hernquist (1990, his
  equation~43).} to describe their inner region.  In order to measure the host
galaxy light, they appropriately advocated for excluding the additional
nuclear components (common in both dwarf and ordinary ETGs, and LTGs:
e.g.\ Rest et al.\ 2001; Balcells et al.\ 2003; B\"oker et al.\ 2004), as can
be seen in the light profiles shown by Lauer et al.\ (1995), Byun et
al.\ (1996), and later K09.  While the {\it HST} data revealed that it was
galaxies with partially depleted cores that were deviating from the
$\mathfrak{M}_B$--$\mu_0$ relation defined by ordinary ETGs without cores ---
as previously reported by Oemler (1973) and Gudehus (1973) --- Faber et
al.\ (1997) claimed that if they had even better spatial resolution then they
would obtain brighter central surface brightnesses for those ETGs without
cores in the magnitude range $-18 > \mathfrak{M}_B > -20.5$~mag, thereby
uniting the core-S\'ersic and S\'ersic galaxies in the $\mathfrak{M}$--$\mu_0$
diagram and separating them from the dwarf ETGs.  Faber et al.\ (1997, their
Figure~4) used the nearby, nucleated, S\'ersic galaxy M32 as an example of
this, including the nuclear star cluster in order to report the higher surface
brightness required to make this galaxy follow the $\mathfrak{M}_B$--$\mu_0$
relation defined by the luminous ETGs with depleted cores.  This was, however,
at odds with their treatment of other galaxies where they excluded additional
nuclear components from the modelling process and obviously meant that
they were no longer working with the properties of the host galaxy but are
subject to the whims of nuclear star cluster formation.

Combining CCD images with deep, wide field-of-view, photographic data, Caon et
al.\ (1993) and D'Onofrio (1994) had shown that the S\'ersic function fits the
brightness profiles of ETGs remarkably well over a large dynamic range.
Graham \& Guzm\'an (2003) and Balcells et al.\ (2003) used the S\'ersic
$R^{1/n}$ model to quantify HST-resolved galaxies and bulges.  In addition,
they {\it simultaneously} modelled the galaxy's discs and nuclear components,
while Graham et al.\ (2003) showed how to unite the inner and outer regions of
galaxies using the core-S\'ersic model, rather than using a disconnected
double power-law for the inner region and an $R^{1/4}$ model for the outer
region.  Furthermore, unlike with the double power-law model, the S\'ersic
$R^{1/n}$ model previously applied to core-less galaxies has a finite central
surface brightness, useful for quantifying the host bulge/galaxy $\mu_0$ in
galaxies without depleted cores, such as M32.  Graham et al.\ (2003) and
Balcells et al.\ (2003) advocated the identification and quantification of
both central deficits (partially depleted cores) and excesses (additional
nuclear components) relative to the inward extrapolation of the bulge's outer
S\'ersic profile, a practice later adopted by Ferrarese et al.\ (2006) and
K09\footnote{One point of distinction with K09 is that they consider ``Core''
  galaxies, i.e.\ the core-S\'ersic galaxies, to be slowly rotating, at odds
  with the known lenticular galaxies with depleted cores (e.g.\ Dullo \&
  Graham 2013, 2014).}.  With these refinements, Graham \& Guzm\'an (2003)
showed that the $\mathfrak{M}_B$--$\mu_0$ diagram reveals no evidence for the
ETG divide at $\mathfrak{M}_B \approx -18$ mag that was advocated by Kormendy
(1985), Faber et al.\ (1997), and in numerous papers since (e.g.\ Kormendy et
al.\ 2016, and references therein).  Instead, Graham \& Guzm\'an (2003) argued
that the only magnitude of importance in this diagram is around $\mathfrak{M}_B =
-20.5$ mag, corresponding to the division between ETGs whose spheroidal
component had a S\'ersic profile versus ETGs whose spheroidal component had a
core-S\'ersic profile.  This magnitude corresponds to a mass of $(1-2)\times
10^{11}\,M_{\odot}$.

\subsection{Formation scenarios}\label{Sec_form}

It had become common to see papers refer to supernova-driven winds as the
physical explanation for creating two disconnected species of ETG, with the
winds blowing the gas out of the dwarf galaxies (e.g.\ Mathews \& Baker 1971;
Saito 1979) and making them different to ordinary ETGs.  Dekel \& Silk (1986)
invoked supernova-driven winds for creating the two distinct classes of
galaxies thought to have been observed in their $\mathfrak{M}_B$--$\langle \mu
\rangle_{\rm e}$ diagram.  However, these studies over-looked the
gravitational binding energy of the dark matter halo (Mac Low \& Ferrara
1999)\footnote{Of course today, AGN winds are invoked to blow away the gas in
  the ordinary ETGs, and both winds are expected to operate on a sliding-scale
  of efficiency unlikely to create two disjoint species of galaxy but rather a
  continuum.}.  A mechanism other than winds was therefore sought to explain
the alleged discontinuity at $\mathfrak{M}_B \approx -18$~mag.

The claim that dwarf ETGs are formed from dIrrs had been challenged by Bothun et
al.\ (1986; see also Hunter \& Gallagher 1985, their Figure~12), who reported
that the irregular galaxies did not have the right magnitudes, surface
brightnesses, nor colours, to evolve into dwarf elliptical galaxies.  
Investigating the structure of dwarf galaxies at
near-infrared wavelengths, which are less affected by dust and the light from
new stars, James (1991, 1994) similarly concluded that the dwarf Irr and dwarf
ETGs
differ markedly and it is unlikely that dwarf Irr galaxies transform into
dwarf ETGs.  
Nonetheless, given the notion that ordinary elliptical galaxies formed from
the merger of spiral galaxies (e.g.\ Toomre \& Toomre 1972), one can
appreciate why the dwarf ETG population may have been suspected to have been
morphologically transformed from something else.  The absence of enough
sufficiently low mass spiral galaxies (e.g.\ Sandage \& Binggeli 1984) 
that could merge and build the dwarf ETG 
population led to thoughts of alternative processes.  Shapley (1951) had
previously suggested that the Small Maggellanic Cloud may be evolving into a
dwarf elliptical galaxy, and Einasto et al.\ (1974), 
Frogel et al.\ (1982), Faber \& Lin (1983), Lin \& Faber (1983),
Wirth \& Gallagher (1984), and others, similarly suggested that removing the gas
content from a dwarf irregular galaxy may produce a dwarf elliptical galaxy,
after-all, they both had exponential light profiles.

Influenced by the shape of the light profile, there were speculations that the
dwarf ETGs were associated with flat discs.  However, the dEs were later shown
to be ellipsoidal, or at least not disc-like (e.g.\ Ichikawa et al.\ 1986).
That is, an exponential-like (S\'ersic $n=1$) light profile does not
necessarily imply a flat disc.  Chen et al.\ (2010) wrote that ``The
distribution of axial ratios of low-mass (``dwarf'') galaxies bears a strong
resemblance to the one observed for the higher-mass (``giant'') galaxies''.
Therefore removing the gas from late-type spiral and Irregular galaxies, via
the ram-pressure stripping from a galaxy cluster's hot X-ray gas, was not
enough to explain the alleged dwarf ETG - Irregular galaxy connection.  In
passing, it is noted that ETGs can contain substantial discs, but as a
population, they do not have the abundance of low bulge-to-total ratios as
seen in late-type spiral and irregular galaxies.

Moore et al.\ (1996) subsequently revealed how multiple fast fly-bys past a
massive perturbing galaxy, coupled with gravitational tidal forces within a
galaxy cluster, could first invoke the formation of a bar within a small
low-mass (Sd-like) disc galaxy, and then strip away the remaining outer disc,
leaving behind a somewhat harassed and puffed-up bar that was proposed to be the
answer to the existence of dwarf ETGs and the lower-mass dSph galaxies in
clusters.  This mechanism maintained the suspected association with spiral galaxies and
removed the connection with brighter ETGs.  However, while the galaxy
harassment scenario (Moore et al.\ 1996, 1998; Mastropietro et al.\ 2005, see
also Mayer et al.\ 2001a, 2001b) likely transforms low mass disc galaxies, it
is unlikely to be capable of creating ETGs with $\mathfrak{M}_B = -16$ to -18
mag by stripping away much more luminous spiral galaxies. Moreover, a
population of dynamically-heated, but still rather flat, bars has not been
observed in galaxy clusters.  Instead, the dwarf ETGs are a population of
multi-component galaxies possessing features that ordinary ($\mathfrak{M}_B
\lesssim -18$ mag) ETGs, not thought to have been built by stripping/harassing
of spiral galaxies, also contain.  The existence of isolated dwarf ETGs, shown
to have the same kinematic properties as dwarf ETGs in clusters (Janz et
al.\ 2017; Graham et al.\ 2017), reveals that dwarf ETGs need not have been
built by ``galaxy harassment''.  They are thus built by Nature
rather than Nurture: they do not require the harsh (nurturing) environment
of a galaxy cluster to form.  

The presence of disc-like features (e.g.\ Graham et
al.\ 2003b; Lisker et al.\ 2006; Lisker \& Fuchs 2009) and rotation in dwarf
ETGs in clusters --- known since Davies et al.\ (1983) --- had been heralded
as evidence that they were previously spiral galaxies (e.g.\ Boselli et
al.\ 2008; De Rijcke et al.\ 2010; Penny et al.\ 2014; Ry\'s et al.\ 2015;
Toloba et al.\ 2015), even though ordinary ETGs display the very same
features.  Such discs, observed in ETGs up to $\sim 10^{11}\,M_{\odot}$
(e.g.\ Emsellem et al.\ 2011; Krajnavi\'c et al.\ 2013), may be built from gas
accretion and minor mergers (e.g.\ Schweizer 1986; Schweizer \& Seitzer 1988; Birnboim \& Dekel
2003; Katz et al.\ 2003; Mei et al.\ 2006; Wei et al.\ 2010; Kaviraj et
al.\ 2011).   Structurally, and kinematically, dwarf and ordinary ETGs (both in
and out of clusters) display a continuum of disk-like features across the
alleged divide at $\mathfrak{M}_B = -18$~mag.

As Tolstoy et al.\ (2009) note, in their Introduction's third paragraph, dwarf
galaxies ($\mathfrak{M}_B > - 16$~mag) form a continuum with brighter
non-dwarf galaxies, in terms of potential well, size, structural, kinematic
and population features. Indeed, the dwarf and ordinary ETGs follow a
continuous luminosity-metallicity relation (Dekel \& Silk 1986, their Figure~2;
Mateo 1998; Tremonti et al.\ 2004; Veilleux et al.\ 2005) which the Sm-Irr
galaxies do not follow: the latter are less metal-rich than dwarf ETGs of the same
luminosity (Mateo 1998; Grebel 2004).  In addition, the $L$--$\sigma$ relation reveals
unity rather than division among the ETGs at $\mathfrak{M}_B = -18$~mag, with a
continuous log-linear relation, having a slope of 2 in the $B$-band, which
encompasses ETGs with $\mathfrak{M}_B \gtrsim -20.5$~mag (Davies et al.\ 1983;
Matkovi\'c \& Guzm\'an 2005; Hyde \& Bernardi 2009; Graham \& Soria 2019,
their Figure~1).

\begin{figure*}[ht]
\begin{center}
\includegraphics[angle=270, trim=10.0cm 2.5cm 1.5cm 2.5cm, width=\textwidth]{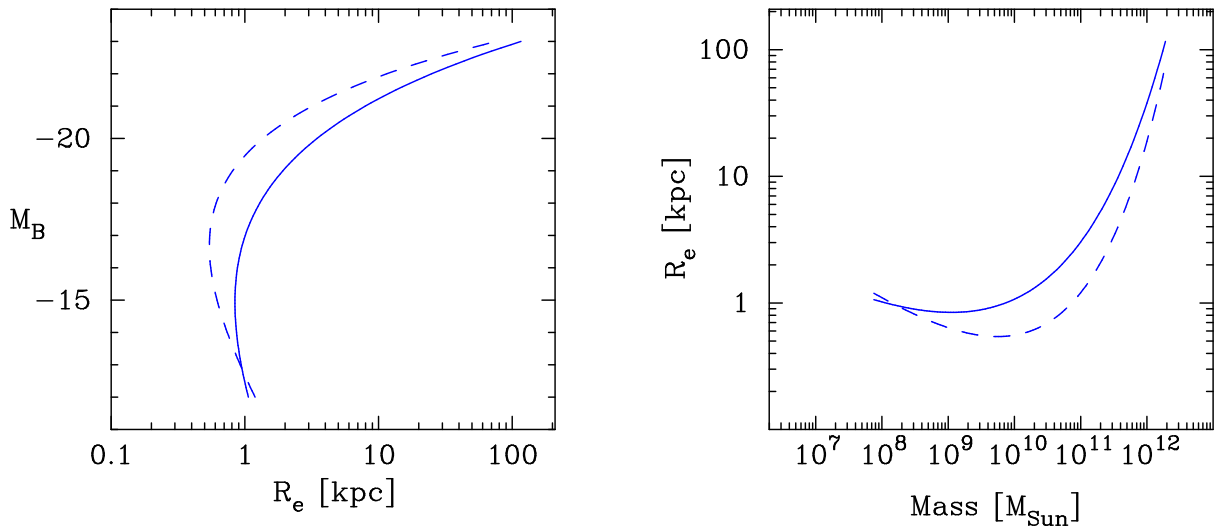}
\caption{Left panel: 
The solid curve to the right is the $Z=50$\% curve from 
Figure~\ref{Fig3}, while the dashed curve to the left is the $Z=50$\% curve for bulges
based upon equations~\ref{Eq_Mn_bul} and \ref{Eq_Mmu_bul}. 
Right panel: Similar to the left-hand panel, but using simple mass estimates as
explained in the text. While there is scope for improvement, one can see that
bulges are naturally expected to be smaller than discy ETGs of the same
mass.  At high masses, the ETGs do not have discs, and the relationships
converge. Perhaps the same is true at the low-mass end, although the upturn in
bulge size seen here at low masses, and in Fisher \& Drory (2016, their
Figure~1.4), is at odds with the bulge data in Figure~\ref{FigKorm3} and
Gadotti (2009). 
}
\label{fig_nugg}
\end{center}
\end{figure*}

\subsection{Bulges and Discs}

A large fraction of ETGs contain discs.  Careful image analysis had previously
revealed overlooked discs in many ETGs (e.g.\ Capaccioli 1987), and
subsequent kinematic studies revealed the prevalence of discs in ETGs
(e.g.\ D'Onofrio et al.\ 1995; Graham et al.\ 1998; Pedraz et al.\ 2002;
Emsellem et al.\ 2011; Scott et al.\ 2014).  In addition, modern structural
decompositions reveal multiple physical\footnote{I use the term ``physical
  component'' to differentiate between blind decompositions involving multiple
  S\'ersic components divorced from physically distinct entities.} 
  components in ETGs (e.g.\ Prieto et al.\ 2001; L\"asker et al.\ 2014; Savorgnan \&
Graham 2016; Sahu et al.\ 2019).  Nowadays, in such studies, S\'ersic's model
provides an accurate description of the spheroidal component of the galaxy.
In some instances, this is the entire galaxy, but ETGs less luminous than
$\mathfrak{M}_B \approx -20.5$~mag invariably contain a disc, and sometimes a
bar, ring, ansae, nuclear star cluster or a more extended nuclear disc
(e.g.\ de Vaucouleurs et al.\ 1991; Ferrarese et al.\ 2006; Saha et al.\ 2019). 
Not surprisingly, those studying galaxy structure have presented the 
$\mathfrak{M}$--$\log\, n$ and $\mathfrak{M}$--$\mu_0$ relations for {\it
  bulges} and observed the offset from ETGs 
(e.g.\ Andredakis et al.\ 1995; Khosroshahi et al.\ 2000;
Graham 2001; M\"ollenhoff \& Heidt 2001; MacArthur et al.\ 2003).

\subsubsection{Galaxy Bulges and Red Nuggets} 

For well over a decade, the galaxy size-luminosity relation, or rather, the $R_{\rm
  e}$--(stellar mass, $M_*$) relation, has been used to argue that the compact
($R_{\rm e} \lesssim 2$ kpc) 
massive ($M_* \gtrsim 10^{11}\,M_{\odot}$) spheroidal-shaped galaxies at
high-redshifts ($z\approx 2\pm1$) --- referred to as ``red nuggets''
by Damjanov et al.\ (2009) --- must
have evolved into large massive {\it elliptical} galaxies in the local ($z=0$)
Universe (e.g.\ Daddi et al.\ 2005; Kriek et al.\ 2006; Trujillo et al.\ 2006;
van Dokkum et al.\ 2008). The argument has been that dry merger events have
increased the size of the three-dimensional spheroidal structures,
building elliptical galaxies rather than lenticular galaxies with large scale-disks. 
However,  most early-type galaxies in 
the local universe contain fast-rotating two-dimensional discs.
In many instances, it is the  
disc which makes these galaxies large.  

Graham et al.\ (2015) identified two dozen ``compact massive spheroids'' as
the bulge component of local lenticular galaxies.  These bulges have the same
distribution of sizes, masses, and S\'ersic indices as the high-$z$ compact
massive galaxies, and comparable number densities per unit volume of space (de
la Rosa et al.\ 2016).  This similarity strongly suggests that the above
popular evolutionary scenario building bigger spheroids is not the complete
picture.  Rather, the evolution of these high-$z$ red nuggets may instead be
connected with the growth of discs (e.g.\ Caldwell 1983b; Morganti et
al.\ 2006; Sancisi et al.\ 2008; Stewart et al.\ 2009; Pichon et al.\ 2011;
Moffett et al.\ 2012, 2015; Stark et al.\ 2013; Graham et al.\ 2015, their
section~4.1; Kleiner et al.\ 2017).  Gas accretion, both preceding and during
``cosmic noon'', is expected to play a key role for the massive galaxies
(Feldmann et al.\ 2016), while the less massive galaxies, i.e.\ smaller
over-densities, take longer to accrete smaller quantities of gas in a
``down-sizing'' (Cowie et al.\ 1996) scenario (Graham et al.\ 2015, 2017).

It is insightful to compare the two key empirical relations for ETGs 
(equations~\ref{Eq_Mn} and \ref{Eq_Mmu}) with the equivalent 
relations for the bulges
of S0 galaxies and LTGs. 
Graham \& Worley (2008) compiled $K$-band data for the bulges of 
$\sim$400 spiral and S0 galaxies,
and Graham (2013) 
presented these two key equations for bulges 
(his 2.19 and 2.20).\footnote{Dust and colour gradients due to stellar
  population differences are expected to alter these equations from band to
  band.  Kelvin et al.\ (2012) and H{\"a}u{\ss}ler et al.\ (2013) provide
  insight into this.}  They are such that  
\begin{eqnarray} 
\mathfrak{M}_K &=& -7.5\log(n) - 20.0 \, {\rm and} \label{Eq_Mn_bul} \\
\mathfrak{M}_K &=& 0.6\mu_{0,K} - 29.7. \label{Eq_Mmu_bul} 
\end{eqnarray}  
We can 
take a quick look at how ETGs and bulges compare by using the following rough
transformation.  Given the current interest in compact massive spheroids, 
a simple $B-K$ colour of 4, and a stellar mass-to-light ratio of 8 is used 
for everything.  The left panel in Figure~\ref{fig_nugg} shows the
$\mathfrak{M}_B$--$R_{\rm e}$ relation for the ETGs (as seen in
Figure~\ref{Fig3}) and the expected $\mathfrak{M}_B$--$R_{\rm e}$ relation for
bulges (using equations~\ref{Eq_Mn_bul} and \ref{Eq_Mmu_bul}, and assuming
$B-K =4$).  Converting $\mathfrak{M}_B$ into a stellar mass using
$\mathfrak{M}_{\odot,B} = 5.44$ mag (Mann \& von Braun 2015) and $M_*/L = 8$,
the right-hand panel of Figure~\ref{fig_nugg} displays the size-mass diagram
using $R_{\rm e}$.  While varying stellar mass-to-light ratios are expected to
alter the low-mass end, and it is not yet clear if the ETG bulge and the LTG
bulge relations differ, one can see that bulges are in general more compact than ETGs
(with discs) of the same mass.  One can also see that at stellar masses 
above $2\times10^{11}~M_{\odot}$, the bulge 
and ETG relations merge as the ETGs no longer have extended discs making them
larger than their bulge component.  This diagram supports the claim that
some/many of the high-$z$ compact massive galaxies are now the bulges of
today's galaxies, having accreted and built a more substantial disc (Graham
2013; Driver et al.\ 2013).  Those stellar systems which only built an
intermediate-scale disc, as opposed to a large-scale disc, are today's
``ellicular'' (ES) galaxies (Liller 1966; Graham et al.\ 2016, 2017).

\subsubsection{Pseudobulges}

Balcells et al.\ (2003) was among the first to clearly call out lenticular and
early-type spiral galaxies for not having bulges with $R^{1/4}$ light
profiles.  Laurikainen et al.\ (2005, 2006) and Graham \& Worley (2008)
further noted that most lenticular and spiral galaxies have near-infrared
($K$-band) bulge-to-total flux ratios less than one-third, thereby placing
uncomfortable constraints on some simulations that were producing big
$R^{1/4}$-like bulges (e.g.\ Abadi et al.\ 2003; Stewart et al.\ 2008).
Weinzirl et al.\ (2009) also noted this absence of galaxies with big bulges
having 
$R^{1/4}$-like profiles and high bulge-to-total flux ratios, and Kormendy et
al.\ (2010) effectively took these trends to mean an abundance of
pseudobulges built from the secular evolution of the discs.  
While some of these galaxies may have pseudobulges built 
from bars, one needs to be careful in identifying such bulges\footnote{See Graham
(2014, his section~5) for a discussion of the difficulties with the assorted
criteria used by some to identify pseudobulges}.  Moreover, many galaxies are
alleged to contain both a pseudobulge and a classical bulge (e.g.\ Erwin et
al.\ 2003; Peletier et al.\ 2007), making a mockery of attempts to divide
galaxies based on whether they have a pseudobulge versus a classical 
bulge.

Parallel lines of thought on the bulges of disk galaxies and ETGs have
contributed to the idea of pseudobulges having (exponential)-like light
profiles and classical bulges built from different processes having
($R^{1/4}$)-like profiles (e.g.\ Carollo et al.\ 2001; Fisher \& Drory 2010,
2016, their Figure~1.4), with some authors using a S\'ersic index of 2 or 2.5
to divide bulges into one bin or the other.  Graham \& Worley (2008, their
Figure~8) showed that the S0, Sa, ...Sbc galaxy bulges (i.e.\ those with, in
general, bright magnitudes and high S\'ersic indices) roughly follow a
linear $\mu_{\rm e}$--$\log R_{\rm e}$ trend, while bulges in later type
galaxies do not.  This same pattern is evident in Fisher \& Drory (2016, their
Figure~1.5, using data from Gadotti 2009 and Fisher \& Drory 2010), who take
this as evidence for a division between bulges.  Gadotti (2009)\footnote{The
  Gadotti (2009) data is not quite as expected, in that their bulges do not
  display a slight upturn in size at masses below $\sim
  3\times10^9\,M_{\odot}$ or $n\lesssim 2$ (cf.\ Fisher \& Drory 2016, their
  Figure~1.5).}  wrote ``We show that pseudo-bulges can be distinguished from
classical bulges as outliers in the Kormendy relation'', and ``our
identification of pseudo-bulges [via this method] is not only more reliable
but also better physically motivated.'' This view was echoed by Blanton \&
Moustakas (2009) in their review of the physical properties of nearby
galaxies.  However, as we have seen, the $\mu_{\rm e}$--$\log R_{\rm e}$
diagram is strongly curved, with the value of both $n$ and $\mathfrak{M}$
at the bend midpoint changing considerably depending on the arbitrary fraction
of light used to define the effective radius.  This is, therefore, clearly not
a signature of different formation physics occurring on either side of the
bend midpoint. 

Figure~\ref{Fig4} reveals what the linear $\log R_{\rm e}$--$\langle \mu
\rangle_{\rm e}$ relation is, why it exists, and why it represents only a
segment of a more encompassing, and unifying, curved distribution shown by the 
$Z=50$ lines in Figure~\ref{Fig4} and given by equations~2.14 and 2.15 in Graham
(2013).  
Figure~\ref{Fig4} also reveals why the lower-luminosity ETGs appear to deviate from
the $R_{\rm e}$--$\langle \mu \rangle_{\rm e}$ relation, and why the scatter
is larger at the low-luminosity end.  It is a result of ETG structure
systematically varying with absolute magnitude (a proxy for stellar mass).
It is not an artifact of the S\'ersic model: $R_{\rm e}$ and $\mu_{\rm e}$
can be measured independently of the S\'ersic model, and the curved trend
remains (e.g\ Trujillo et al.\ 2001; Figure~\ref{FigKorm2}).  
As revealed in Section~\ref{Sec_eff}, the absolute magnitude and
S\'ersic index corresponding to the bend midpoint is not a physically
meaningful quantity that can be used to separate dwarf ETGs from ordinary
ETGs, nor should it be used to divide pseudo-bulges from classical bulges.
This becomes increasingly apparent when one realises that the bend midpoint
occurs at a range of different magnitudes (and S\'ersic indices) simply
depending on what arbitrary fraction was used to measure the radii.

\subsection{Fundamental Planes}\label{Sec_planes}

Bright ETGs, with S\'ersic indices $n \gtrsim 3$, have similar light profiles
over their inner 0.01--1.0 kpc, once excluding the presence of their depleted
core or additional small scale nuclear components such as nuclear discs,
nuclear bars, AGN, or star clusters.  One will therefore observe, from a
sample of such bright ETGs with $n \gtrsim 3$, a clear relation between core
radius and the associated surface brightness, even if there is no depleted
core or if one dramatically over-estimates the core radius (so long as the
estimate is $\lesssim$1 kpc).  This was pointed out by Dullo \& Graham (2012,
see their Figure~18).

Among the ETGs with $n \gtrsim 3$, the ETGs with higher (lower) S\'ersic
indices, which are known to have higher (lower) stellar velocity dispersions
(Graham et al.\ 2001), can be seen to have slightly higher (lower) surface
densities in Figure~\ref{Fig1}.  One can readily appreciate how folding in the
stellar velocity dispersions with the (correct or incorrect) core radii and
associated core surface brightnesses, to create a ``core fundamental plane''
(Faber et al.\ 1997), will result in less scatter than that about the
$R_{\rm core}$--$\mu_{\rm core}$ relation.  This is because ETGs with $n
\gtrsim 3$ that also have light profiles brighter (fainter) than the median
light profile of ETGs with $n \gtrsim 3$, will have larger (smaller) velocity
dispersions.  How the slope of this ``core fundamental plane'' 
might (not) be related to {\it dark matter}
will be dealt with in a forthcoming paper, as will an investigation into the
original ``Fundamental Plane'' (Fish 1963; 
Djorgovski \& Davis 1987).\footnote{Fish (1963) reported that,
  if homology holds, then $(L/R_{\rm e})^{1/2} \propto \sigma$, or $R_{\rm e}
  \propto \sigma^2 \langle I \rangle_{\rm e}^{-1}$.}$^,$\footnote{Djorgovski \& Davis (1987), also assuming
  $R^{1/4}$-model homology, reported $R_{\rm e} \propto \sigma^{1.39} \langle
  I \rangle_{\rm e}^{-0.90}$.}
A related question regarding what radius is
optimal for obtaining a tight plane, for both understanding ETGs and their
use as a distance estimator for improved studies of peculiar
velocity flows and cosmological parameter constraints (e.g.\ Johnson et
al.\ 2014; da Cunha et al.\ 2017),  will similarly be addressed, 
although some further insight is provided below.

Figure~\ref{Fig4} reveals why the Fundamental Plane works; the reason is
the same as why the ``core Fundamental Plane'' (Faber et al.\ 1997) works.  As
noted above, for $\mathfrak{M}_B \lesssim -19$~mag, the ETG light profiles are
such that they result in a bunching up of the $\mu_z$--$R_z$ and $\langle \mu
\rangle_z$--$R_z$ relations for different fractions $z$.  At $\mathfrak{M}_B
\lesssim -19$~mag, for a given mean surface brightness, the ETGs with brighter
magnitudes --- and thus higher S\'ersic indices $n$, and also larger stellar
velocity dispersions $\sigma$, plus redder colours and stronger metallicities
(see de Vaucouleurs \& Olson 1982 and de Carvalho \& Djorgovski 1989) --- will
have larger scale radii.  The introduction of such a third parameter ($n$ or
$\sigma$ or colour or Mg$_2$ index) acts to counter this trend and thus the
``Fundamental Plane'' has a reduced scatter than the $\mu_{\rm e}$--$R_{\rm
  e}$ and $\langle \mu \rangle_{\rm e}$--$R_{\rm e}$ relations.  Graham
(2002a) introduced a ``Photometric Plane'' in which $n$ was used rather than
$\sigma$ or Mg$_2$.

Contradictory to the claims in Bender et al.\ (1992) and Burstein et
al.\ (1997), the offset position of low-luminosity ETGs from the ``Fundamental
Plane'' defined by brighter ETGs (e.g.\ Kourkchi et al.\ 2012, their Figure~9)
is not clear evidence that dwarf galaxies are a different species of galaxy to
ordinary ETGs with a divide occurring at $\mathfrak{M}_B \approx -18$ mag.
The proof is simple: use of radii, and mean surface brightnesses, enclosing a
different percentage of the galaxy light will result in a different mass range
of ETGs appearing included or offset from the revised plane defined by the
brighter galaxies.

More recently, a 
division between bulges with $n\approx4$ and $n\approx1$ has been carried
over into research into massive black holes, with several authors referring to
low mass bulges as pseudobulges (e.g.\ Kormendy \& Ho 2013; Ho \& Kim 2014;
Heckman \& Best 2014; Somerville \& Dav\'e 2015).  As Davis et al.\ (2018)
points out, almost every spiral galaxy with a directly measured black hole mass
has been alleged to contain a pseudobulge.  A warning is therefore issued here
for studies searching for a plane within three dimensional spaces involving
black hole mass and bulge (or galaxy) effective radii (e.g.\ Marconi \& Hunt
2003; de Francesco et al.\ 2006; Barway \& Kembhavi 2007; Saglia et al.\ 2016;
van den Bosch et al.\ 2016).  The use of half light parameters is going to
result in a curved manifold rather than a plane, and the slopes of reported
planes will be a function of one's sample selection boundary.  Departures at
low black hole masses, from the plane constructed by galaxies with high black
hole masses, need not be a sign of different formation physics.  This should
be independent of the ($R_{\rm e}$-independent)-observation that the bulges of
late-type galaxies follow a different $M_{\rm bh}$--(bulge luminosity)
relation than the bulges of ETGs (Savorgnan et al.\ 2016; Davis et al.\ 2018,
2019; Sahu et al.\ 2019).

\section{Conclusions}

In the late 1980s, a growing body of increasingly
accurate photometry revealed that the $R^{1/4}$ model was not a universal law
for bright ETGs, and nor was the exponential model universal for faint ETGs
(see Capaccioli 1985, 1987, 1989; Schombert 1986; Davies et al.\ 1988).
Subsequent analyses revealed that S\'ersic's $R^{1/n}$ model both provided a
superior description of the galaxy light and unified the faint and bright ETGs
(e.g.\ Caon et al.\ 1993).  This not only eliminated the prime reason 
($R^{1/4}$ model versus exponential ellipsoid model) 
that had
led to thoughts and theories dividing faint and bright ETGs, but it also provided
the necessary clue for understanding the (curved) distributions in the
``effective'' parameter scaling diagrams.  However, this revolution ($R^{1/n}$
model) came too
late, experienced 10-15 years of considerable opposition due to perceptions
that the $R^{1/4}$ model was some kind of physical law of nature, and its
implications are yet to permeate much of the field. 
Nonetheless, it should be noted that
in the early 1980s, Binggeli, Bothun, Caldwell, Sandage and their
collaborators were correct in their understanding of the 
continuum between dwarf and normal ETGs, as understood 
by Gudehus and Oemler a decade before that. 

$R_{\rm e}$, or not $R_{\rm e}$?  Perhaps that  should be the question.  Given the
nature of ETG light profiles, those using
$R_{\rm e}$, or $\mu_{\rm e}$ or $\langle \mu \rangle_{\rm e}$, need to have
an awareness of what these quantities are, beyond the obvious superficial
meaning related to the radius containing 50 percent of the galaxies'
light.  Studies attaching a physical meaning to slopes or bends in diagrams
using these ``effective parameters'' need to be carefully considered 
given that a radius containing a different percentage of the galaxies’ light 
will yield different (equally physically-meaningless) slopes and bend-points.

\begin{acknowledgements}

I thank the Australian Research Council's funding scheme DP17012923 which
enabled me the opportunity to explain here why $R_{\rm
  e}$ has been such a misleading quantity.  I am additionally indebted to the
Swinburne Research Sabbatical Scheme.

\end{acknowledgements}

\end{document}